\documentclass[%
 reprint,
 table,
superscriptaddress,
 amsmath,
 amssymb,
 aps,
 prapplied,
]{revtex4-2}

\usepackage{graphicx}

\makeatletter
\newcommand*\bigcdot{\mathpalette\bigcdot@{.5}}
\newcommand*\bigcdot@[2]{\mathbin{\vcenter{\hbox{\scalebox{#2}{$\m@th#1\bullet$}}}}}
\makeatother

\usepackage{dcolumn}
\usepackage{bm}

\usepackage[T1]{fontenc}
\usepackage{amssymb}
\usepackage{mathtools}
\usepackage{MnSymbol}
\usepackage{bbold}
\usepackage[position=top, caption=false]{subfig}
\usepackage[table,x11names,dvipsnames,table]{xcolor}
\usepackage{tikz}
\usetikzlibrary {shapes.geometric}

\usepackage{amsthm}
\usepackage{amsmath}
\usepackage{mathrsfs}
\usepackage{dsfont}
\usepackage{physics}
\usepackage[thinc]{esdiff}
\usepackage{nicefrac}
\usepackage{xfrac}
\usepackage{algorithm}
\usepackage{algpseudocode}

\usepackage[pdftex, pdftitle={Article}, pdfauthor={Author}]{hyperref}
\usepackage[capitalise]{cleveref}
\creflabelformat{equation}{#2\textup{#1}#3}

\newtheorem{innercustomthm}{Result}
\newenvironment{customthm}[1]
  {\renewcommand\theinnercustomthm{#1}\innercustomthm}
  {\endinnercustomthm}

\begin{document}

\preprint{APS/123-QED}

\title{Time--adaptive single--shot crosstalk detector on superconducting quantum computer}

\author{Haiyue Kang}
 \email{haiyuek@student.unimelb.edu.au}
 \affiliation{School of Physics, University of Melbourne, VIC, Parkville, 3010, Australia}
\author{Benjamin Harper}
 \affiliation{School of Physics, University of Melbourne, VIC, Parkville, 3010, Australia}
 \affiliation{Quantum Systems, Data61, CSIRO, Clayton, 3168, Victoria Australia}
\author{Muhammad Usman}
 \affiliation{School of Physics, University of Melbourne, VIC, Parkville, 3010, Australia}
 \affiliation{Quantum Systems, Data61, CSIRO, Clayton, 3168, Victoria Australia}
\author{Martin Sevior}
 \affiliation{School of Physics, University of Melbourne, VIC, Parkville, 3010, Australia}
\date{\today}

\begin{abstract}
Quantum crosstalk which stems from unwanted interference of quantum operations with nearby qubits is a major source of noise or errors in a quantum processor. In the context of shared quantum computing, it is challenging to mitigate the crosstalk effect between quantum computations being simultaneously run by multiple users since the exact spatio-temporal gate distributions are not apparent due to privacy concerns. It is therefore important to develop techniques for accurate detection and mitigation of crosstalk to enable high-fidelity quantum computing. Assuming prior knowledge of crosstalk parameters, we propose a time-adaptive detection method leveraging spectator qubits and multiple quantum coherence to amplify crosstalk-induced perturbations. We demonstrate its utility in detecting random sparsely distributed crosstalk within a time window. Our work evaluates its performance in two scenarios: simulation using an artificial noise model with gate-induced crosstalk and always-on idlings channels; and the simulation using noise sampled from an IBM quantum computer parametrised by the reduced HSA error model. The presented results show our method’s efficacy hinges on the dominance of single-qubit coherent noise across channels, and the impact of angle mismatching is suppressed as spectator qubits increase. From simulation using real-device noise parameters, our strategy outperforms the previous constant-frequency detection method of Harper et. al.  [arXiv: 2402.02753 (2024)] in the detection success rate, achieving an average detection success probability of $0.852\pm 0.022$ (equally scaled noise channels) and $0.933\pm0.024$ (asymmetrically scaled noise channels) from 1 to 7 crosstalk counts.

\end{abstract}

\maketitle


\section{Introduction}\label{sec:introduction}
Quantum computing is an emerging paradigm for computing and various quantum algorithms have been proposed to be highly efficient tools to reduce the computational complexity of problems that were impossible or expensive to solve in classical computing, as the size of the problem increases \cite{shor, grover, VQE, QPE, QML, QML2, quantum_chemistry, quantum_bio, quantum_finance, quantum_algorithms_summary, nielsen_chuang, Max_West_adv_QML_1, Max_West_adv_QML_2, Zehang_Wang_QML_1}. Despite theoretical proofs confirming the computational advantage of quantum algorithms, their successful implementation in practice is heavily dependent on the performance of quantum computers \cite{surface_codes, wallman2016noise}, which are currently under development. The current generation of quantum processors offers access to a limited number of qubits and their performance is plagued by errors or noise which reduce their capability to execute large-scale computational problems. In terms of physical representation, the noise that arises during quantum computation can be categorised into different channels, such as coherent errors, depolarising / Pauli errors, dephasing errors, and damping errors \cite{nielsen_chuang, noise_channels_summary_2, HSA_model_paper}. A major source of noise that causes such errors is crosstalk \cite{Xtalk_mitigation_via_circuit_optimisation, zhou2023quantum}, which by definition is the unwanted correlated dynamics between qubits as a result of operations that violate \textit{independence} (context-dependent classical correlation) or \textit{locality} (unwanted entanglement) \cite{crosstalk_definition}. This work introduces a new technique for the detection and mitigation of crosstalk noise enables higher fidelity in quantum gates and improves the scalability of quantum algorithms on real quantum computers.

Although crosstalk can cover various phenomena \cite{crosstalk_definition}, in this work, we use `crosstalk' to refer to the gate-induced errors that violate \textit{independence}, which occurs when an operation acting on some action qubit(s) ($\mathcal{H}_A$ to denote its Hilbert space) of a quantum system leaks its control signal to other target qubits ($\mathcal{H}_T$ to denote its Hilbert space) and induces an unwanted evolution. This type of crosstalk is observed in atomic-architecture quantum processors \cite{gate_ctrl_leak3} and superconducting qubits \cite{gate_ctrl_leak, gate_ctrl_leak2}, which was exhibited as the combinations of coherent rotations, contractions, and transformations of Bloch vectors on the other qubits. Additionally, as the number of qubits in a quantum processor grows, it is expected that tasks owned by different users are likely to be executed simultaneously on the same quantum processor in different subsets of the device. In such a shared environment, crosstalk becomes a serious problem \cite{Ben_crosstalk, crosstalk_adverserial_attack_1, crosstalk_adverserial_attack_2}. Addressing the problem of crosstalk in such scenarios is a challenging task because standard crosstalk-noise mitigation protocols cannot be implemented, as users have no prior knowledge of when and where other users would place the crosstalk-inducing gates relative to their circuits. Furthermore, the non-local property of crosstalk noise allows potential hostile users to perform adversarial fault injection into its neighbouring circuits by taking advantage of the gate-induced crosstalk, making it vulnerable to hazardous attacks. Consequently, it is important to develop new methods to efficiently and robustly detect and mitigate crosstalk in quantum processors.

Several strategies have been proposed to characterise and suppress crosstalk noise within a quantum circuit. From the hardware perspective, crosstalk noise can be minimised by optimising control systems such as careful tuning of couplers' frequencies to achieve destructive interference \cite{Xtalk_mitigation_via_tunable_coupling}, and combining qubits with opposite anharmonicity \cite{Xtalk_mitigation_via_opposite_anharmonicity}. On the software level, solutions include the optimisation of the gate sequence compiler \cite{Xtalk_mitigation_via_circuit_optimisation}, gate cancellation \cite{gate_ctrl_leak3}, dynamical decoupling \cite{Xtalk_mitigation_via_dd}, qubit mapping via reinforcement learning or constant-period measurements on a spectator qubit \cite{Ben_crosstalk}. However, despite their utility, they all assume that the exact spatiotemporal distribution of the quantum gates that emit crosstalk noise is known, which is generally not the case, especially under privacy constraints.\vspace{7pt}

In this paper, we present an adaptive strategy to address the aforementioned challenges by amplifying and detecting weak crosstalk perturbations -- a common trick in quantum sensing \cite{quantum_sensing_summary} -- through multiple quantum coherences (MQC)\cite{MQC, MQC2, Fidel_paper} of modified GHZ states, which we call Crosstalk Spectating Multiple Quantum Coherences (CSMQC). To minimise disturbance of data qubits that we are protecting from measurements, detection is achieved through spectator qubits, which is also widely used in general error correction \cite{phased_array_spectator_qubits, RTP_noise_spectator_qubits}. In principle, within a single shot of the circuit, this design allows users to detect the presence of crosstalk given crosstalking gates distributed arbitrarily over time on nearby qubits. The remaining contents are structured as follows: In \Cref{sec: Crosstalk noise amplification}, we highlight the novelty of this strategy and outline the assumptions, especially the dominance of the single-qubit Hamiltonian channel as an important condition. In \Cref{sec: artificial model results}, we evaluate the performance of such a technique subject to idle noise and other subdominant crosstalk channels using a synthetic model. In \Cref{sec: results}, we investigate its performance using a classical simulator with the reduced HSA noise model \cite{HSA_model} with the noise parameters generated from benchmarking of IBM quantum devices. In the simulation, we find an average detection success probability of $0.852\pm0.0224$ for interference counts ranging from 1 to 7 using equally-amplified noise parameters benchmarked from $ibm\_hanoi$. We also evaluate the performance of CSMQC in improving the fidelity of data qubits by filtering the shots from circuits where crosstalk is detected using spectator qubits. In both idle tomography and random circuit experiments, we report significant improvements in the final fidelities of the computation after applying CSMQC and outperform the technique of constant-period measurement  \cite{Ben_crosstalk} for crosstalk counts less than the number of delayed layers between two resets. This method offers a potential solution for detecting crosstalk with relatively high precision compared to previous strategies on near-term superconducting transmon qubit quantum computers, regardless of whether the perturbation is intentionally produced. However, in the long term, the utility of such a technique is expected to decrease due to expected improvements in quantum computers. However, this also reduces the necessity of crosstalk detection as the impact of the crosstalk interference is reduced relative to other sources of noise.\raggedbottom
\begin{figure*}[hbtp]
    \subfloat[\label{fig:MQC detector}Crosstalk-Spectating Multiple Quantum Coherence (CSMQC)]{
    \:\includegraphics[width=1\linewidth]{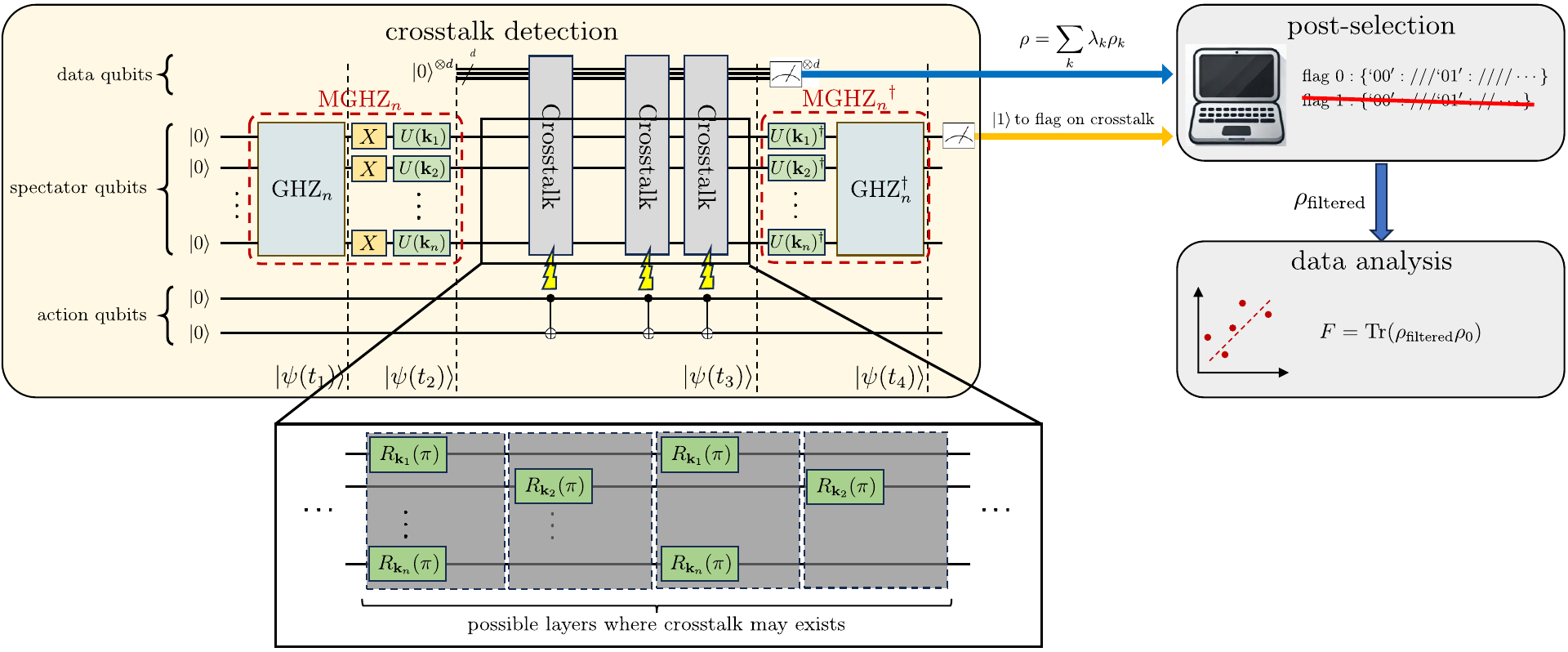}\:
    }\\
    \subfloat[\label{fig:rot axis mapping}rotation axis mapping]{
    \:\includegraphics[width=0.28\linewidth]{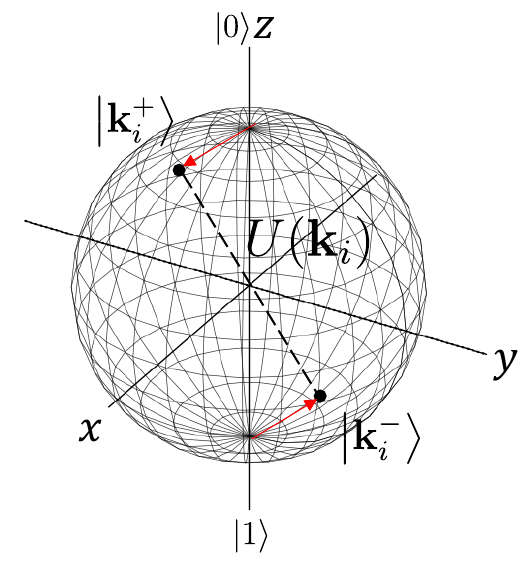}\:
    }
    \subfloat[\label{fig:multiple ghz cycles}multiple simultaneous spectator qubits detectors with multiple detection time frames]{
    \:\includegraphics[width=0.7\linewidth]{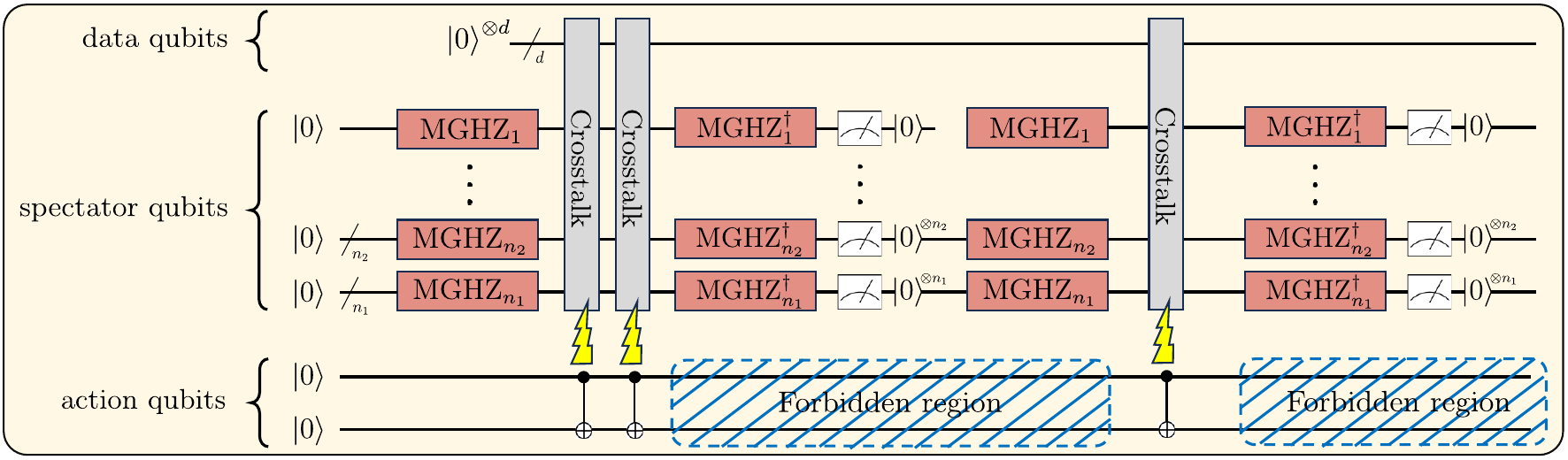}\:
    }
    \caption{\textbf{(a)} General workflow of the CSMQC circuit. Before performing any calculation in data qubits, a set of $n$ spectator qubits satisfying the condition \Cref{eq: total angle condition} is prepared into a GHZ state at time $t_1$, which is further modified into $\ket{\text{MGHZ}_n(\bm{K})}=\left(\bigotimes\limits_{i=1}^{n}{U(\bm{k}_i)}\right)\ket{\text{GHZ}_n}$ so that \textbf{(b)} each qubit $i$ has its component aligned ($\ket{\bm{k}_i^{+}}$) or anti-aligned ($\ket{\bm{k}_i^{-}}$) with the predetermined crosstalk noise rotation axis $\bm{k}_i$ after $t_2$(see \Cref{eq: MGHZ} in \Cref{appendix: GHZ evolution}). The layer of $X$ gates inserted to $\text{MGHZ}_n$ acts as a single dynamical decoupling pulse that stabilises the state and suppresses the noise generated during the preparation and inversion that is not due to crosstalk generated in action qubits (see details in \Cref{appendix: DD}). After the modified GHZ state preparation, crosstalk generated by CNOT gates from neighbouring action circuits are sensed by the spectator qubits and data qubits in the form of rotations around the predetermined axes. One may implement additional layers of dynamical decoupling (DD) pulses after the preparation circuit (denoted by the enlarged circuit diagram of spectator qubits) during the crosstalk detection to suppress the noise contribution not from single-qubit coherent rotations. As such, the axes of these pulses are chosen to be the same as the axes $\bm{k_i}\in\{\bm{K}\}$. Offsets are added to maximise the level of anti-commutation with nearest-neighbour two-qubit coherent rotations. Once the computation is done on the data qubits after $t_3$, a set of inversion gates that undo the preparation of $\ket{\text{MGHZ}_n(\bm{K})}$. Since the final state $\ket{\psi(t_4)}$ directly depends on the collective net rotated angle induced by the crosstalk (see \Cref{eq:|1>probability}), an alarm on the existence of crosstalk is then determined from the measurement of the first spectator qubit for an odd number of crosstalk gate operations. 
    After the detection, the measured outcomes from the data qubits are post-selected by discarding the shots with $\ket{1}_n$ flagged on the last spectator qubit. The filtered outcomes can be used to reconstruct its density operator for fidelity tests against the ideal state or used in other applications. \textbf{(c)} Simultaneous sets of spectator qubits with different numbers $n_1, n_2,\cdots, 1$ can be placed in parallel. The spectator qubits chosen for each set have the total rotated angle per crosstalk perturbation equal to $\pi,\frac{\pi}{2},\cdots$ responsible for detecting the crosstalk with gate counts in $\{1,3,5\cdots\}, \{2,6,,10\},\cdots$ respectively. These together cover all possible numbers of perturbations up to a certain threshold until there is only one spectator qubit in one set. The $\text{MGHZ}$ preparation and inversion circuits can be performed multiple times with resetting after each cycle, thus enabling it to detect crosstalk in multiple time frames between the preparation and inversion circuits. However, exactly during the preparations and inversions, the execution of crosstalking gates must be forbidden or cleverly shifted/compiled at the machine level as any crosstalk they induce cannot be detected properly in the prescribed way.}
    \label{fig: fig1}
\end{figure*}

\section{Noise amplification and detection}\label{sec: Crosstalk noise amplification}
\subsection{Conditions and Assumptions}\label{subsec: conditions and assumptions}
Before exploring the details of CSMQC, it is necessary to establish a few assumptions and conditions regarding crosstalk noise, which are essential for our crosstalk mitigation protocol to function normally \cite{crosstalk_definition}.

First, we require that the noise generated by the crosstalk is dominated by single-qubit Hamiltonian error channels, a common assumption in recent literature \cite{phased_array_spectator_qubits, zhou2023quantum, majumder2020real}, in a quantum system consisting of source and target qubits of crosstalk in the Hilbert space $\mathcal{H}_A\otimes\mathcal{H}_T$. We are not claiming that our assumption
applies universally, but it is suitable in our context. Other channels will be discussed alongside single-qubit Hamiltonian error channels, because it is the benchmark for evaluating the significance of the other channels. Secondly, the noise must be ``global'' for all qubits regardless of their distance from the crosstalk source. In other words, the radius of the noise effect must be at least comparable to the size of the entire quantum system. However, the magnitude of the noise does not need to be completely uniform. Therefore, the evolution of the reduced density operator $\rho\in \mathcal{H}_T$ of the quantum system on target qubits can be expressed as rotation angles for each qubit.
Mathematically,
\begin{equation} \label{eq: mixing channels}
    \mathcal{E}_{\text{Xtalk}}(\rho)=(1-\epsilon)\mathcal{E}_{\text{single-H}}(\rho)+\epsilon \mathcal{E}_{\text{other}}(\rho)
\end{equation}
where $\mathcal{E}_{\text{single-H}}$ and $\mathcal{E}_{\text{other}}$ are both completely-positive, trace-preserving (CPTP) mappings for single-qubit Hamiltonians and other channels respectively. $\epsilon$ is the mixing parameter between noise channels and we expect $\epsilon \ll 1$ for modern quantum computers. The single-qubit Hamiltonian channel explicitly gives the unitary evolution as
\begin{equation}
    \mathcal{E}_{\text{single-H}}(\rho)=e^{-iH}\rho e^{iH},
\end{equation}
where
\begin{equation}
    H=\sum_{i}{\bm{k}_{i}\cdot\bm{\sigma}_i}
\end{equation}
is the instantaneous Hamiltonian generated by crosstalk that results in a rotation around the Bloch sphere of qubit $i$ around the axis $\bm{k}_i=(k_x,k_y,k_z)_i$, with  rotation angle $\theta=2\abs{\bm{k}_i}$, and $\bm{\sigma}_i=(X,Y,Z)_{i}^{T}$ is the Pauli-vector. Therefore, the evolution of the total quantum system $\zeta\otimes\rho\in\mathcal{H}_A\otimes\mathcal{H}_T$ is expressed as
\begin{equation}
    \zeta\otimes\rho\rightarrow U\zeta U^{\dagger}\otimes\mathcal{E}_{\text{Xtalk}}(\rho),
\end{equation}
where $U$ is the quantum gate acting on the source qubits that cause crosstalk. Despite being necessary for implementing CSMQC, this might not be satisfied on real quantum computers. For example, a large number of crosstalk models assume that the noise decays exponentially with respect to its topological distance from the source \cite{gate_ctrl_leak3, pogorelov2021compact}, which conflicts with the requirement on the radius of the noise in our work. This implies that a small upper bound on the number of spectator qubits useful for noise amplification might be inevitable in practice.

In addition, to set the spectator qubits in the desired orientation, we require that the crosstalk noise is modelled and defined with an accurate parameterization from prior benchmarks. It is important to note that this is not the same as knowing when crosstalk will occur during circuit execution.
\subsection{Noise sensing via Modified GHZ state on spectator qubits}\label{subsec: GHZ noise sensing}
Capturing the existence of crosstalk noise inevitably requires one to measure the qubits subject to such perturbation. However, the principle of measurement-collapse and non-cloning theorem prevents one from conducting multiple measurements on the same quantum state without irreversably altering the state. To overcome such a dilemma, spectator qubits, which are separated or only weakly coupled to the actual data qubits, serve as sensors to the noise without disturbing the data qubits \cite{takita2016demonstration, majumder2020real}. However, the rotations generated by a single instance of crosstalk are often small \cite{Ben_crosstalk}, making it difficult to distinguish with a single measurement. Therefore, we design Crosstalk Spectating Multiple Quantum Coherences (CSMQC) as a software-level approach to detect this noise using the amplified sensitivity of the GHZ state prepared on spectator qubits inspired by Multiple Quantum Coherences (MQC) \cite{MQC, MQC2, Fidel_paper}. 

We demonstrate the overall process of CSMQC as outlined in \Cref{fig:MQC detector}. Initially, there is a set of $d$ data qubits ($T_D$), $n$ spectator qubits ($T_S$), and two action qubits ($A$) that use the CNOT gate as the source of crosstalk. The number $n$ is chosen to satisfy the condition
\begin{equation}\label{eq: total angle condition}
    \theta\coloneqq\sum\limits_{i=1}^{n}{\delta_i}\approx \pi,
\end{equation}
where $\delta_i$ is the angle rotated on qubit $i$ if an instance of crosstalk is triggered from action qubits. In practice, due to the limited choice of spectator qubits and the randomness of $\delta_i$, it is impossible for the sum to be exactly equal to $\pi$. Instead, we select the spectator qubits in such a way that their total rotation angle best matches \Cref{eq: total angle condition}. The detailed evolution of the procedures in \Cref{fig:MQC detector} is explained in \Cref{appendix: GHZ evolution}. To address the issue of total angle mismatching that could occur in practice, we quantitatively simulate its potential impact in \Cref{sec: artificial model results}.

At the end of the circuit, by leveraging multiple quantum coherence that enhances the perturbation signal by summing up the rotated phases applied to each spectator qubit, we find that the probability of measuring $\ket{1}$ on the last qubit of spectator qubits is a simple cosine function of the number of perturbations $m$ and the sum of angles $\theta$ rotated by crosstalk on all spectator qubits. Specifically, the state on the last spectator qubit with label $n$ has the probability of measuring 1 equal to
\begin{equation}\label{eq:|1>probability}
    P(\ket{1}_n)= \frac{1-\cos(m\theta)}{2}.
\end{equation}
In the ideal situation where $m\theta=\pi \mod{2\pi}$, the above equation reduces to
\begin{equation}
 P(\ket{1}_n)=
    \begin{cases}
    1, m\in \mathbb{N}_{\text{odd}}  \\
    0, m\in \mathbb{N}_{\text{even}}.
    \end{cases}
\end{equation}
We quantitatively addressed the impact of other noise channels in the simulation completed in \Cref{sec: results}.

In the ideal situation, crosstalk perturbations with an odd number of gate counts can be fully detected regardless of their frequency or temporal distribution. Under this setting, the results measured on the data qubits after multiple shots can be treated as measurements from a mixed state constituting the ideal crosstalk-free state $\rho_0$ and the contaminated states affected by crosstalk:
\begin{equation}
    \rho_{T_D}=\lambda_{0}\rho_0+\sum_{k\in X}{\lambda_k\mathcal{E}_{\text{Xtalk}}(\rho_0)_k},
\end{equation}
where $X$ is the set of possible time configurations of crosstalk with odd counts, $\mathcal{E}_{\text{Xtalk}}(\rho_0)_k$ is the underlying state density operator perturbed by crosstalk given configuration $k$, and $\lambda_k$ is the probability of $k$ satisfying the Born probability rule $\sum_{k\in X\cup\{0\}}{\lambda_k}=1$. In an ideal situation, the crosstalk-free shots can be distinguished from the noisy shots by the result measured from the last spectator qubit, in which the state of the data qubit depends directly on the state of the last spectator qubit,
\begin{equation}
    \rho_{T_{S_n}+T_D}=\ket{0}\bra{0}\otimes\lambda_{0}\rho_0+\ket{1}\bra{1}\otimes\sum_{k\in X}{\lambda_k\mathcal{E}_{\text{Xtalk}}(\rho_0)_k}.
\end{equation} 
Therefore, the counts measured from the crosstalk-free pure state on the data qubits can be obtained via post-selection that filters the results flagged by $\ket{1}_n$:
\begin{equation}
    \rho_{T_D (\text{filtered})}=\rho_0.
\end{equation}
As outlined in \Cref{fig:MQC detector}, the precision of the filtering process can be tested by computing the fidelity of the filtered state against the ideal state of the data qubits. For the case where the number of crosstalk gate counts is even, one can add additional sets of spectator qubits to the circuit with a total rotated angle per gate tuned to equal to $\frac{\pi}{2}, \frac{\pi}{4},\cdots$ as in \Cref{fig:multiple ghz cycles}. This enables the detector to report the existence of crosstalk noise for perturbation counts in sets such as $\{2,6,10,\cdots\}, \{4,12\cdots\},\cdots$ etc. In practice, the total number of available layers to insert a crosstalk gate between the modified GHZ preparation and inversion circuit must be limited, as the choice of spectator qubits reaches the threshold when the required angle of rotation to detect a particular number of perturbations is smaller than the smallest possible angle rotated by a single perturbation. To access more circuit depth, a possible solution is to reset the spectator qubits and allow the preparation/inversion cycle to be performed multiple times to extend detection time frames, while any gates acting on action qubits that cause crosstalk are forbidden exactly during the $\text{MGHZ}_n$ preparation and inversion process (see details in \Cref{fig:multiple ghz cycles}). This can be achieved at the machine level where a set of spectator qubits for crosstalk detection is assigned to each user and crosstalk gates are compiled to stagger with the preparation and inversion circuits on the spectator qubits.


\section{Simulation with artificial model}\label{sec: artificial model results}
\subsection{Setup}\label{subsec: setup}
To examine the performance of our CSMQC strategy under other perturbations except single-qubit Hamiltonian channels induced by crosstalk, we set up a synthetic noise model to mimic the noise behaviour of the IBM superconducting quantum computers. The model is comprised of two parts: idle noise that is always activated in each circuit layer iteration, which consists of dephasing, amplitude damping, always-on nearest-neighbour ZZ coupling, and crosstalk noise that is only activated when it is generated via a corresponding circuit layer, which consists of single and two-qubit coherent rotations. In the Kraus-operators form, the channels that contribute to the $n$-qubit idle noise are given by
\begin{equation}
\begin{aligned}
    \mathcal{E}_{\text{Dephase}}[\bigcdot] &= \sum\limits_{\bm{k}}{D_{\bm{k}}\bigcdot D_{\bm{k}}^{\dagger}}\\
    \mathcal{E}_{\text{Amplitude damping}}[\bigcdot] &= \sum\limits_{\bm{k}}{A_{\bm{k}}\bigcdot A_{\bm{k}}^{\dagger}}\\
    \mathcal{E}_{\text{ZZ coupling}}[\bigcdot] &= U_{ZZ} \bigcdot U_{ZZ}^{\dagger}
\end{aligned}
\end{equation}
where $\bm{k}=(k_1,k_2\cdots,k_n)\in\{0,1\}^{\otimes n}$, $D_{\bm{k}}=\bigotimes\limits_{i=1}^{n}{D_{k_i}}$, such that $D_0=\sqrt{1-p_{\text{D}}}\mathds{1}$ and  $D_1=\sqrt{p_D}Z$ are the operation elements that maps a state $\rho$ into the noisy state $\mathcal{E}_{\text{Dephase}}[\rho]$ with probability $p_D$. Similarly for the amplitude damping channel, $A_{\bm{k}}=\bigotimes\limits_{i=1}^{n}{A_{k_i}}$, $A_0=\left(\begin{matrix}
   1 & 0  \\
   0 & \sqrt{1-p_A}  \\
\end{matrix}\right)$ and $A_1=\left(\begin{matrix}
   0 & \sqrt{p_A}  \\
   0 & 0  \\
\end{matrix}\right)$. The transition probability $p_A$ and $p_D$ are calculated based on the measured relaxation time $T_1$ and coherence time $T_2$ from an IBM quantum computer, as well as the timestep $\tau$ of the two-qubit gate time. Hence, the mappings $\mathcal{E}_{\text{Dephase}}$ and $\mathcal{E}_{\text{ZZ coupling}}$ precisely describe the idle noise in the timeframe equivalent to the two-qubit gate time.  Specifically,
\begin{equation}
\begin{aligned}
    p_A &= 1-e^{-2\gamma_A t}=1-e^{-t/T_1}\\
    p_D &= \frac{1-e^{-2\gamma_D t}}{2}=\frac{1-e^{-t/T_1}}{2}
\end{aligned}
\end{equation}
are the standard relationships \cite{Pauli_Lindblad_paper, long_range_gates, nielsen_chuang} between the transition probability $p$ and the coefficient of the error generators $\gamma$ in the Lindbladian representation given evolution timestep $\tau$. In this model, we arrange the qubits in a one-dimensional Ising chain with nearest-neighbour couplings. Thus, the operator responsible for the always-on nearest-neighbour ZZ coupling is denoted as 
\begin{equation}
    U_{ZZ} = \exp\left(-i\alpha\sum\limits_{j=1}^{n-1}{Z_j Z_{j+1}}\right),
\end{equation}
where $\{1,2\cdots,n\}$ labels neighbouring qubits with consecutive indices, and $\alpha$ is the rotation rate. These idle noise channels can be mapped to the super-operator representation in the following form, 
\begin{equation}
    \mathcal{E}[\bigcdot]\rightarrow e^{\mathcal{L}}[\bigcdot]\coloneqq e^{\mathcal{L}}\vert\bigcdot\rrangle,
\end{equation}
where $\vert\bigcdot\rrangle=\text{vec}(\bigcdot)$ stands for vectorising the density operator into a column vector with dimension $4^n\cross 1$, and $\mathcal{L}$ is the Lindbladian generator of the error channel $\mathcal{E}$ in the representation of $\vert\bigcdot\rrangle$. Hence, the overall mapping subject to the idle noise is given by
\begin{equation}
    \vert\rho\rrangle\rightarrow e^{\mathcal{L_\text{Dephase}}+\mathcal{L_\text{Amplitude damping}}+\mathcal{L_\text{ZZ coupling}}}\vert\rho\rrangle.
\end{equation}

On the other hand, the Kraus operators of the channels that contribute to the crosstalk noise are simply described by
\begin{equation}
    \mathcal{E}_{\text{Xtalk}}[\bigcdot]= U_{\text{Xtalk}} \bigcdot U_{\text{Xtalk}}^{\dagger},
\end{equation}
where $U_{\text{Xtalk}} = \exp\left(-i\theta\sum\limits_{i=1}^{n}{\bm{\hat{n}}_i\cdot\bm{\sigma}_i}-i\phi\sum\limits_{i,j}{\bm{\hat{m}}_{ij}\cdot\bm{\sigma}_{ij}}\right)$, $\bm{\sigma}_i=(X_i,Y_i,Z_i)$, $\bm{\sigma}_{ij}=(X_i X_j,X_i Y_j,\cdots, Z_i Z_j)$ with subscripts denoting the label of the qubits is acting on. Each $\bm{\hat{n}}_i$ and $\bm{\hat{m}}_{ij}$ is a random unit vector with cardinality 3 and 9 respectively. $\theta$ and $\phi$ denote the rate of single-qubit and two-qubit coherent rotation, respectively. In our simulation, we set $\theta=\pi/\Tilde{n}$ where $\Tilde{n}$ is the number of spectator qubits used to satisfy the condition in \Cref{eq: total angle condition}. We use $\bm{\hat{n}}_i$ as rotation axes when preparing for $U(\bm{k}_i)$ and $\ket{\text{MGHZ}_{\Tilde{n}}(\bm{K})}$ as in \Cref{fig:MQC detector}. 

We also implement additional layers of dynamical decouplings into the CSMQC circuit to optimise its performance. As shown in \Cref{fig:MQC detector}, to stabilise the modified GHZ state from other perturbations while keeping the phases induced by single-qubit coherent rotations intact, the axes of decoupling pulses are chosen to be the same as the axes $\{\bm{K}\}$ and the pulse counts must be even per spectator qubit. To maximise the chance of anti-commutation with the noise represented by nearest-neighbour two-qubit operators, we also set an offset such that the pulses never appear simultaneously on any nearest-neighbour pairs \cite{dd_offset}. 

\begin{figure*}[t!]
    \subfloat[\label{fig: false alarm simulation a}]{\:\includegraphics[width=0.46\linewidth]{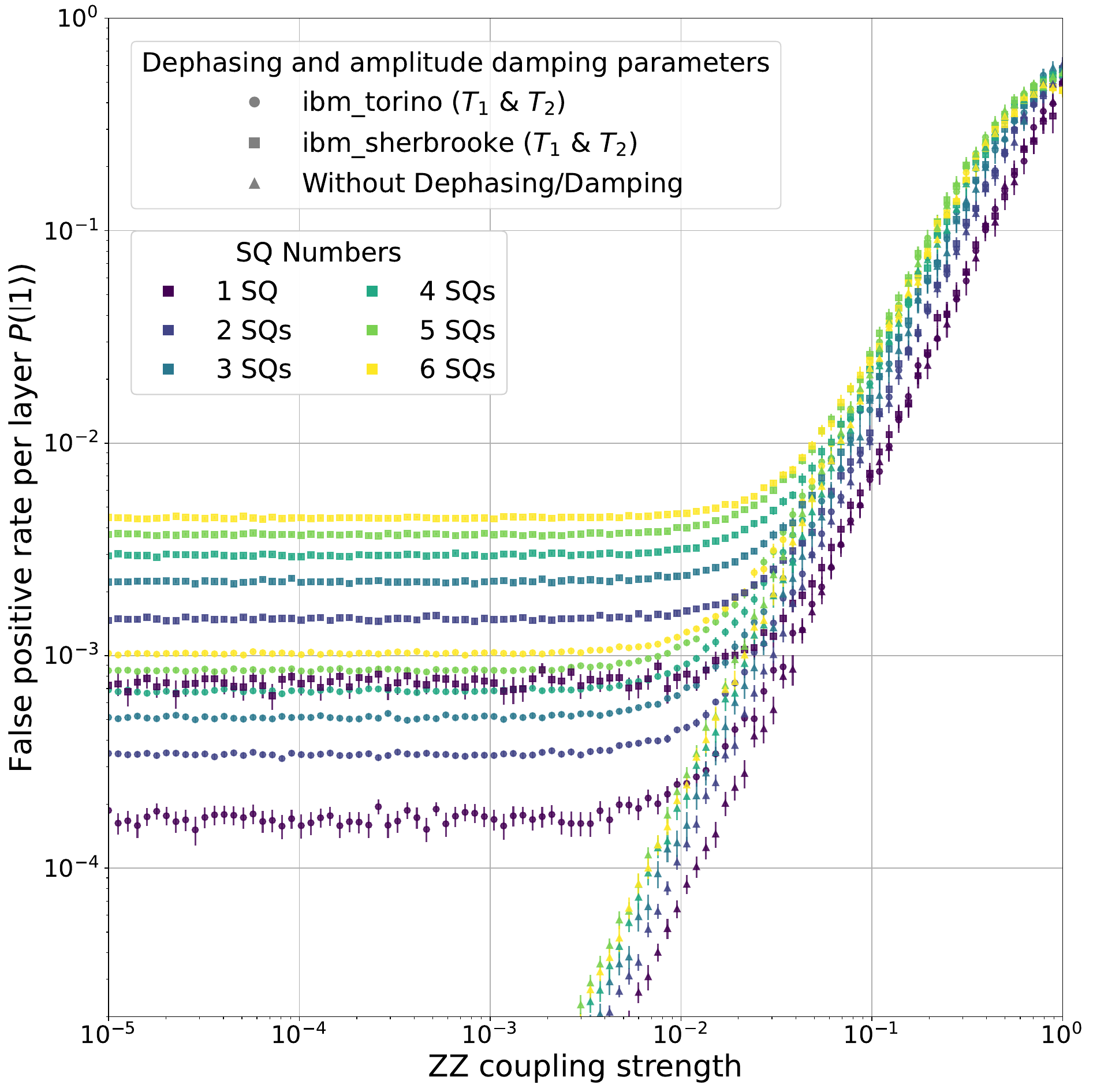}\:}\subfloat[\label{fig: false alarm simulation b}]{\:\includegraphics[width=0.46\linewidth]{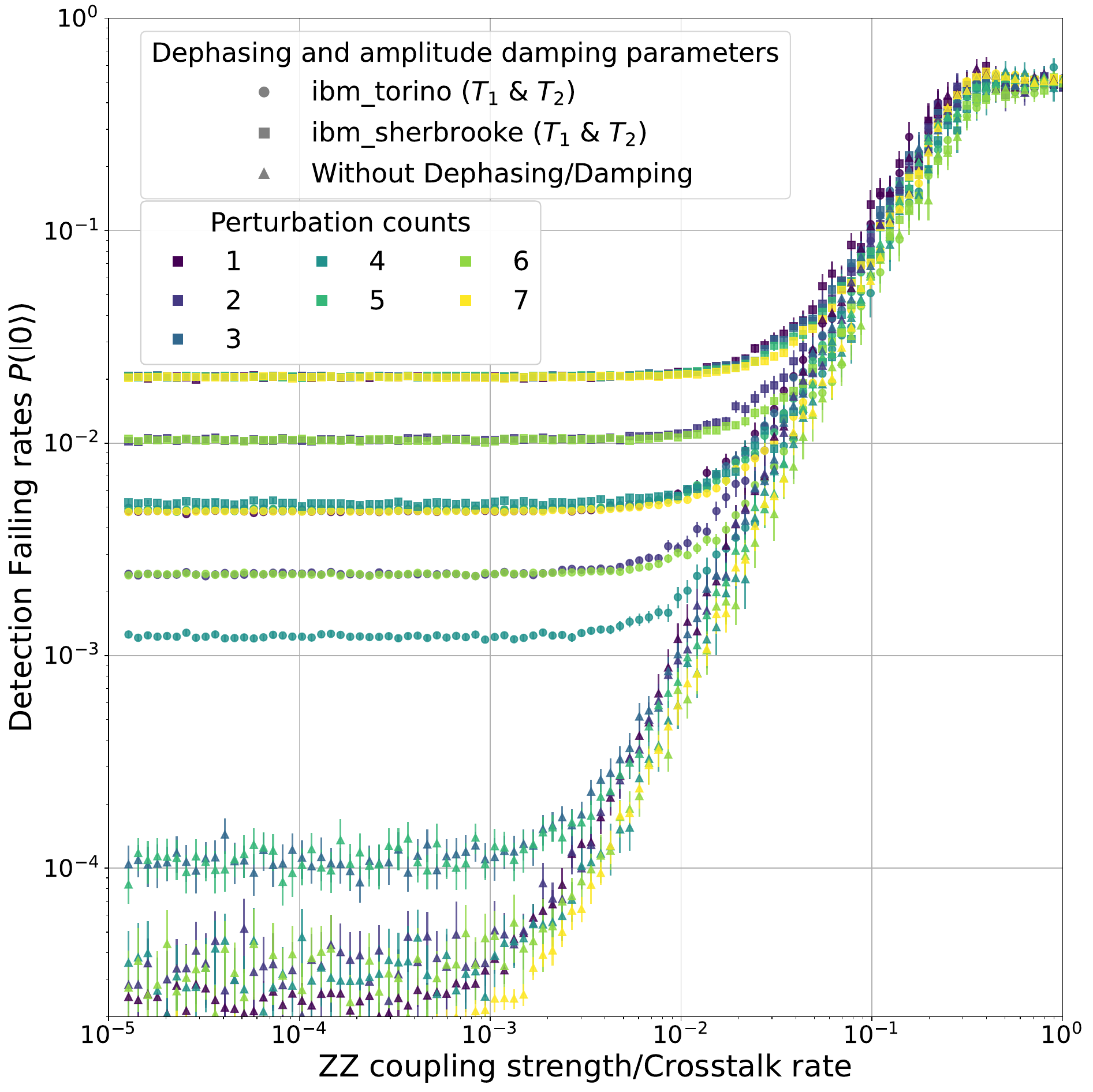}\:}\\
    
    \subfloat[\label{fig: false alarm simulation c}]{\:\includegraphics[width=0.45\linewidth]{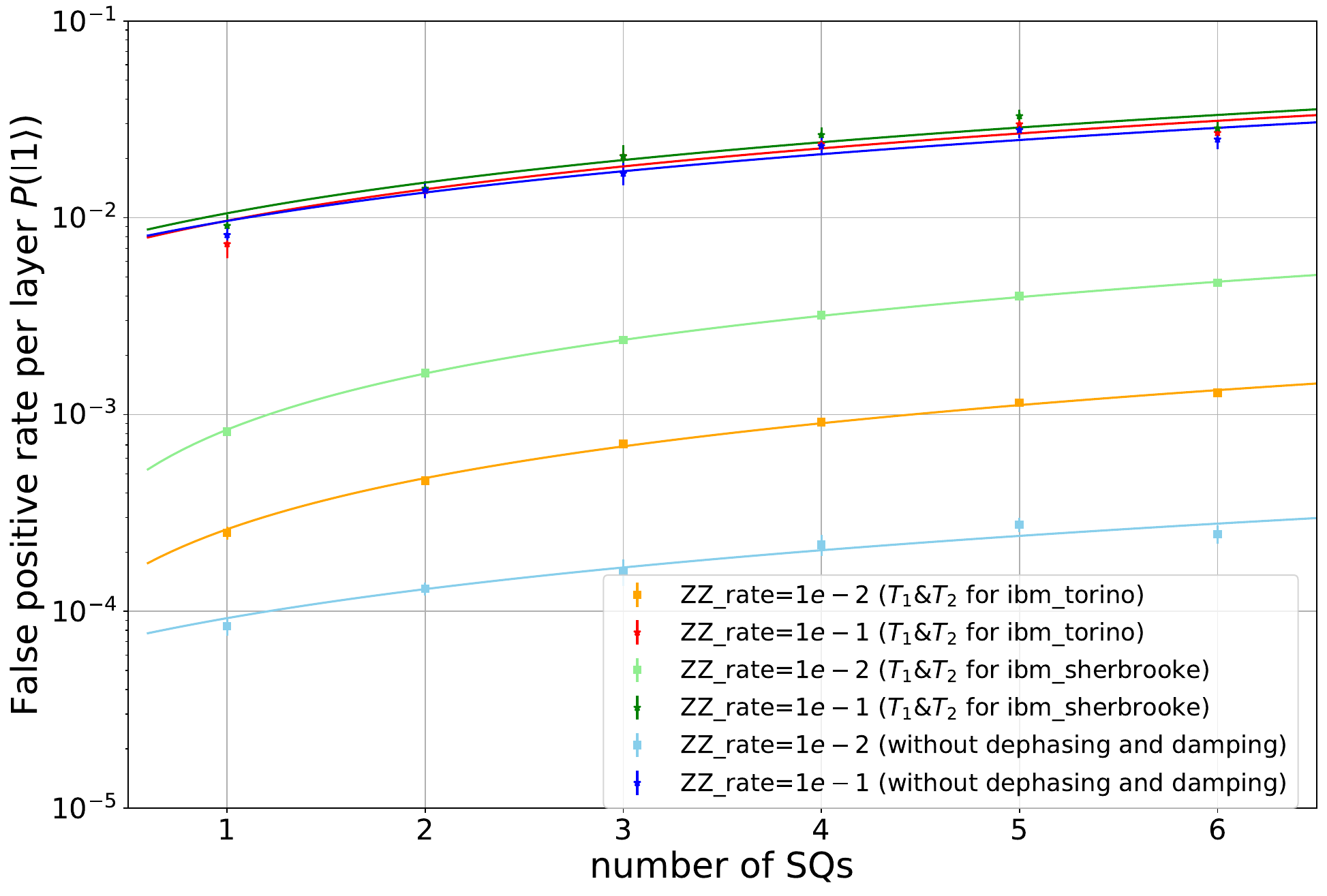}\:}
    \subfloat[\label{fig: false alarm simulation d}]{\:\includegraphics[width=0.45\linewidth]{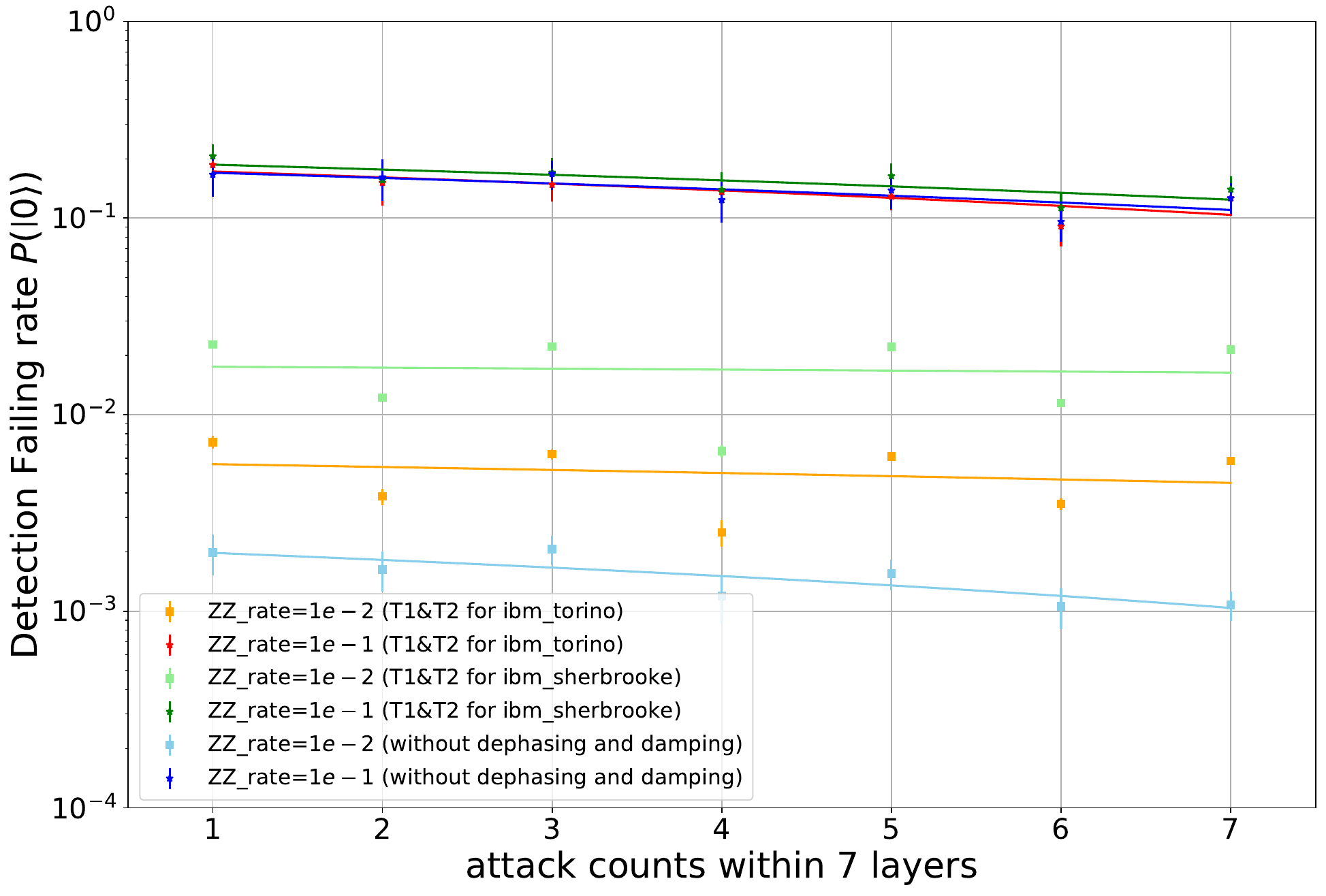}\:}
    
    \caption{\textbf{(a)} Graphics showing the false positive rate of CSMQC versus the always-on ZZ coupling strength typically in the IBM quantum computers on a log-log plot. Each colour represents a given number of spectator qubits using the IBM quantum computer's dephasing and amplitude damping parameters. Each data point averages over 50 samples and the error bars use two times the standard error. \textbf{(b)} Graphics showing the false negative probability of CSMQC, but plotted against the ratio between the ZZ coupling strength to the angle of rotation produced per crosstalk. \textbf{(c)} The false positive rate plotted against the number of spectator qubits given specific ZZ coupling strength with the linear fit ($y$-axis shown in logarithmic scale). Two typical ZZ coupling strengths ($1\cross 10^{-2}$ and $1\cross 10^{-1}$) are chosen for comparison. \textbf{(d)} The false negative probability versus the number of perturbations from 1 to 7. Similar to \textbf{(a)} and \textbf{(c)}, \textbf{(b)} and \textbf{(d)} uses two times the standard error as the half-length of the error bars.}
    \label{fig: false alarm simulation}
\end{figure*}

\begin{figure*}[t]
    \subfloat[\label{fig: torino 3 attacks Xtalk pi4 6q colorplot} Dephasing and amplitude damping using $T_1,T_2$ from $ibm\_torino$, $\theta=\frac{\pi}{4}$]{\includegraphics[width=0.33\linewidth]{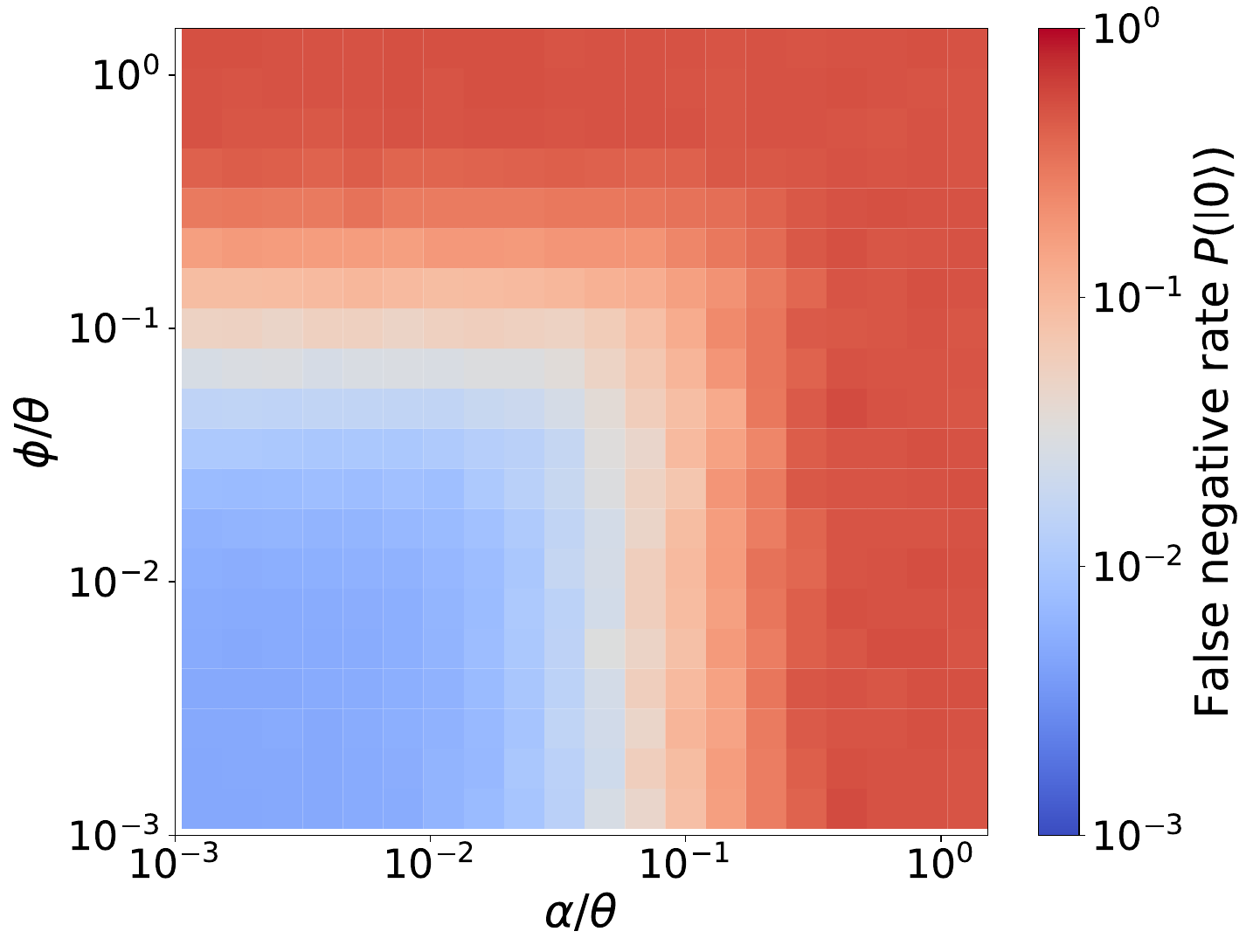}}
    \subfloat[\label{fig: pure 3 attacks Xtalk pi4 6q colorplot}No depahsing or amplitude damping, $\theta=\frac{\pi}{4}$]{\includegraphics[width=0.33\linewidth]{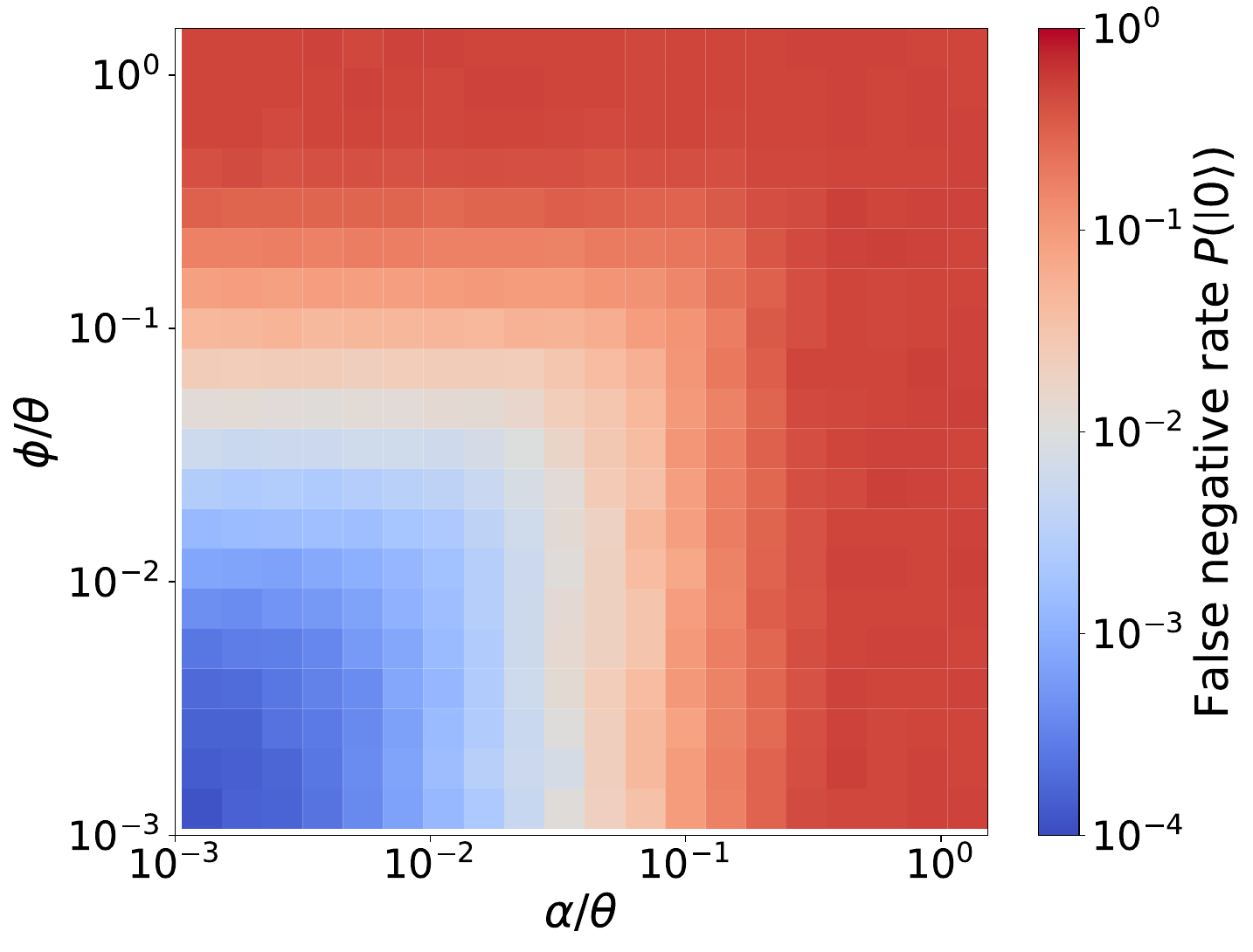}}
    \subfloat[\label{fig: pure 3 attacks Xtalk pi8 8q colorplot}No depahsing or amplitude damping, $\theta=\frac{\pi}{8}$]{\includegraphics[width=0.33\linewidth]{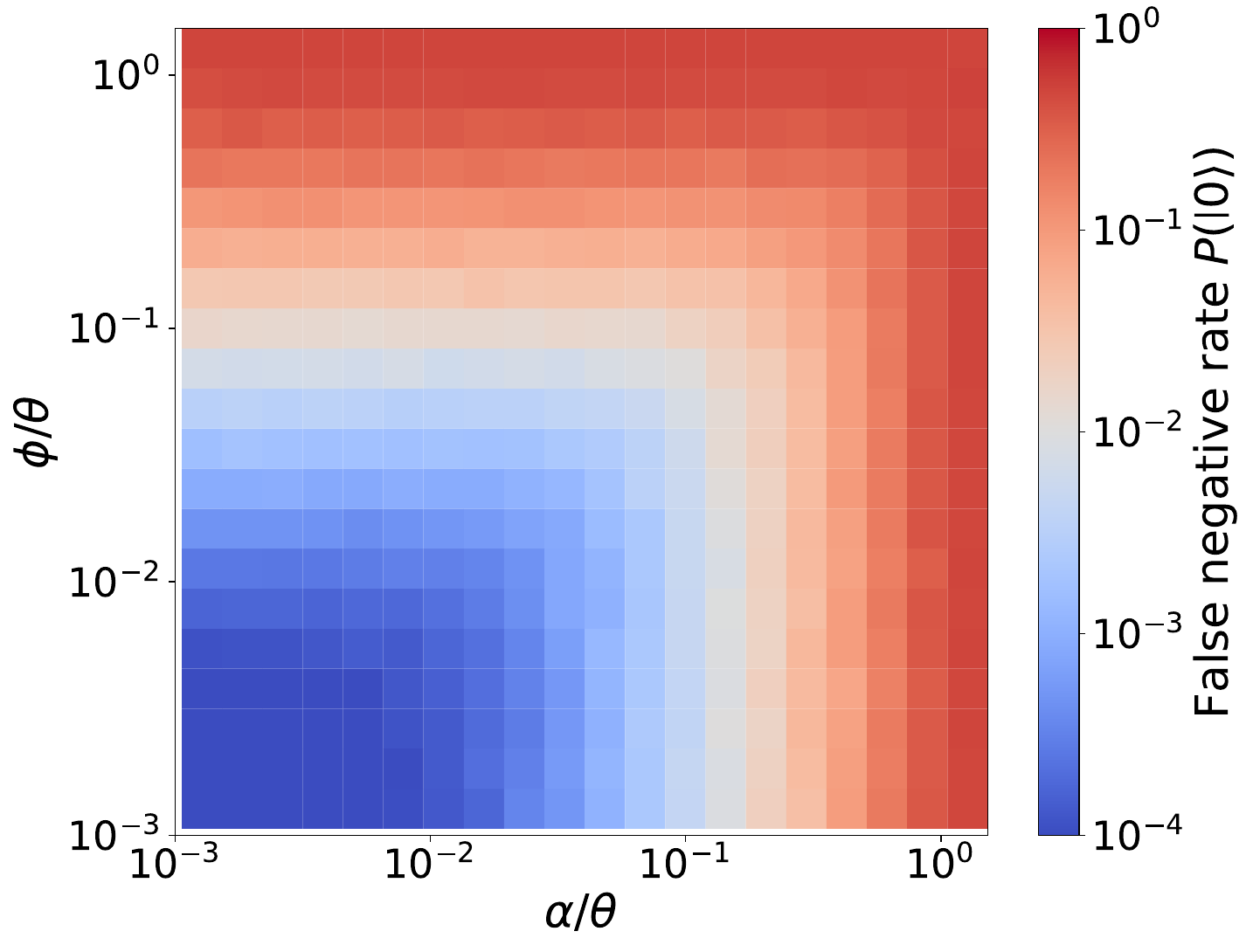}}
    \\
    \subfloat[\label{fig: torino 3 attacks Xtalk 2pi9 6q colorplot}Dephasing and amplitude damping using $T_1,T_2$ from $ibm\_torino$, $\theta=\frac{2\pi}{9}$]{\includegraphics[width=0.33\linewidth]{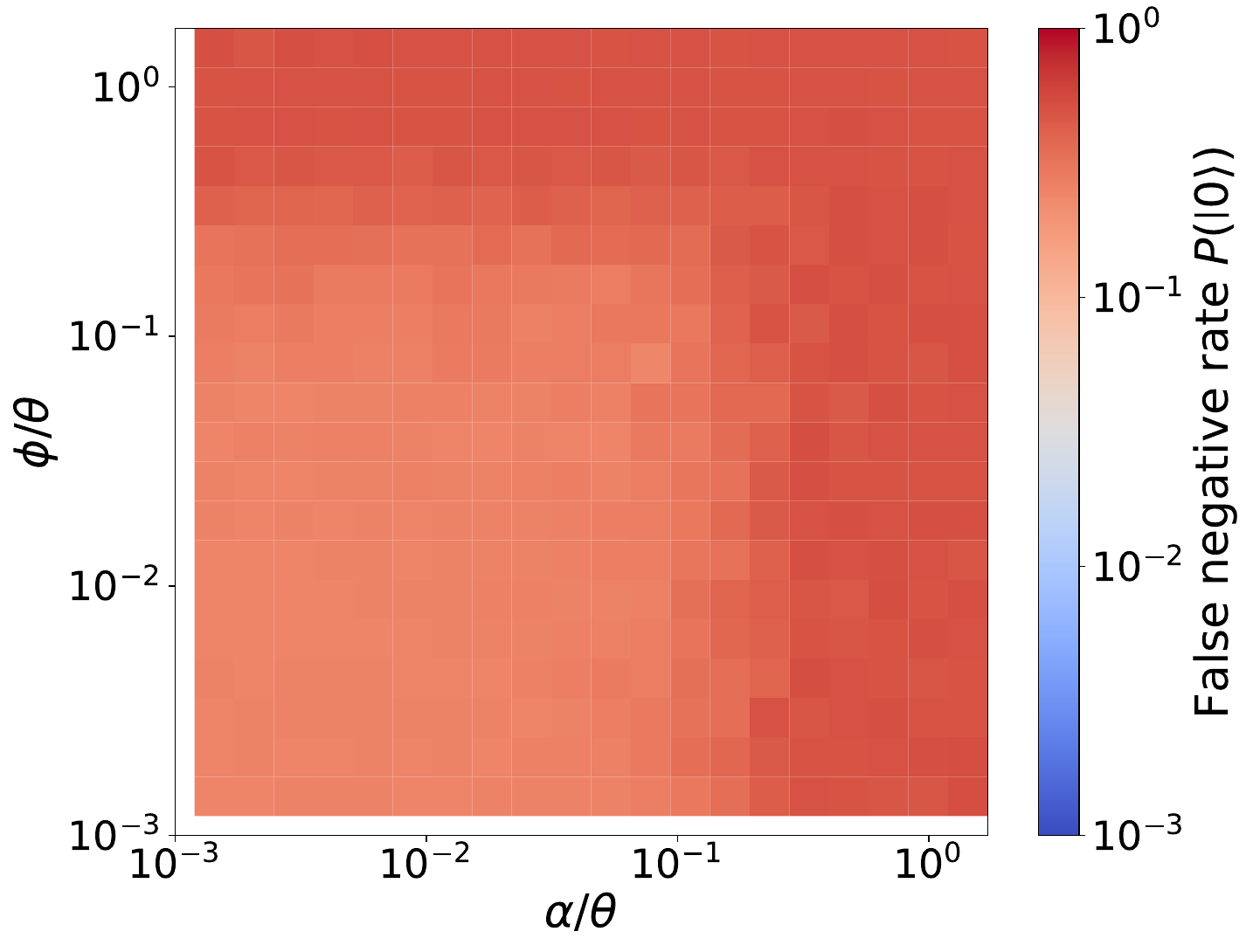}}
    \subfloat[\label{fig: pure 3 attacks Xtalk 2pi9 6q colorplot}No dephasing or amplitude damping, $\theta=\frac{2\pi}{9}$]{\includegraphics[width=0.33\linewidth]{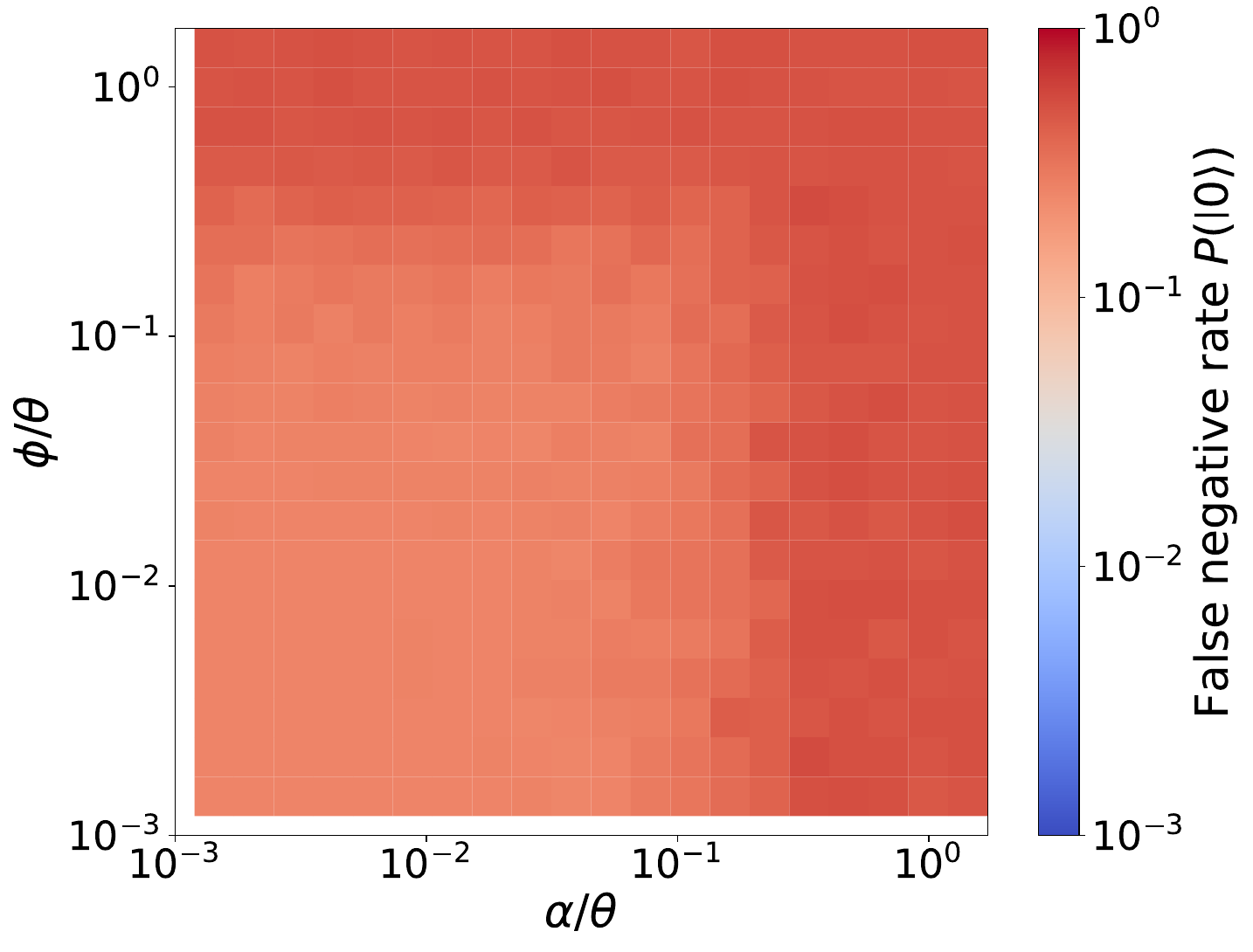}}
    \subfloat[\label{fig: pure 3 attacks Xtalk 2pi17 8q colorplot}No dephasing or amplitude damping, $\theta=\frac{2\pi}{17}$]{\includegraphics[width=0.33\linewidth]{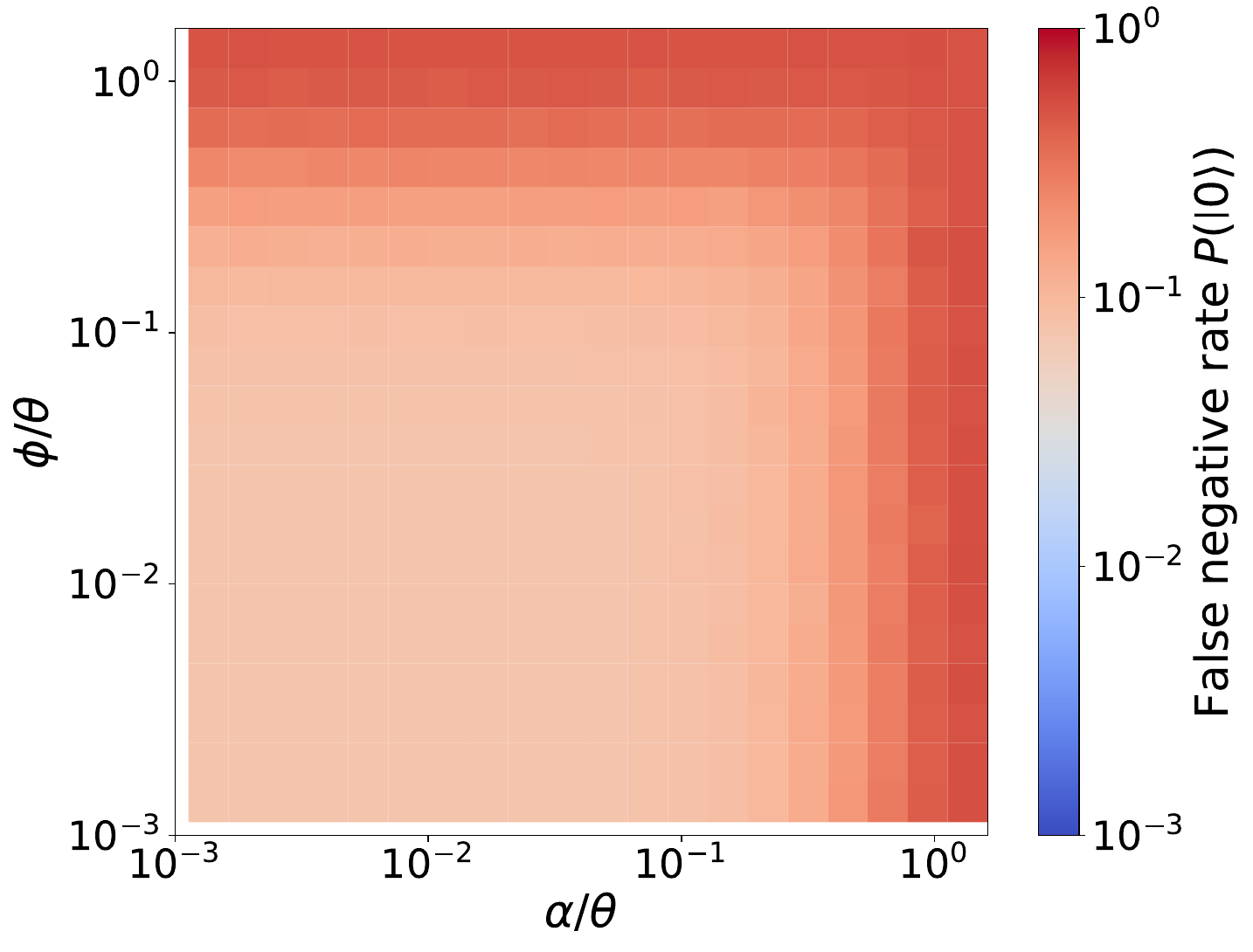}}
    \caption{Examples of the false negative rate when it is perturbed three times plotted against the ratio of the always-on ZZ coupling contributed by the idle noise to the single-qubit crosstalk rotation rate ($\alpha/\theta$), as well as the ratio between the two-qubit and single-qubit crosstalk rate ($\phi/\theta$) in a colour plot. All axes are shown in logarithmic scale. \textbf{(a)} Single qubit crosstalk rotation angle $\theta=\frac{\pi}{4}$, with 4 spectator qubits. Dephasing and amplitude damping channels with strengths derived from the $T_1$ and $T_2$ parameters of $ibm\_torino$ quantum computer as part of the idling noise are included. \textbf{(b)} Same as (a), but without dephasing or amplitude damping noise. \textbf{(c)} Same as (b), with the single-qubit rotation angle set as $\theta=\frac{\pi}{8}$ (thus with 8 spectator qubits). \textbf{(d)}, \textbf{(e)} and \textbf{(f)} have the same settings as (a), (b) and (c), except that the rotation angles are set as $\frac{2\pi}{8}, \frac{2\pi}{8}, \frac{2\pi}{17}$ which satisfies the condition of the worst-case scenario of the angle mismatching in \Cref{eq: general angle value}.}
    \label{fig: 3 attacks Xtalk colorplot}
\end{figure*}

\subsection{Results}
Based on the model proposed in \Cref{subsec: setup}, we first evaluate the false positive rate of CSMQC per every two-qubit gate timestep. That is, the probability of falsely reporting crosstalk when it does not actually exist. \Cref{fig: false alarm simulation a} shows the false positive rate when the system is evolved after a timestep equal to the two-qubit gate time plotted against the always-on ZZ coupling strength $\alpha$. When the ZZ coupling strength is low, the noise is dominated by constant dephasing and amplitude damping, thus producing an asymptotically flattened graph. However, as the ZZ coupling strength increases and becomes the dominant factor, we observe a phase transition in the false positive rate around $1\cross10^{-2}$ using parameters from $ibm\_torino$ and $3\cross10^{-2}$ from $ibm\_sherbrooke$. In general, the ZZ coupling strength such that the false positive rate becomes significant is approximately $10^{-1}$ for $ibm\_torino$ and $ibm\_sherbrooke$. This result highlights the critical value for when ZZ coupling noise becomes significant/harmful to crosstalk detection for the current IBM quantum computers. In the case without dephasing and amplitude damping noise, the phase transition is not observed, and the false positive rate scales polynomially, which matches the linear trend on the log-log plot. Meanwhile, as in \Cref{fig: false alarm simulation c}, the asymptotic false positive rate scales linearly 
with respect to the number of spectator qubits involved in the modified GHZ state. This provides evidence to underpin the scalability of CSMQC. 

Next we also examine the false negative rate subject to idle noise only. That is, the probability of falsely reporting that there is no crosstalk. Due to limited simulation power, we set the rotation angle produced by crosstalk as $\theta=\frac{\pi}{4}$ so that the largest number of spectator qubits is 4. In this case, we also set $\phi=0$ without coherent rotations of two qubits for simplicity. 
This test 
highlights the relationship between the ZZ coupling strength and the size of the crosstalk angle such that the false negative rate becomes detrimental to crosstalk detection when their ratio is too high. In \Cref{fig: false alarm simulation b}, we show the probability of not detecting crosstalk versus the ratio of the ZZ coupling strength to the crosstalk angle. Similarly to \Cref{fig: false alarm simulation a}, we observe an overall trend in which the false negative probability flattens when idle noise is dominated by dephasing and damping channels, while a phase transition occurs when the ZZ coupling becomes dominant. In particular, we find that the ratio such that this transition occurs is in the range between $1\cross 10^{-2}$ to $1\cross 10^{-1}$, which defines the usefulness of this technique. Because there are different numbers of spectator qubits $\{1,2,4\}$ responsible for detecting different numbers of crosstalk perturbations, there are splittings between different numbers of spectator qubits, especially when the ZZ coupling strength is low. This demonstrates a scenario where more spectator qubits provides greater exposure to global idle noise. \Cref{fig: false alarm simulation d} shows the false negative probability plotted against the number of perturbations. Surprisingly, as the number of perturbations increases, the false negative probability slightly decreases. We interpret this phenomenon as the consequence of noise generated from the ZZ coupling channel contributing less to the total perturbations as the rotations caused by crosstalk increase proportionally, thus improving the quality of detection.

We further evaluate the behaviour of the false negative rate by sampling from different combinations of always-on ZZ coupling strengths as part of idle noise, and the rate of the two-qubit coherent rotations $\phi$ as part of the noise induced by crosstalk. In \Cref{fig: 3 attacks Xtalk colorplot}, we show the false negative rate with the ratios of the ZZ-coupling strength and the two-qubit crosstalk rate $\phi$ to the single-qubit crosstalk angle $\theta$ as the two variables. Full plots with various numbers of crosstalk perturbations, spectator qubits and angle $\theta$ are included in \Cref{appendix: colorplots}. In the scenarios where dephasing and damping noise are included and the single-qubit crosstalk rotation angle $\theta$ satisfies \Cref{eq: total angle condition} exactly (see \Cref{fig: torino 3 attacks Xtalk pi4 6q colorplot}, \Cref{fig: torino Xtalk pi4 6q colorplot}, \Cref{fig: sherbrooke Xtalk pi4 6q colorplot}), the transition in the false negative rate occurs when $\phi$ becomes large enough that the noise caused by the always-on ZZ coupling and two-qubit crosstalk is no longer overwhelmed by dephasing and amplitude damping. By examining different numbers of perturbations, we observe a general trend that the value of $\phi$ where the transition happens decreases as the number of perturbations increases, as more noise is generated when there are more crosstalking gates. We also find that the transition position for the ZZ-coupling strength ratio ($\alpha/\theta$) increases slightly as the number of perturbations increases, which is a result of its decreasing contribution to the overall noise as there are more perturbations. Apart from this, since dynamical decoupling was applied to stabilise the noise which includes nearest-neighbour couplings, the change in the transition position is relatively small for different number of spectator qubits. In most cases, the transition happens when the ratio between other noise channels to the single-qubit crosstalk rotation rate is around $10^{-1}$. The splittings between different numbers of spectator qubits at low ZZ coupling strengths/low two-qubit crosstalk rate also persist in a linear trend (see \Cref{fig: detection failure rate pi4 flat vs SQs}) with respect to the number of spectator qubits due to the same reason as in \Cref{fig: false alarm simulation d} and \Cref{fig: false alarm simulation b} respectively. However, as from the shallow gradient in \Cref{fig: detection failure rate pi4 flat vs SQs}, the impact of dephasing and damping noise would only become significant (i.e. $\text{false negative rate}=0.1$) when the number of spectator qubits is $>19$ and $>76$ for devices with dephasing and damping noise similar to $ibm\_sherbrooke$ and $ibm\_torino$ respectively. In the case without any dephasing and damping noise, however, there is no obvious transition nor splittings as in \Cref{fig: pure 3 attacks Xtalk pi4 6q colorplot}, \Cref{fig: pure Xtalk pi4 6q colorplot}.

To address the problem of mismatching of the total angle condition in \Cref{eq: total angle condition}, we also simulate the detection rate in the worst-case scenario where the difference between the actual rotated angle and the ideal angle $\pi$ is the maximum regardless of adding/subtracting one spectator qubit. Specifically, we set
\begin{equation}
    \theta=\frac{2\pi}{2kn+1}
\end{equation}
as the angle of rotation per spectator qubit, where $n\in\mathbb{Z}^{+}$ is the threshold number of spectator qubits for which $\pi-kn\theta >0$, and $k$ is the smallest number in a chosen set of crosstalk perturbation numbers (see \Cref{appendix: total angle mismatching}). Using this formula, we simulate the false negative rate again for $kn=4$ and $kn=8$ (corresponding to $\theta=\frac{2\pi}{9},\frac{2\pi}{17}$). Due to memory scaling and limited resources, dephasing and amplitude damping noise are not included in the simulation for $\theta=\frac{2\pi}{17}$ 
\Cref{fig: torino 3 attacks Xtalk 2pi9 6q colorplot} and \Cref{fig: pure 3 attacks Xtalk 2pi9 6q colorplot} with 4 spectator qubits, shows the mismatching severely impacts the detection rate, 
However, when there are 8 spectator qubits involved 
, the performance at low ZZ-coupling strength and the two-qubit crosstalk rate improves significantly (\Cref{fig: pure 3 attacks Xtalk 2pi17 8q colorplot}) from an average of 0.250 to 0.075, which agrees with our finding in \Cref{fig: delta_theta} and \Cref{fig: detection failure rate plat vs mk} that $\text{false negative rate}\propto \frac{1-\cos\left(n^{-1}\right)}{2}$ given a fixed number of perturbations. We also observe that the transition positions $\alpha/\theta$ and $\phi/\theta$ are almost doubled when the single-qubit rotation strengths $\theta$ are halved. This means that the values of two-qubit coupling strengths $\alpha$ and $\phi$ at transition positions remain almost unchanged thanks to the good performance of dynamical decoupling even when more spectator qubits are involved (as shown in more comparisons in \Cref{fig: pure Xtalk 2pi9 6q colorplot} and \Cref{fig: pure Xtalk 2pi17 8q colorplot}).  Even after considering the dephasing and damping noise that was not included in this case, their net contribution to the false negative rate is at most 0.02 for 8 spectator qubits according to \Cref{fig: detection failure rate pi4 flat vs SQs}. Therefore, we anticipate that there will a wide range of sweet regions where the false negative probability is insignificant (<$10^{-1}$). In practice, the average angle of rotation induced by crosstalk per spectator qubit is almost certain to be smaller than $\pi/8$, which results in an even smaller $\Delta\theta$ and thus false negative rate. Also, considering that the simulation tests the worst-case scenario of maximal deviation in the total angle, it is reasonable to expect the performance to be better than simulated above.

\section{Simulation with benchmarked device parameters}\label{sec: results}
\subsection{Crosstalk noise characterisation for simulation}
In this work, we implement a typical crosstalk noise model --- the reduced HSA model \cite{HSA_model_paper, HSA_model} equipped with real IBM quantum device error parameters determined via idle tomography experiments --- to emulate the action of CNOT crosstalk on target qubit(s) in the IBM quantum computers. We use the python package \texttt{pyGSTi}, which provides an implementation of the reduced HSA model to conduct the simulations. Due to limited resources, we use the same crosstalk error parameters obtained from \cite{Ben_crosstalk}, which were derived from experiments conducted on $\textit{ibm\_hanoi}$, an IBM 27-qubit device with CNOT gate as the two-qubit gate.

\begin{figure}[t]
    \centering
    \includegraphics[width=1\linewidth]{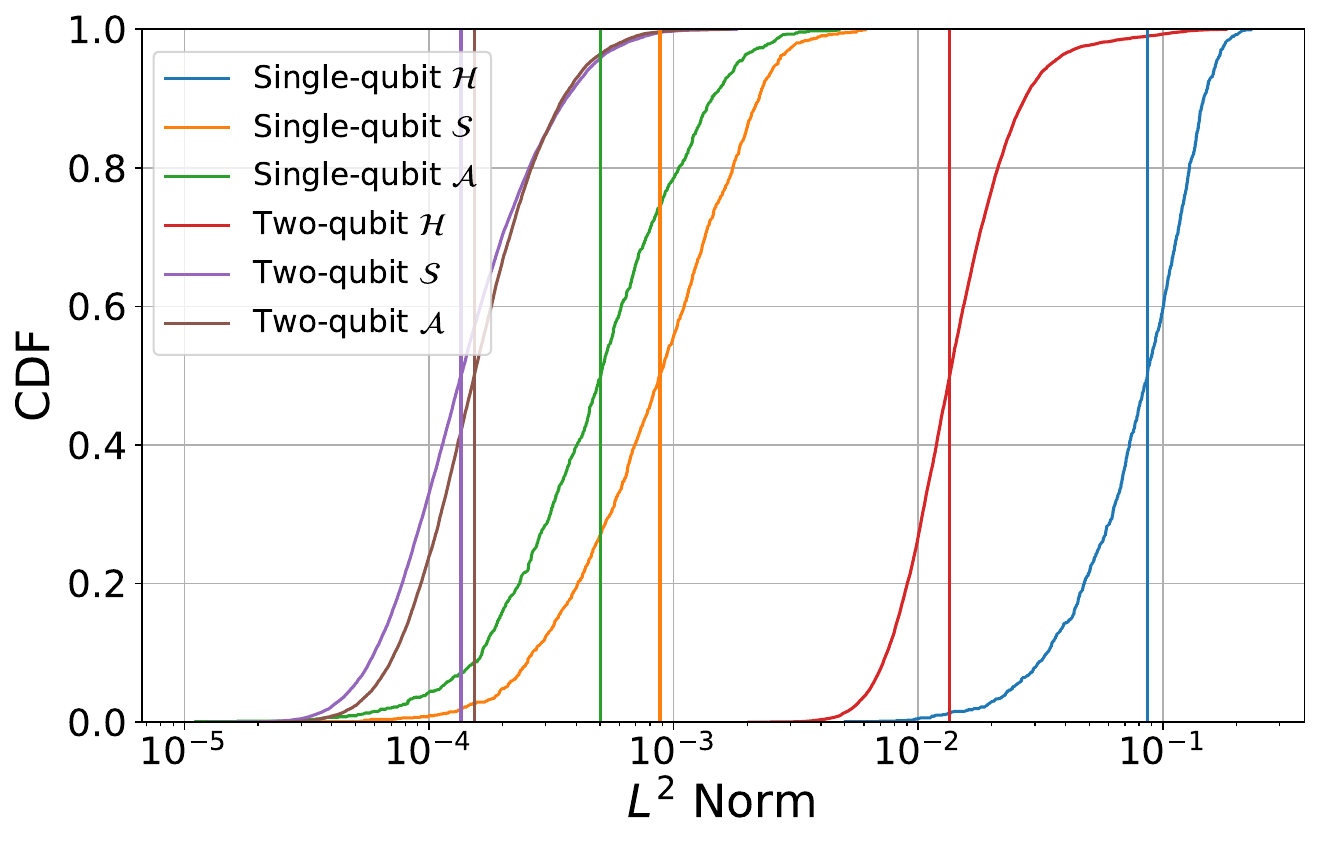}
    \caption{Cumulative density function (CDF) in logarithmic scale demonstrating the $L^2$ norm of the error parameters sampled from $\textit{ibm\_hanoi}$ on different noise channels. Vertical lines indicate the median magnitude (0.5 on CDF) for each of the error parameters. Noise is dominated by single-qubit Hamiltonian errors with a median magnitude of 0.086, followed by double-qubit Hamiltonian errors with a median magnitude of 0.013, as well as other negligible noise channels.}
    \label{fig: HSA error comparison}
\end{figure}

The reduced HSA model consists of three noise channels: Hamiltonian (coherent evolution of the quantum state under the Pauli-basis Hamiltonian), Stochastic (random Pauli errors which depolarise the quantum state in some directions) and Affine (nonunital noise that pushes the quantum state into a certain direction in the Bloch sphere that is not the maximally mixed state) errors. The three channels are explicitly represented by the Lindbladian error generators,
\begin{equation}\label{eq: noise generators}
\begin{aligned}
    \mathcal{H}_P[\bigcdot] &= -i\left[P,\bigcdot\right]\\
    \mathcal{S}_P[\bigcdot] &= P\bigcdot P - \mathds{1}\bigcdot\mathds{1}\\
    \mathcal{A}[\bigcdot] &= \frac{i}{2}\{[P,Q],\bigcdot\} + iP\bigcdot Q -iQ\bigcdot P,
\end{aligned}
\end{equation}
where $P,Q\in\{I,X,Y,Z\}^{\otimes n}$ are an arbitrary Pauli-basis of $n$ target qubits of interest, among which the $P$s with only one non-identity operator in the Hamiltonian channel has its Kraus representation the same as the unitary rotation described in \Cref{subsec: conditions and assumptions}, while others contribute to $\mathcal{E}_{\text{other}}$. Therefore, the dominance of single-qubit Hamiltonian errors must serve as an important condition for the success of CSMQC. By combining these with their respective weights, the net generator is given by
\begin{equation}\label{eq: Lindbladian op}
    \mathcal{L}[\bigcdot] = \sum\limits_{P}{\epsilon_{HP}\mathcal{H}_P[\bigcdot]+\epsilon_{SP}\mathcal{S}_P[\bigcdot]+\epsilon_{AP}\mathcal{A}_P[\bigcdot]},
\end{equation}
where $\epsilon_{HP}, \epsilon_{SP}, \epsilon_{AP}$ are the coefficients of the Lindbladian error generators. Therefore, the mapping which describes the evolution of a state under these channels is given in the super-operator form:
\begin{equation}
    \vert\rho\rrangle\rightarrow e^{\mathcal{L}}\vert\rho\rrangle,
\end{equation}
where $\mathcal{L}$ is the Lindbladian generator in the representation of $\vert\rho\rrangle\in\mathcal{H}_T$. \Cref{eq: Lindbladian op} can also be thought of as a discretised case of a continuous-time dynamics of the quantum state described by the following master equation,
\begin{equation}
    \frac{d\rho}{dt}=\sum\limits_{P}{\epsilon_{HP}\mathcal{H}_P[\rho]+\epsilon_{SP}\mathcal{S}_P[\rho]+\epsilon_{AP}\mathcal{A}_P[\rho]}.
\end{equation}
As a result, we draw comparison between the magnitude of coefficients for different error generators, benchmarked using data from $\textit{ibm\_hanoi}$ in \Cref{fig: HSA error comparison} using cumulative statistics. We use the $L^2$ norm of the error parameters in the corresponding Pauli basis space to determine their magnitude,
\begin{equation}
    \norm{\epsilon}_2=\sqrt{\sum_{\vec{v}}{\epsilon_{\sigma_{\vec{v}}}^2}}
\end{equation}
where $\vec{v}$ is the Pauli basis space with 3 elements for single-qubit errors and 9 elements for double-qubit errors. $\epsilon_{\sigma_{\vec{v}}}$ denotes the corresponding error parameter in the $\sigma_{\vec{v}}$ basis.

From the statistics sampled as in \Cref{fig: HSA error comparison}, we find that the crosstalk on \textit{ibm\_hanoi} is dominated by single-qubit Hamiltonian errors in terms of their $L^2$ norm, which satisfies our assumption. However, as denoted by the red curve, a significant amount of two-qubit Hamiltonian errors still persist, which may disrupt the performance of CSMQC. Apart from these, all other stochastic and affine noise channels are negligible, which is in agreement with the artificial model of crosstalk noises in \Cref{sec: artificial model results}.  In later sections, we will discuss and compare the results simulated using the error parameters scaled equally relative to their $L^2$ norm as in \Cref{fig: HSA error comparison}, as well as those simulated using inhomogeneously adjusted error parameters in which only single-qubit Hamiltonian errors are amplified. 

\begin{figure}[t]
    \subfloat[\label{fig: HSA error radii single}Single-qubit error parameters ($ibm\_hanoi$)]{\includegraphics[width=1\linewidth]{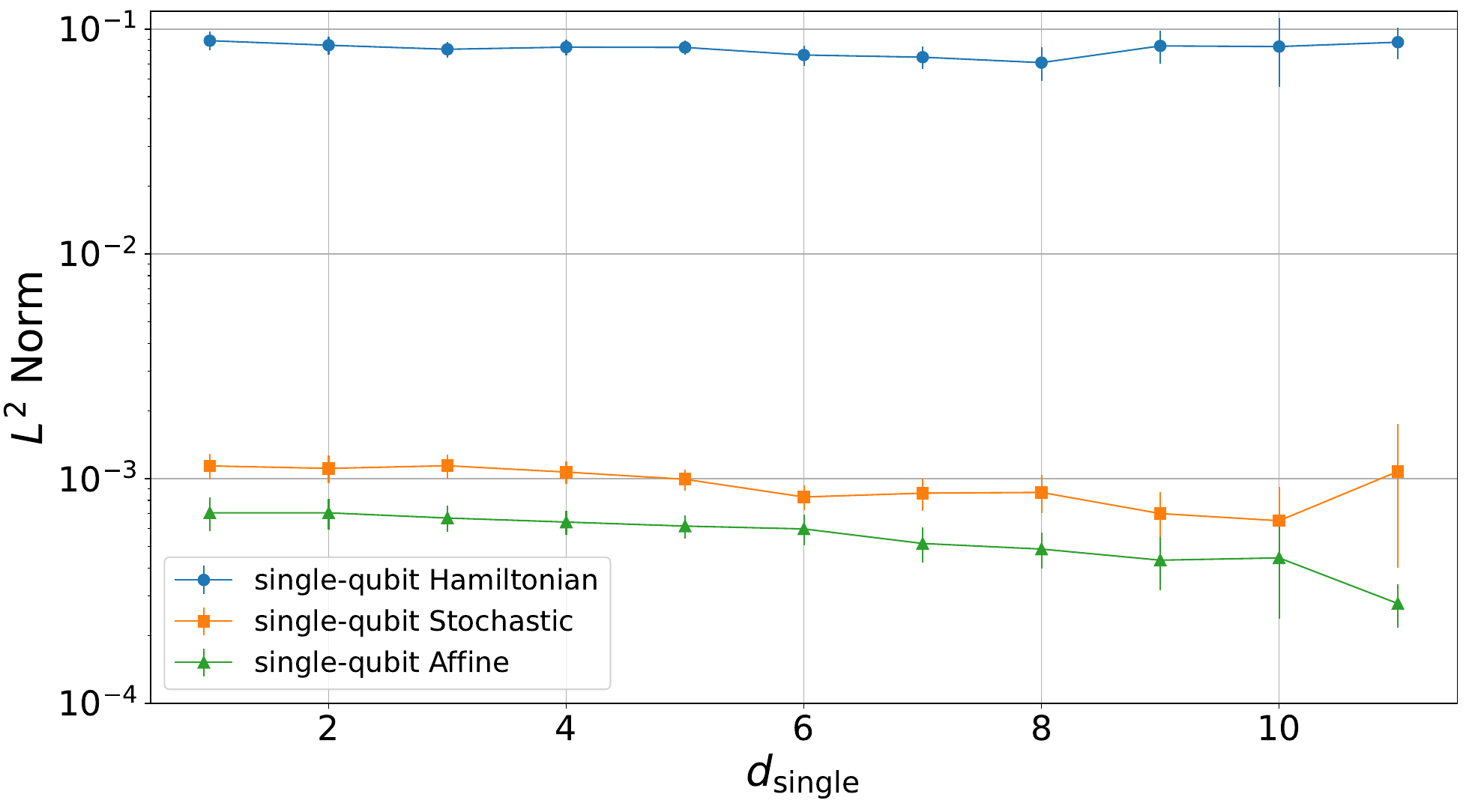}}\\
    \subfloat[\label{fig: HSA error radii double}Two-qubit error parameters ($ibm\_hanoi$)]{\includegraphics[width=1\linewidth]{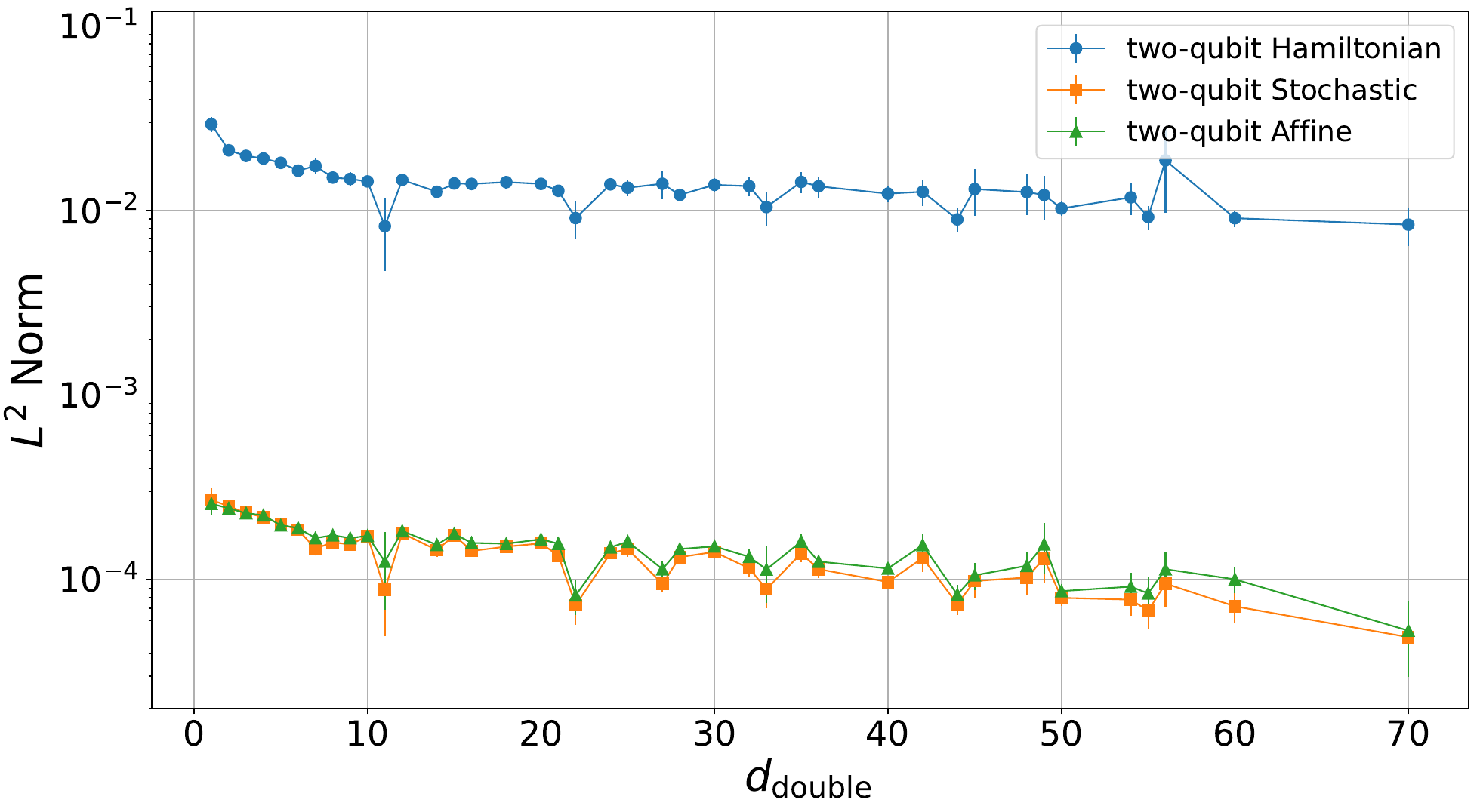}}
    
    \caption{\textbf{(a)} Magnitude of the single-qubit crosstalk noise parameters on a logarithmic scale as a function of the distance (minimum shortest path length) from the source to the qubit(s) that experience the noise. We use the $L^2$ norm of the error parameters for a particular noise channel in the space of the corresponding Pauli basis of one or two qubits as its magnitude. \textbf{(b)} Magnitude of the two-qubit crosstalk noise parameters (also in logarithmic scale) as a function of weighted distance, which accounts for both the distance from the sources of the noise to the nearest qubit experiencing the noise and the distance between the two target qubits.
    Evidently, the noise is dominated by the single-qubit Hamiltonian channel, followed by contributions from the two-qubit Hamiltonian channel. The magnitude of the dominating noise barely changes as the distance increases. While other noise channels exhibit decaying trends in magnitude, their contributions are minimal and have negligible impact on the overall noise.}
    \label{fig: HSA error radii}
\end{figure}

\begin{figure}[t]
    \subfloat[\label{fig: torino HSA error radii single}Single-qubit error parameters ($ibm\_torino$)]{\includegraphics[width=1\linewidth]{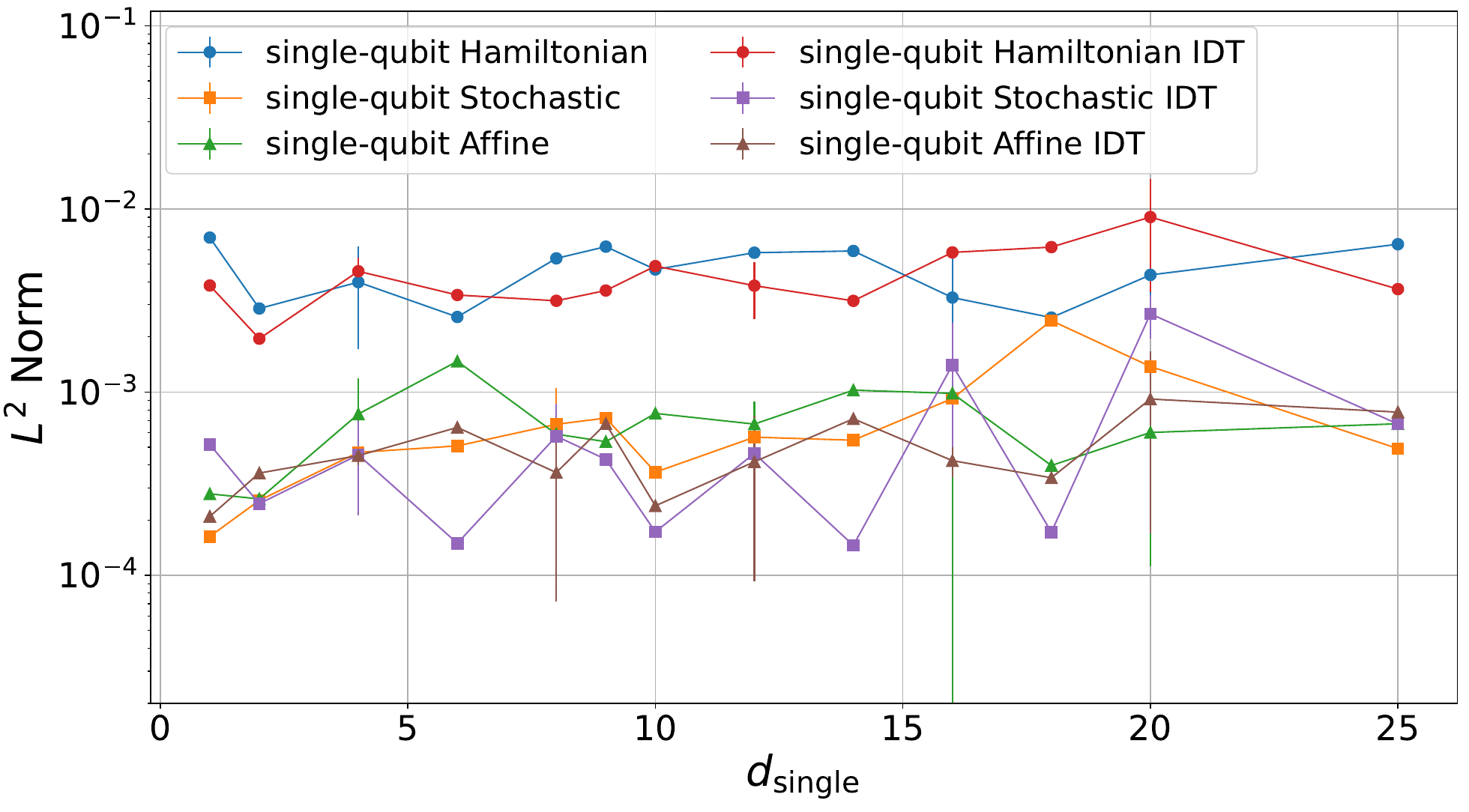}}\\
    \subfloat[\label{fig: torino HSA error radii double}Two-qubit error parameters ($ibm\_torino$)]{\includegraphics[width=1\linewidth]{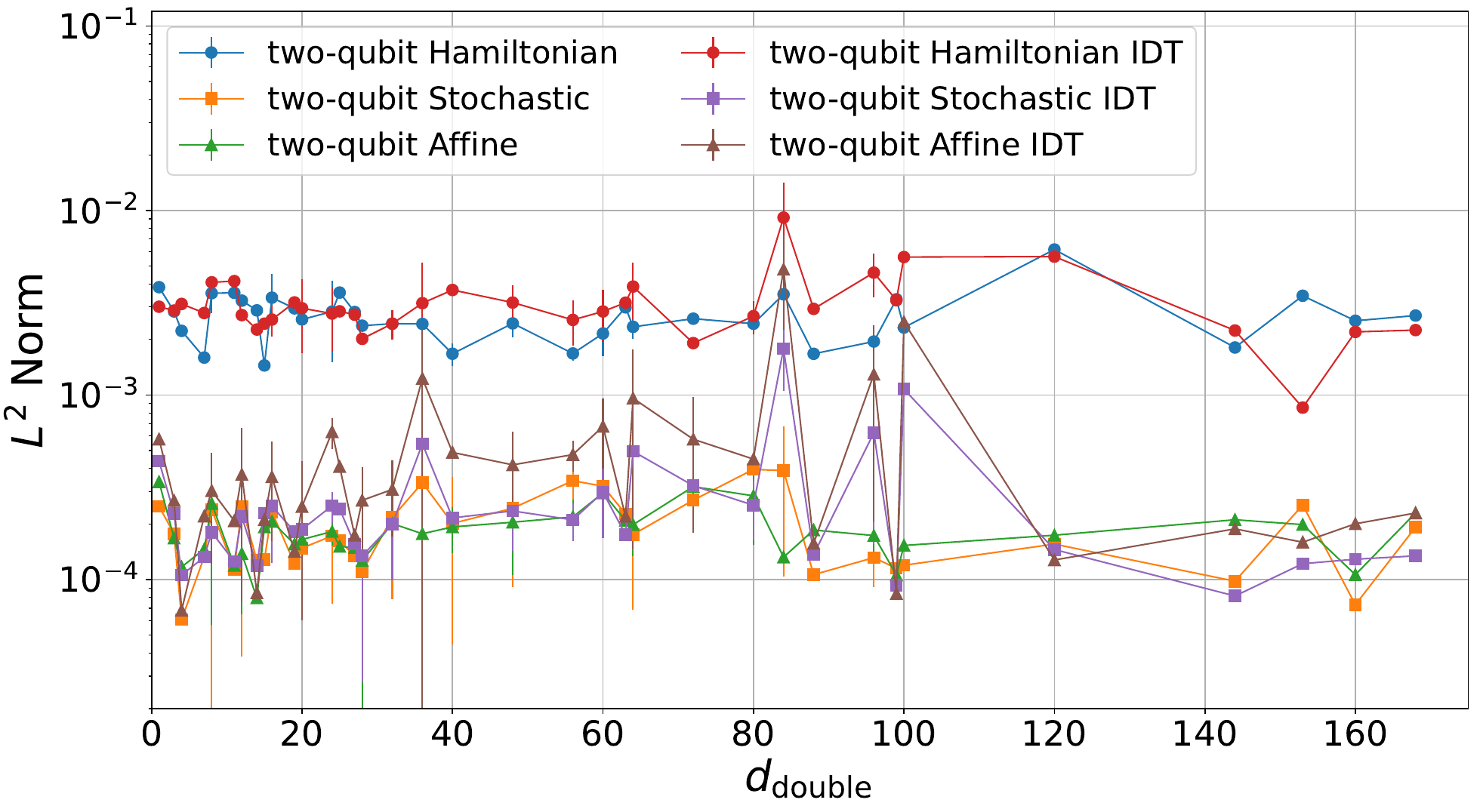}}
    
    \caption{\textbf{(a)} Single-qubit error parameters magnitude versus distance, but sampled from $ibm\_torino$. The legends with `IDT' labelling indicate the noise parameters sampled from circuits without any crosstalking gates, or idle tomography. \textbf{(b)} The corresponding two-qubit error parameters. The definition of the magnitude, distance, weighted distance and scale of the vertical axis are all the same as \Cref{fig: HSA error radii}.}
    \label{fig:torino HSA error radii}
\end{figure}

Similarly, we also verify the feasibility of our assumption on the range of influence of the noise by evaluating the average magnitude of the noise at different distances from the crosstalk source. For single-qubit errors, the distance is defined as the minimum shortest path length from the qubit $v$ of interest that experiences the noise to one of the two qubits $a\in A$ ($\abs{A}=2$) that produces such noise,
\begin{equation}
    d_{\text{single}}\coloneqq \min_{a\in A}{l(P_{a,t})},
\end{equation}
where $P_{a,t}$ is the shortest path from qubit $s$ to qubit $t$. For the two-qubit errors, as they describe correlated evolution experienced by two qubits, we define their distances as the product of the minimum shortest path length between one of the two qubits $t\in T$ ($\abs{T}=2$) experiencing the noise to a qubit $a\in A$ producing the noise, and the shortest path length between the two target qubits $t_1,t_2\in T$:
\begin{equation}
    d_{\text{double}}\coloneqq \left(\min_{a\in A, t\in T}{l(P_{a,t})}\right)l(P_{t_1,t_2}).
\end{equation}
In \Cref{fig: HSA error radii}, we plot the average $L^2$ norm of the error parameters for different noise channels as a function of their distance from the source of crosstalk. Despite observations showing that the magnitude of the two-qubit error parameters decays as the distance increases, it is clear that the magnitude of the dominant error parameter: the single-qubit Hamiltonian channel maintains at the same level as the distance changes. In general, the noise parameters sampled from the IBM quantum device $ibm\_hanoi$ align well with the global range assumption for noise influence between action and target qubits. We also investigate the scaling of the error parameters on specific qubits of the newer IBM quantum processors $ibm\_torino$ in \Cref{fig: torino HSA error radii double} which employs Controlled-$Z$ ($CZ$) gate as the native two-qubit gate while exhibiting improved performance in error reduction \cite{heron_performance}. Notably, the correlations between the noise magnitude and the distance/weighted distance are still minimal across all channels, which reinforces that the globalness condition can hold true across different devices. However, the magnitude of the single-qubit Hamiltonian error is reduced even further compared to $ibm\_hanoi$, making it very close to the background noise when idle without activating the CZ gate, as shown in \Cref{fig:torino HSA error radii}. This reduces the influence of crosstalk noise on data qubits and requires a larger number of spectator qubits in the CSMQC circuit to detect its effect. 

\subsection{Approximation as single-qubit coherent rotation}
To set up the modified GHZ state with better encapsulated rotation axes, we propose a method to approximate net crosstalk noise as a coherent single-qubit Hamiltonian error along a specific axis in each qubit. Based on the previous assumption of minimal contribution from other noise channels, we approximate the action of the noise on the reduced density operator as a single-qubit Hamiltonian error only and characterise the corresponding noise parameters accordingly. We make this approximation so that the rotation axes capture and utilise the rotation from both the single- and two-qubit channels. This turns the two-qubit channel away from a purely disruptive factor and reduces the overall disruption from the two-qubit channels if the axes are taken directly from the single-qubit Hamiltonian component.

By treating the action of the crosstalk noise as single-qubit errors, the evolution operator can be decomposed onto each qubit, and thus the rotation axis affected by crosstalk can be determined independently for each qubit. To infer the rotation axis for a qubit subjected to crosstalk, we sample the coordinates of the single-qubit state’s path on its Bloch sphere for various crosstalk gate counts, with $g$ gates per timestep, using quantum state tomography (QST).  The coordinates are sampled from the evolving state, where the parameters are determined from a real device. 
Consequently, for the qubit with label $i$, its Cartesian coordinates on the Bloch sphere are given by the expectation values of the corresponding Pauli operators at site $i$:
\begin{equation}\label{eq: xyz coord expected value}
    (x,y,z)_i=\left(\text{tr}( X_i\rho), \text{tr}( Y_i\rho), \text{tr}( Z_i\rho)\right),
\end{equation}
where $\rho$ is the density operator of the quantum system. After some algebra (see details in \Cref{result: C1} in \Cref{appendix: single qubit Pauli operator expected value}), \Cref{eq: xyz coord expected value} is shown to be equivalent to
\begin{equation}\label{eq: xyz coord expected value reduced}
    (x,y,z)_i=\left(\text{tr}( X_i\rho_i), \text{tr}( Y_i\rho_i), \text{tr}( Z_i\rho_i)\right),
\end{equation}
where $\rho_i\coloneqq\text{tr}_{\mathcal{H}_{A\cup T\setminus i}}(\rho)$ is the reduced density operator of the quantum subsystem on qubit $i$. Hence, the Pauli operator expectation values for qubit $i$ can be determined solely from its measured outcomes, independent from measured outcomes of other qubits. In this method, to minimise the cost of noise characterisation, we approximate the rotation axis as being the normal to the plane of a circle drawn through the three points. 
Specifically, we obtain the Barycentric coordinates $\bm{b}_{\mu}$ of the circumcentre with respect to each of the three points with label $\mu\in\{1,2,3\}$,
\begin{equation}
    \bm{b}_{\mu}=\bm{h}_{\mu}^2\left((1-\bm{\delta}_{\mu}^{\nu})\bm{h}_{\nu}^2-\bm{h}_{\mu}^2\right),
\end{equation}
where $\bm{h}$ is a rank-1 tensor with each element $\bm{h}_{\mu}$ representing the length of the side on the opposite of the point $\mu$ of the triangle formed by the three points. $\bm{\delta}$ is the Kronecker-$\delta$ with $\bm{\delta}_{\mu}^{\nu}=1$ if and only if $\mu=\nu$ and 0 otherwise.  Therefore, Cartesian coordinates of the circumcentre is given by
\begin{equation}
    \bm{O}_{\nu} = \frac{\bm{b}_{\mu}\bm{X}_{\nu}^{\mu}}{\norm{\bm{b}_{\gamma}}_1},
\end{equation}
where the Barycentric coordinates $\bm{b}_{\mu}$ serves as the weight contribution for the sampled point $\mu$, $\norm{\bm{b}_{\gamma}}_1$ is the $l^1$ norm of $\bm{b}$ and $\bm{X}$ is a rank-2 tensor with columns equal to the Cartesian coordinates of each of the three sampled points. Hence, the Cartesian coordinates of centre of the circle $\bm{O}$, is expressed as the weighted sum of the three coordinates. Finally, the normal vector or rotation axis can be deduced from the cross-product between the vectors connecting the sampled points to the circumcentre, 
\begin{equation}
    \bm{\hat{k}} = \frac{(\bm{x}_i-\bm{O})\cross(\bm{x}_j-\bm{O})}{\norm{(\bm{x}_i-\bm{O})\cross(\bm{x}_j-\bm{O})}_2},
\end{equation}
where $x_i$ is the Cartesian coordinates of the $i^{\text{th}}$ timestep sampled points. In practice, we always set $j>i$ and constrain the number of crosstalk gates $g$ per timestep to ensure alignment between the cross-product vector and the actual rotation axis. Moreover, $\delta$, the average angle of rotation generated per crosstalk gate can also be evaluated as
\begin{equation}\label{eq: angle of rotation per crosstalk}
    \delta = \frac{\arccos\left((\bm{x}_i-\bm{O})\cdot(\bm{x}_j-\bm{O})\right)}{g(j-i)\norm{\bm{x}_i-\bm{O}}\norm{\bm{x}_j-\bm{O}}},
\end{equation}
which is derived from the rule of dot product between vectors $\bm{x}_i-\bm{O}$ and $\bm{x}_j-\bm{O}$, with $\delta g(j-i)$ as the angle between them. The further normalisation factor $g(j-i)$ takes into account re-scaling of the total rotated angle caused by multiple gates into the average angle of rotation per gate.

We note that this approximation process is only designed as a proof-of-concept. In a real application, QST has to be performed for each spectator qubit for each pair of CNOTs in a crosstalk circuit instead of just a single pair. This requires the cost of the QST characterisation to scale in $O(mn)$, where $m$ is the number of available CNOT control-target pairs and $n$ is the number of spectator qubits. It is reasonable to assume $m$ also scales polynomially with $n$, and leads to the result that the complexity is only quadratic and scalable. In addition, the cost of performing Idle Tomography and obtaining the single-qubit Hamiltonian channel parameters from the reduced HSA model directly only scales with $O(m\log(n))$. However, this may result in a bad approximation as we explained earlier in this subsection.

\subsection{CSMQC Detection success rate analysis}\label{subsec: CSMQC Detection success rate analysis}
Following the discussion in \Cref{subsec: conditions and assumptions}, we find that there are non-negligible contributions to the crosstalk noise that are not classified as the single-qubit Hamiltonian errors. In this subsection, we numerically evaluate the performance of CSMQC in terms of its crosstalk detection success rate when simulated with the full HSA model using the noise parameters found on $ibm\_hanoi$. Due to the simulation cost scaling exponentially with the number of qubits, the number of spectator qubits accessible in simulation is limited. As a result, we artificially amplify the crosstalk noise parameters on the corresponding spectator qubits to fulfil \Cref{eq: total angle condition}, 
Also, because the single-qubit rotation axes are characterised by approximation, dynamical decoupling is not implemented in this part as it introduces more errors than it mitigates. In this work, we test the success rate of crosstalk detection by simulating crosstalk originating from qubit-0 and qubit-1, with up to four spectator qubits, using the parameters characterised from $ibm\_hanoi$. 

\begin{figure}[t!]
    \subfloat[\label{fig: all parameters amplified}all parameters amplified]{
    \includegraphics[width=1\linewidth]{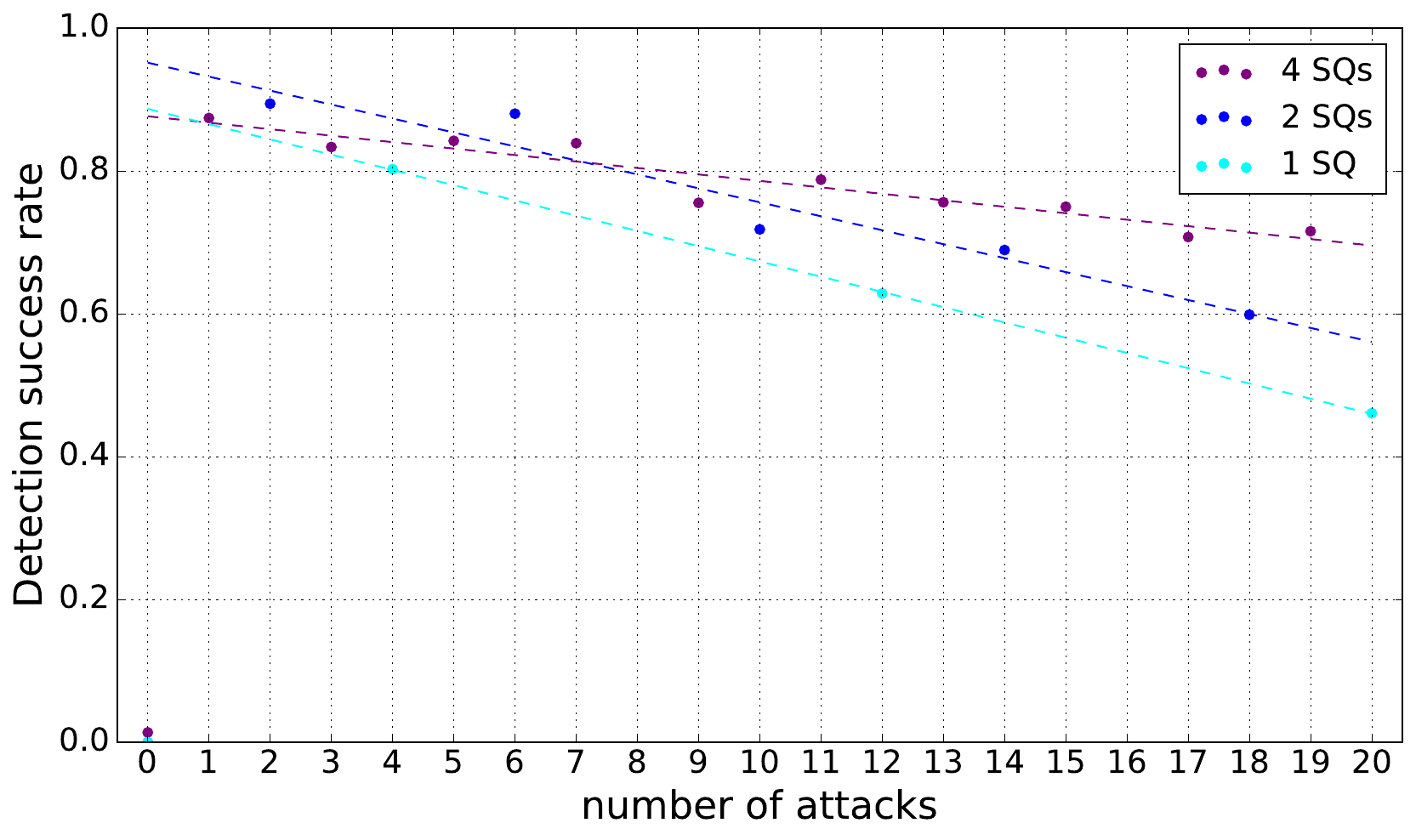}
    }\\
    \subfloat[\label{fig: single-qubit Hamiltonian channel amplified}single-qubit Hamiltonian channel amplified]{\includegraphics[width=1\linewidth]{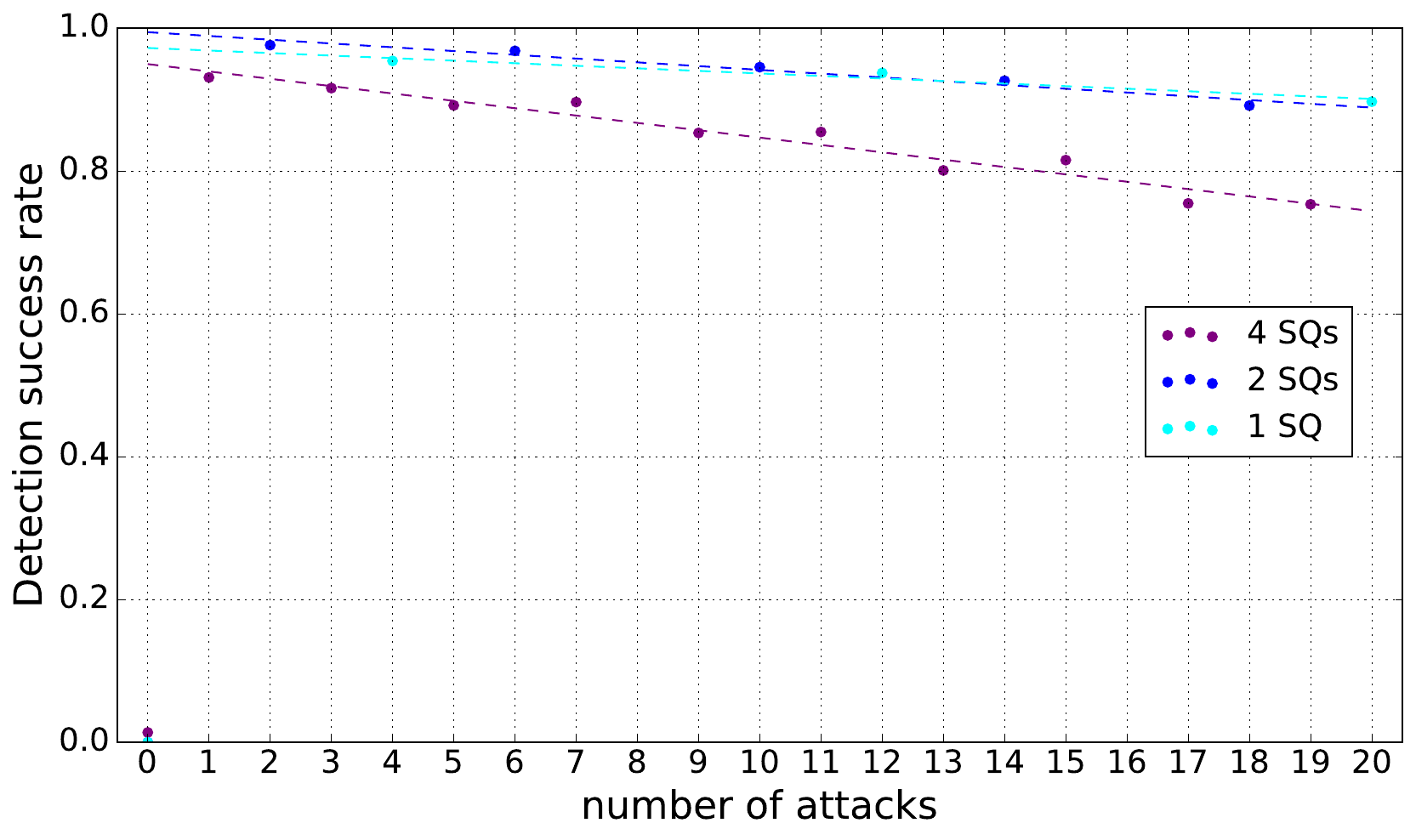}}
    \caption{\textbf{(a)} The CSMQC crosstalk detection success rate plotted against the number of gates applied to qubit-0 and qubit-1. We used four (with label 2,3,5,8), two (with label 2,3) and one (label 2) spectator qubit(s) to detect $2m+1$, $4m+2$ and $8m+2$ ($m\in\mathbb{N}$) gate counts respectively. In each of the experiments, the parameters for all channels in the HSA model are amplified by the same scale to emulate the behaviour of the original noise with factors 2.368, 2.345 and 2.173 for four, two and one spectator qubit(s) respectively. These numbers are calculated such that after scaling, the average angle of rotation per crosstalk is $\delta=\pi/4$ as in \Cref{eq: angle of rotation per crosstalk} . For consistency and to better simulate real hardware, qubit 5,8 in two spectator qubits experiment and qubit 3,5,8 in one spectator qubit experiment are the latency qubits for idling, which has no gate(s) acted on them but still included so that the dimension of $\mathcal{H}_{T}$ is unchanged. \textbf{(b)} The same plot with only single-qubit Hamiltonian noise parameter amplified (with factors 2.478, 2.412 and 2.22 for four, two and one spectator qubit(s)), articulating the dominance of single-qubit Hamiltonian noise channel and its importance in improving the performance of the detected crosstalk.}
    \label{fig: detection rates}
\end{figure}

As shown in \Cref{fig: all parameters amplified}, we demonstrate the detection rate plotted against the number of crosstalking gates using a different number of spectator qubits. The parameters for all channels in the HSA model are uniformly amplified for a given set of spectator qubits to emulate original noise behavior. 
To make the demonstration more realistic, the crosstalk induced by CNOT gates in the construction and inversion of $\ket{\text{MGHZ}_n(\bm{K})}$ on spectator qubits are also simulated. 
As the number of gates increases, we observe that the detection success rate decreases in a linear trend, which is a direct consequence of the accumulated noise disruptions that are not from the single-qubit Hamiltonian channel. In the plot, the success rate is above 0.8 for the number of gates less than 8, indicating the width of the time window where CSMQC performs well. Notably, when there is no crosstalk generated, it reports the detection success rate close to zero as it should,
hence few false positives. 
Similarly, we show the same detection success rate plot in \Cref{fig: single-qubit Hamiltonian channel amplified} with only the single-qubit Hamiltonian noise parameters amplified with factors 2.22, 2.412 and 2.418 for one, two and four spectator qubit(s) respectively. In this case, the overall success rate improves as the other noise channels play a less important role in disturbing the rotations around the characterised rotation axes. Our simulations with theoretical noise model in \Cref{fig: torino Xtalk pi4 6q colorplot} and \ref{fig: sherbrooke Xtalk pi4 6q colorplot} show the detection success rates range between 0.748 to 0.974 (0.926 to 0.991) from 1 to 7 gates given the ratio between the single and two-qubit Hamiltonian error at 0.15 (0.061 to 0.068) and negligible ZZ coupling strength. This result agrees with both of the results in \Cref{fig: detection rates} to a large extent, which uses the noise parameters sampled from $ibm\_hanoi$ with the average ratio between single and two-qubit Hamiltonian error at 0.15 as shown in \Cref{fig: HSA error comparison} and 0.061 to 0.068 after amplifying the single-qubit Hamiltonian error only. Despite these agreements with the previous model, the detection success rate is poorer with fewer spectator qubits as in \Cref{fig: all parameters amplified} when all parameters are amplified. This may arise from stochastic or asymmetric noise sources that were not taken into account in \Cref{sec: artificial model results}.

\begin{figure*}[t!]
    \centering
    \subfloat[\label{fig: Bell state IDT circuit} ]{\includegraphics[width=0.55\linewidth]{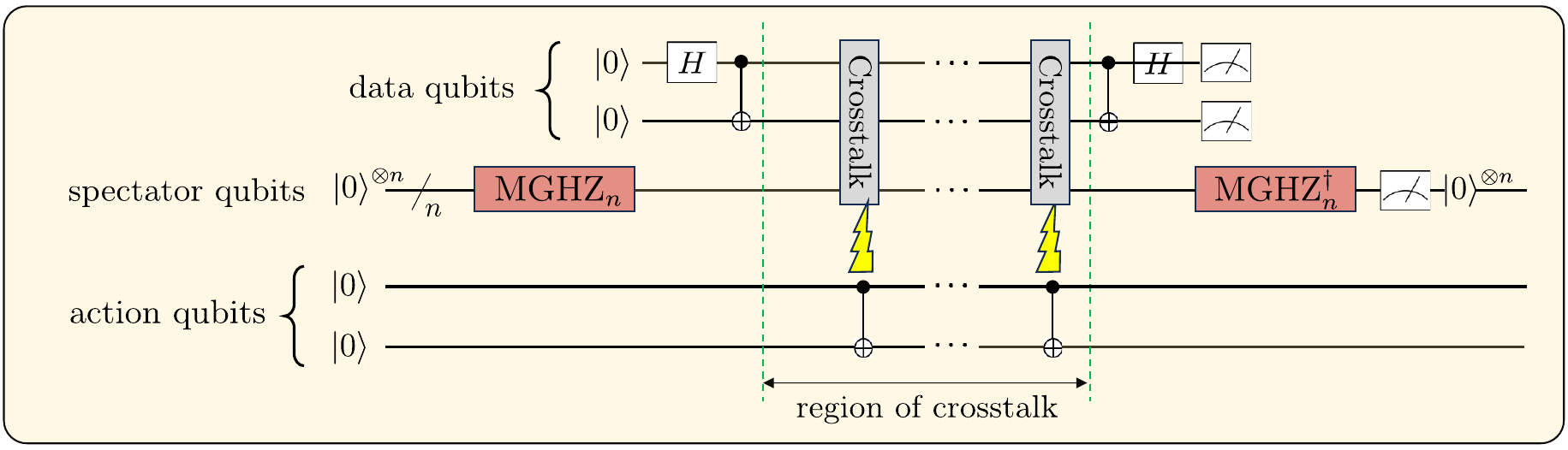}}
    \subfloat[\label{fig: Bell state IDT circuit constant freq} ]{\includegraphics[width=0.38\linewidth]{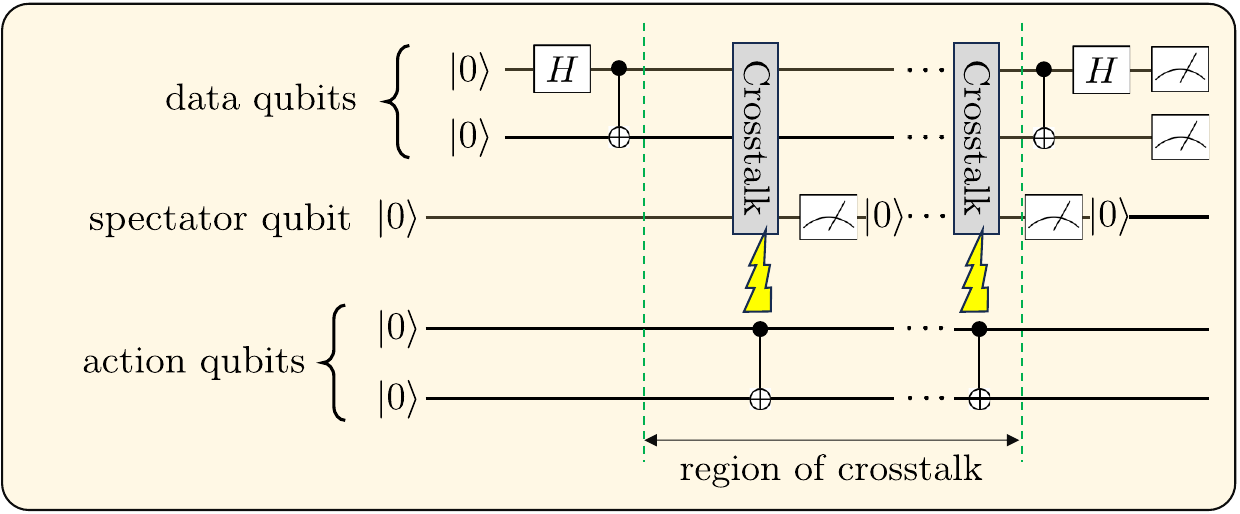}}\\
    \subfloat[\label{fig: all params amplified Bell state IDT} ]{\includegraphics[width=0.5\linewidth]{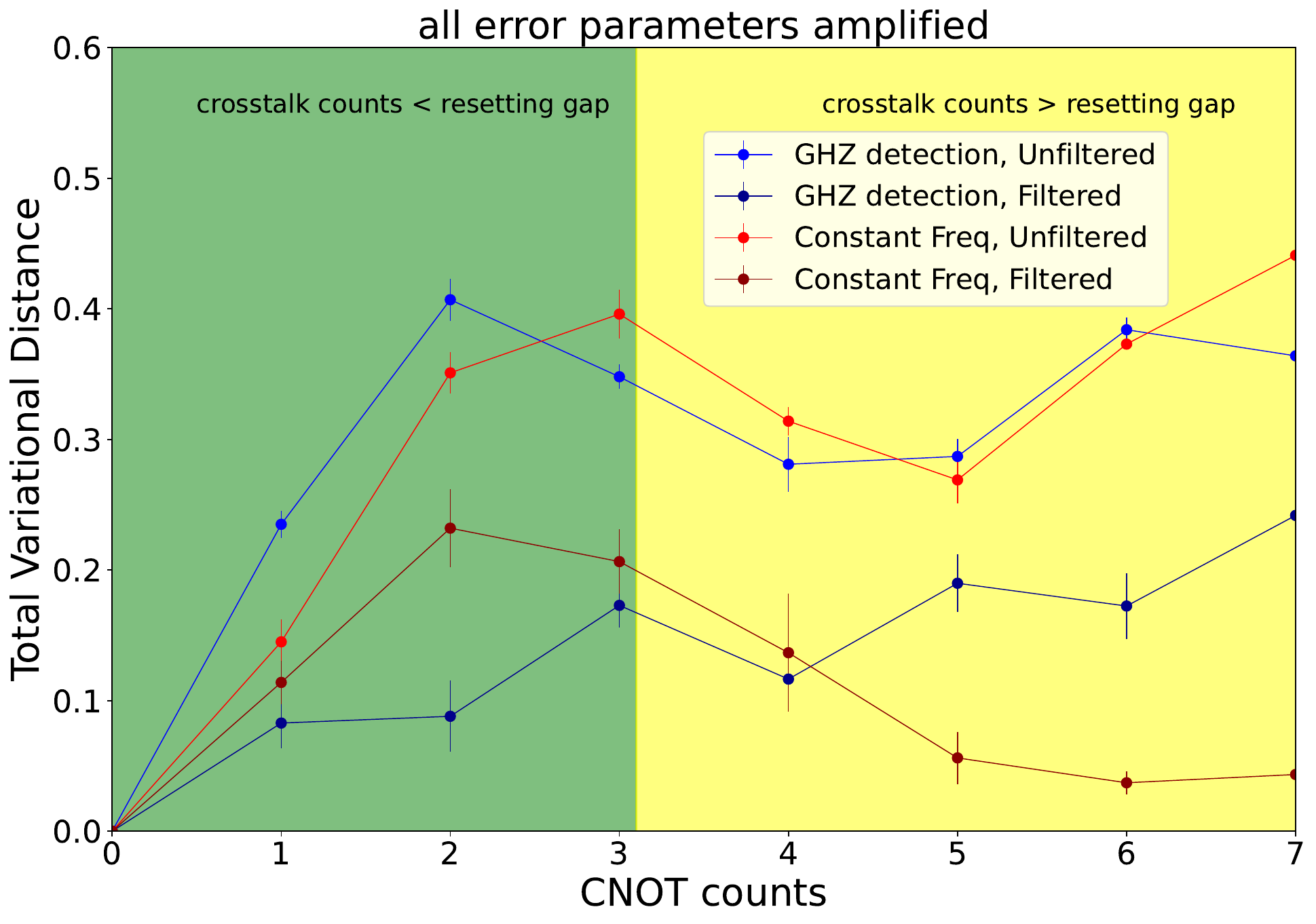}}
    \subfloat[\label{fig: single H amplified Bell state IDT} ]{\includegraphics[width=0.5\linewidth]{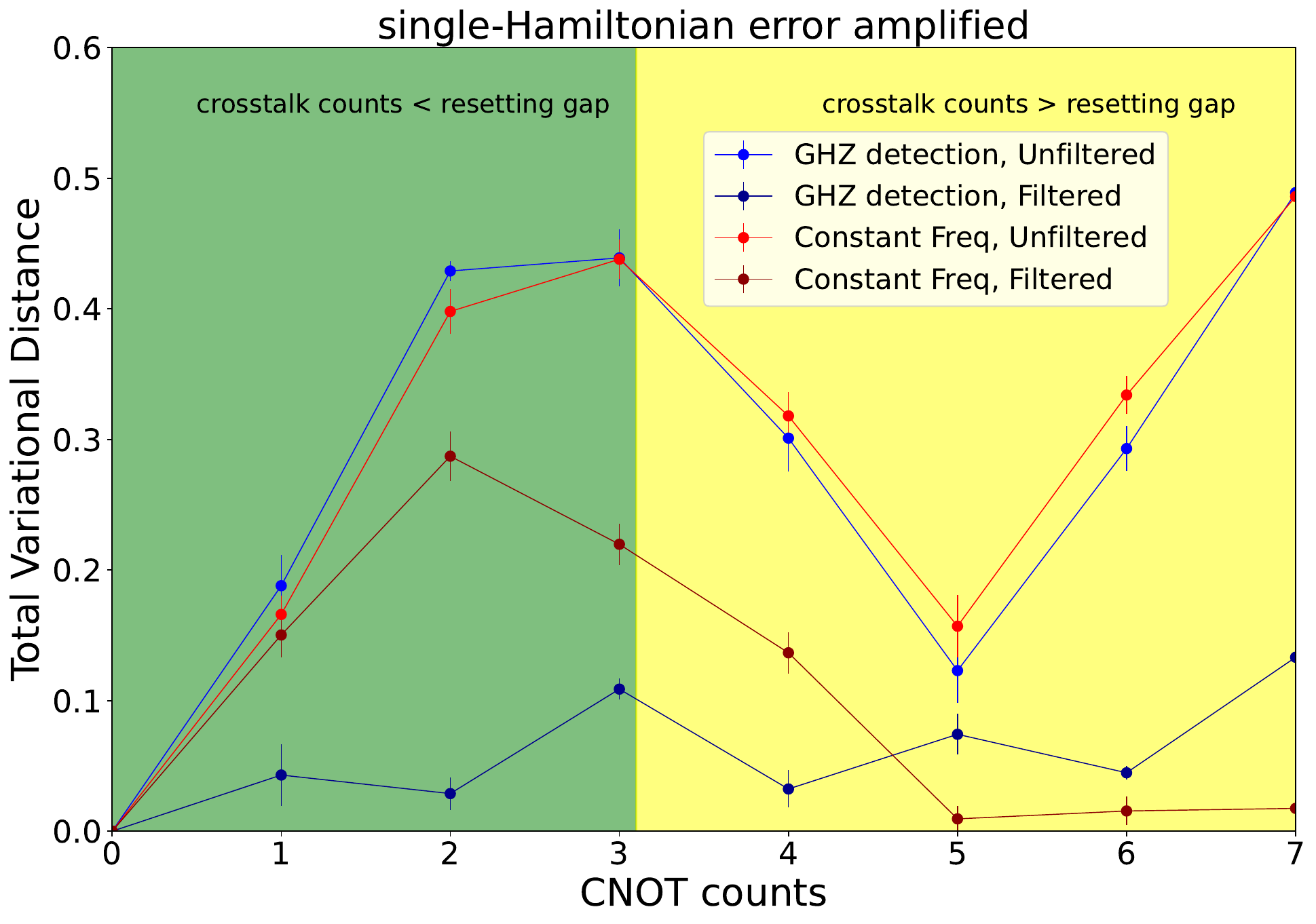}}
    \caption{\textbf{(a)} Overview of the quantum circuit layout where two data qubits are prepared in a Bell state for idle tomography, with spectator qubits placed next to them in the modified GHZ state. Crosstalk is generated only within the region of crosstalk from nearby action qubits after both data and spectator qubits are prepared. Shots where crosstalk is detected are discarded from the total counts. The green-shaded region ($\le4$ counts) has the significance that the perturbation counts below this threshold are unlikely to be detected with high probability by the constant-period strategy. \textbf{(b)} The same circuit layout with only one spectator qubit measured and reset after a constant period, which contains the number of available layers that gives the highest probability of detecting the crosstalk by measuring the result $\ket{1}$. \textbf{(c)} Plots showing the total variational distance (TVD) between the ideal measured result (all $\ket{00}$s) and the bit-string counts measured from the noisy state versus the number of gates, which are distributed randomly within the `region of crosstalk'. All noise parameters in the HSA model are amplified to match the condition in \Cref{eq: total angle condition} (by the same factor as in \Cref{fig: detection rates}). The proportion of the shots that are perturbed at least once is 0.5. The half-length of the error bars is calculated based on two times the standard error sampled from 12 trials. \textbf{(d)} Identical to (c), except that only the single-qubit Hamiltonian noise parameters are amplified to match \Cref{eq: total angle condition}}
    \label{fig: Bell state IDT}
\end{figure*}

\subsection{Improving data qubits fidelity using CSMQC}
Apart from the detection success rate, we further investigate the performance of CSMQC in a realistic scenario where the data qubits are also included as described in \Cref{fig: fig1}. We also compare the results of CSMQC with the traditional constant-frequency detection strategy using a single spectator qubit as in \cite{Ben_crosstalk}, thus demonstrating its utility in filtering the shots contaminated by crosstalk perturbations.

\begin{figure*}[t!]
    \centering
    \subfloat[\label{fig: random circuit} ]{\includegraphics[width=0.52\linewidth]{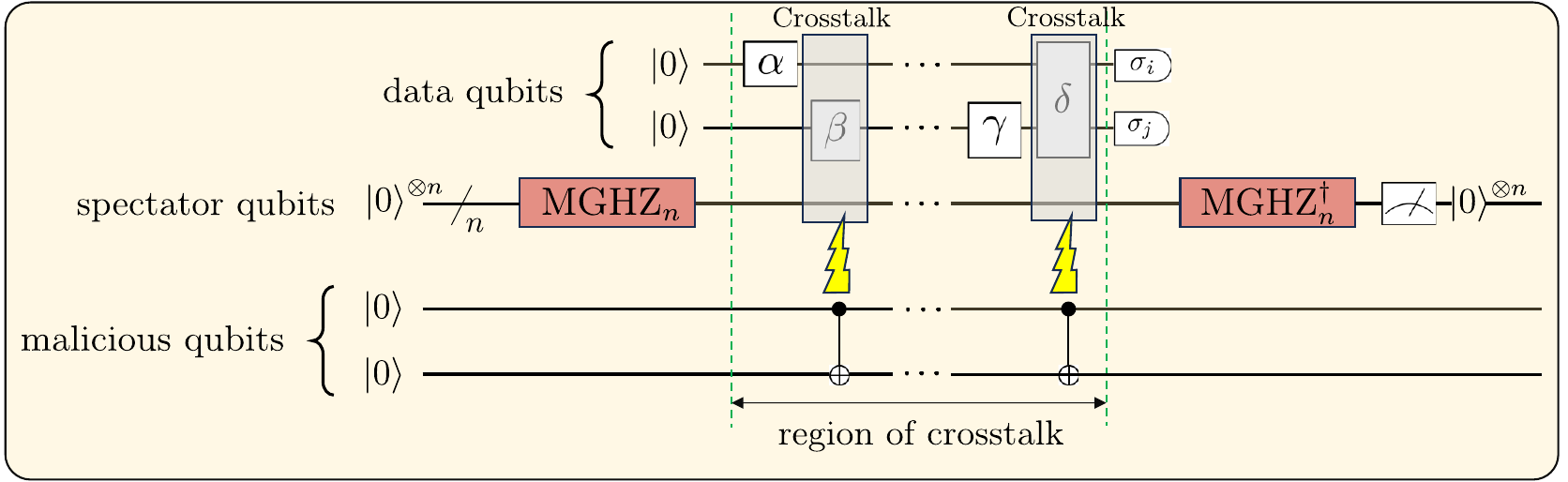}}
    \subfloat[\label{fig: random circuit constant freq} ]{\includegraphics[width=0.425\linewidth]{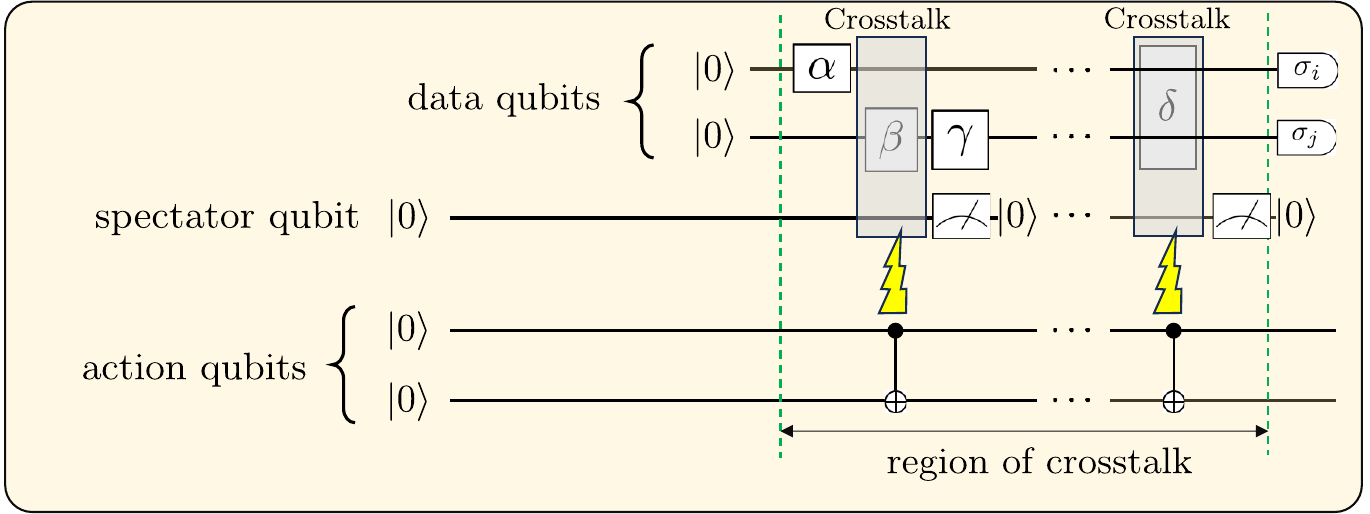}}\\
    \subfloat[\label{fig: all error parameters amplified random circuits} ]{\includegraphics[width=0.5\linewidth]{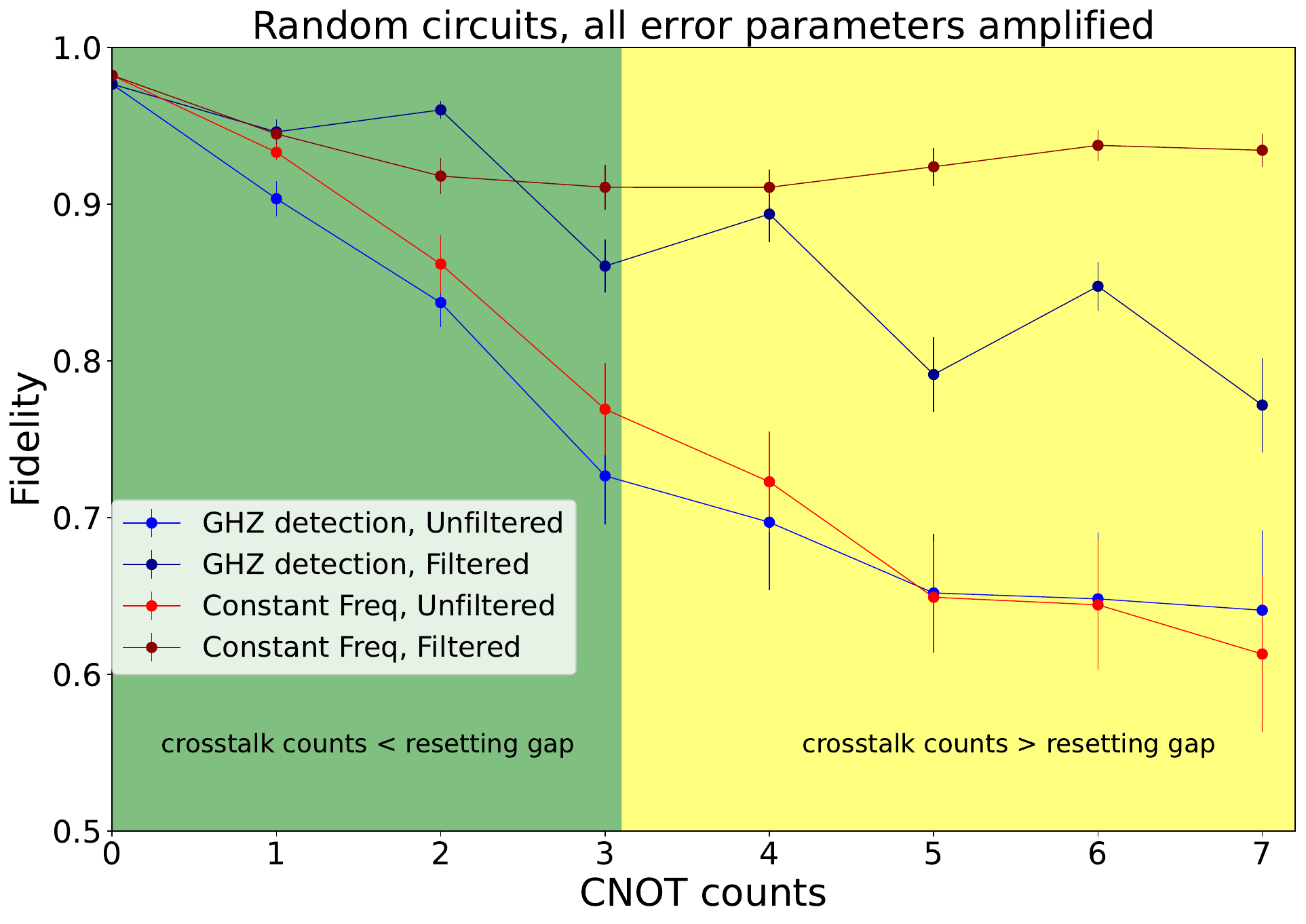}}
    \subfloat[\label{fig: single H amplified random circuits} ]{\includegraphics[width=0.5\linewidth]{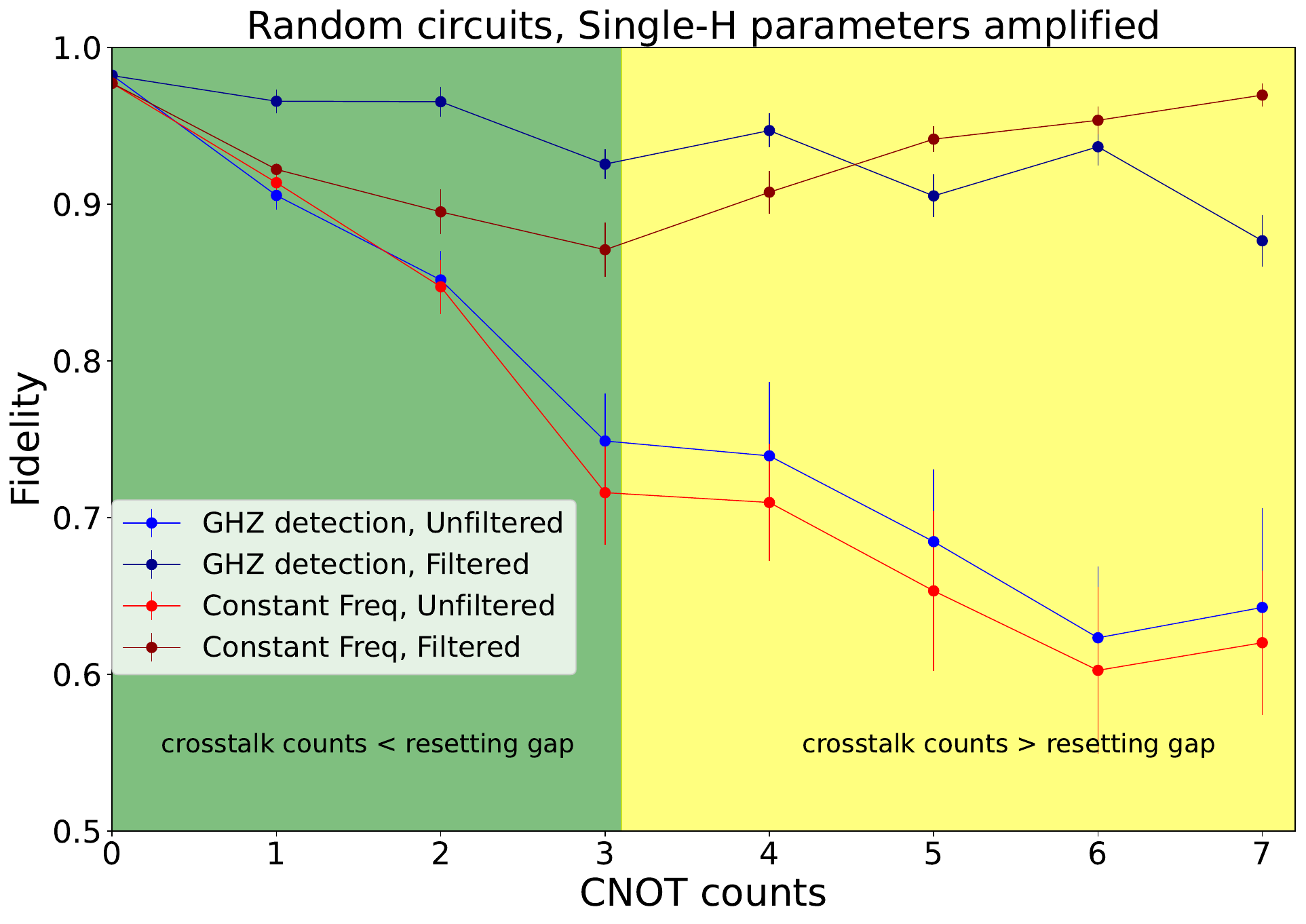}}
    \caption{\textbf{(a)} Overview of the quantum circuit layout when a random circuit is generated on the data qubits, with spectator qubits placed next to it in the modified GHZ state. Same as in \Cref{fig: Bell state IDT}, crosstalk is generated from nearby action qubits after both data and spectator qubits are prepared. The shots with crosstalk detected are discarded from the overall counts. When a gate in the random circuit coincides with the crosstalk evolution in the same layer, the crosstalk is applied first. \textbf{(b)} Same circuit layout, with only one spectator qubit measured and reset at a constant period, which is the optimal number of available layers that gives the highest probability of detecting the crosstalk\cite{Ben_crosstalk} by measuring the result $\ket{1}$. \textbf{(c)} Plots showing the average fidelity between every ideal final state generated by the corresponding random circuit to the corresponding noisy state versus the number of perturbations distributed arbitrarily within the `region of crosstalk'. All noise parameters in the HSA model are amplified to match the condition in \Cref{eq: total angle condition} (by the same factor as in \Cref{fig: detection rates}). The proportion of the shots that are perturbed by crosstalk at least once is 0.5. The half-length of the error bars is calculated based on two times the standard error sampled in 12 trials (Hence 12 distinct random circuits). \textbf{(d)} Same as (c), but only single-qubit Hamiltonian noise parameters are amplified to match \Cref{eq: total angle condition}}
    \label{fig: random circuit benchmark}
\end{figure*}

In the first demonstration, we place the data qubits together with the spectator qubits and prepare them in a Bell state for idle tomography as shown in \Cref{fig: Bell state IDT circuit} and \Cref{fig: Bell state IDT circuit constant freq}. After the crosstalk, the Bell state is unprepared and its `orthogonality' is determined through the total variational distance (TVD) of the measured bit-string counts to the bit-string counts of the ideal state:
\begin{equation}
    \text{TVD}(P,P_{\text{ideal}})= \sup\limits_{E\in \Omega}{\abs{P(E)-P_{\text{ideal}}(E)}},
\end{equation}
where in the context of this work, $P$ is the normalised bit-string counts distribution of the sample and $P_{\text{ideal}}$ is the ideal normalised bit-string counts distribution. Since the ideal state of the data qubits before the measurement is $\ket{00}$, the TVD can be simplified to the proportion of the measured results that do not belong to $\ket{00}$. For comparison, the experiments are conducted with spectator qubits prepared in the modified GHZ state $\ket{\text{MGHZ}_n(\bm{K})}$ (\Cref{fig: Bell state IDT circuit}) as introduced in this work and with just one spectator qubit that is measured and reset after a constant period \cite{Ben_crosstalk} (\Cref{fig: Bell state IDT circuit constant freq}). This method assumes crosstalk occurs in every possible circuit layer between two measurement-and-reset pairs, with the resetting period optimized for the highest probability of measuring $\ket{1}$. Similar to \Cref{subsec: CSMQC Detection success rate analysis}, the crosstalk induced by CNOT gates in the construction and inversion of $\ket{\text{MGHZ}_n(\bm{K})}$ is also included. However, we set the CNOT gate on data qubits to be crosstalk-free so that the impact of crosstalk comes from action and spectator qubits only. Nevertheless, in practice, the perturbation on $\ket{\text{MGHZ}_n(\bm{K})}$ caused by crosstalk from data qubits would be relatively easy to cancel by applying the inverse of the benchmarked crosstalk rotation. In both \Cref{fig: all params amplified Bell state IDT} with all parameters in reduced HSA noise model amplified and \Cref{fig: single H amplified Bell state IDT} with only the single-qubit Hamiltonian noise parameters being amplified the same as in \Cref{fig: detection rates}, the TVD of the state after filtering the shots with detected crosstalk using CSMQC outperforms the constant-period strategy for the number of perturbations smaller than the resetting period of 4 layers. Thus, the total number of measures and resets is two within 7 layers. The significance of this comes from the fact that 4 layers of crosstalk perturbations yield the highest probability of measuring the state $\ket{1}$, and numbers below this threshold are unlikely to be detected with high probability by the constant-period strategy, as indicated by the green-shaded region. Of course, in a realistic scenario without amplifying the noises, the number of available layers in the resetting period would roughly increase proportionally, extending the domain where CSMQC outperforms the constant-frequency strategy. Therefore, this result demonstrates the utility of CSMQC in recognising crosstalk with noise from IBM superconducting quantum computers, especially when the number of perturbations is sparse yet still harmful to the quantum states of the data qubits. When the number of perturbations increases, the performance of CSMQC largely depends on how dominant the single-qubit crosstalk rate is compared to other channels: by comparing the results in \Cref{fig: all parameters amplified} with \Cref{fig: single H amplified Bell state IDT}, it reveals that the TVD after filtering scales much more slowly when the ratio between the single-qubit Hamiltonian noise parameters to other noises are amplified. 

In the next demonstration, we conduct similar experiments as in the earlier ones, comparing the performance of CSMQC (\Cref{fig: random circuit}) and the traditional constant-frequency strategy (\Cref{fig: random circuit constant freq}) using the amplified crosstalk noise parameters. However, instead of evaluating the TVD of the Bell state idle tomography, we apply random circuits to the data qubits and evaluate the final state fidelity, both filtered when crosstalk is detected and unfiltered, against the ideal state. The random circuits are constructed by randomly selecting a gate from the Clifford set generators $\{H,S,\text{CNOT}\}$ plus the $T$ gate to a random qubit (for CNOT we choose a random orientation) to every available layer in the region of crosstalk. The fidelity of the final state is computed as follows: 
\begin{equation}
    F(\rho,\rho^{\text{ideal}})=\Tr\left(\rho\rho^{\text{ideal}}\right),
\end{equation}
where $\rho$ is the noisy state density matrix
\begin{equation}
    \rho = (1-\lambda)\rho^{\text{ideal}}+\lambda\sum\limits_{i\in\text{crosstalk patterns}}p_i\mathcal{E}_i(\rho^{\text{ideal}})
\end{equation}
and $\rho^{\text{ideal}}$ is the ideal density matrix of a pure state generated by a random circuit. In the unfiltered case, $\lambda=0.5$ is consistent with the proportion of shots containing crosstalk the same as in \Cref{fig: Bell state IDT}. In the filtered case, where the shots with detected crosstalk are removed, we expect $\lambda<0.5$. The noisy state density matrix $\rho$ is reconstructed via Quantum State Tomography, a canonical way of obtaining a full picture of any quantum state by measuring the state in different Pauli basis \cite{nielsen_chuang}:
\begin{equation}
    \rho =\sum\limits_{\sigma_{v}\in\{I,X,Y,Z\}^{\otimes n_{D}}}{\frac{\Tr\left({\sigma}_{v}\rho  \right){\sigma}_{v}}{{2}^{n_{D}}}},
\end{equation}
where $n_D$ is the number of data qubits involved. In \Cref{fig: all error parameters amplified random circuits} and \Cref{fig: single H amplified random circuits}, we observe a trend that further reinforces our observation from \Cref{fig: Bell state IDT}: The fidelity of the state after filtering shots with detected crosstalk using CSMQC outperforms the constant-frequency strategy when the number of crosstalk perturbations is sparse or fewer than the available layers in the resetting period, and the rate of fidelity decaying is strongly correlated with the dominance of the single-qubit crosstalk rotation rate relative to other noise channels.



\section{Discussion}
In this paper, we have presented a novel strategy, Crosstalk-Spectating Quantum Multiple Coherences (CSQMC), to indicate the presence of crosstalk in a shared quantum computing environment. This work focused on the context where crosstalk could interfere with the quantum circuits executed by multiple users, who lack the knowledge about when this is going to happen due to privacy constraints. By amplifying the perturbation signal generated by crosstalk, even sparsely distributed crosstalk over the time domain can be detected. 

We constructed an artificial model to quantitatively analyse the detection success and false negative probabilities of CSMQC under different challenges of excessive noise. These included the idle noise generated by the always-on ZZ couplings, dephasing and amplitude damping channels; other sub-dominant crosstalk channels such as the two-qubit coherent rotations; as well as the total angle mismatching in the worst-case scenario. Simulations showed that CSMQC performs effectively when the ratio between the two-qubit and single-qubit crosstalk coherent rotation noise is less than $10^{-1}$, and increasing the number of spectator qubits mitigates the impact of total angle mismatching. 

Beyond the artificial model, we examined the performance of CSQMC's detection success using the crosstalk noise parameters benchmarked from real IBM quantum computers. These results agreed with the prediction from the artificial model. In the context of attack and defence, where malicious users launch the crosstalk attacks using crosstalking gates, CSMQC demonstrated clear utility over the traditional constant-frequency strategy, particularly when the number of attacks is smaller than the available number of layers in the resetting period. When the number of attacks is larger than the number of layers in the resetting period, it is the constant-frequency strategy that demonstrates utility. However, in a real scenario, it is very unlikely that a user will apply the CNOTs at a frequency close to the level considered to be in the region of utility of the constant-frequency strategy. Even in the context of attack and defence, performing an attack with such frequency would significantly increase the chance of being compromised. Due to the complementarity between the two methods in their region of utility, our proposed CSQMC and constant-frequency can be adopted together to improve the overall performance. 

In an application, the entanglement in the modified GHZ  state is itself noise-sensitive, which is a drawback of this strategy. However, as long as reasonable coherence is preserved, the probability of detecting the effects of sparse crosstalk will be amplified, and we are confident that this has been achieved. We also noted that CSMQC's performance still heavily depends on how dominant the single-qubit crosstalk rate is, which is less favourable in one of the latest quantum computers. From another perspective, however, the impact of crosstalk becomes less critical or even negligible, as the noise caused by crosstalk is indistinguishable from the background idle noise. Moreover, as the protocol depends on interactions between physical qubits only, it can run in parallel on the boundary of two operating error-correcting codes owned by different users. However, whether this can outperform the error-correcting codes themselves is still an open question for future research.


\begin{acknowledgments}
The authors greatly acknowledge the University of Melbourne through the establishment of an IBM Quantum Network Hub at the University. Computational resources were provided by the National Computing Infrastructure (NCI) through the National Computational Merit Allocation Scheme (NCMAS). H.K. was supported by the Australian Government Research Training Program Scholarship. M.S. was supported by Australian Research Council Discovery Project DP210102831. Both M.S. and M.U. were supported by funding through the Australian Army Quantum Technology Challenge.
\end{acknowledgments}

\textbf{Data Availability Statement} \par
The code and datasets used in the current study are openly available \cite{Program_Codes}.

\appendix
\section{State evolution in CSMQC}\label{appendix: GHZ evolution}
First, before executing any computation on the data qubits, a GHZ state is prepared on the $n$ spectator qubits, which makes the overall state comprised of $d$ data ($T_D$), $n$ spectator qubits ($T_S$) and two action qubits ($A$) into 
\begin{equation}
\begin{aligned}
    \ket{\psi(t_1)}&= \ket{0}^{\otimes d} \otimes \frac{\ket{0}^{\otimes n}+ \ket{1}^{\otimes n}}{\sqrt{2}}\otimes\ket{00}\\
    &=\ket{0}^{\otimes d} \otimes \ket{\text{GHZ}_n}\otimes \ket{00},
\end{aligned}
\end{equation}
where $\ket{\psi(t_1)}\in \mathcal{H}_{T_D}\otimes \mathcal{H}_{T_S} \otimes \mathcal{H}_A$, and $\dim{\mathcal{H}_{T_D}}=2^d$, $\dim{\mathcal{H}_{T_S}}=2^n$, $\dim{\mathcal{H}_A}=4$. Then, the GHZ state prepared on the spectator qubits is modified into $\ket{\text{MGHZ}_n(\bm{K})}$ by rotating each qubit to match the rotation axes for each qubit found in the characterisation process as outlined in the assumption made in \Cref{subsec: conditions and assumptions},
\begin{equation}\label{eq: MGHZ}
\begin{aligned}
    \ket{\text{MGHZ}_n(\bm{K})} &= \left(\bigotimes\limits_{i=1}^{n}{U(\bm{k}_i)}\right)\ket{\text{GHZ}_n}\\
    &=\frac{\ket{\bm{k}_1^{+},\bm{k}_2^{+},\cdots,\bm{k}_n^{+}}+ \ket{\bm{k}_1^{-},\bm{k}_2^{-},\cdots,\bm{k}_n^{-}}}{\sqrt{2}},
\end{aligned}
\end{equation}
where $\bm{K}=(\bm{k}_1,\bm{k}_2,\cdots,\bm{k}_n)$ is the $n\cross 3$ tensor such that $\bm{k}_i$ is the rotation axis found on spectator qubit $i$. The unitary $U(\bm{k}_i)$ denotes the operation that maps a qubit from state $\ket{0}$ to state $\ket{\bm{k}_i^{+}}$ that lies on position aligned with vector $\bm{k}_i$ on the Bloch sphere, so that
\begin{equation}
    U(\bm{k}_i)\ket{0} = \ket{\bm{k}_i^{+}}, U(\bm{k}_i)\ket{1} = \ket{\bm{k}_i^{-}},
\end{equation}
and it is easy to show that $\ket{\bm{k}_i^{-}}$ corresponds to the state lies on the position anti-aligned with the vector $\bm{k}_i$ on the Bloch sphere. For conciseness, we will denote $\mathcal{U}\coloneqq\bigotimes\limits_{i=1}^{n}{U(\bm{k}_i)}$ in the remaining parts of this paper. We also apply the modified stabiliser for $\ket{\text{MGHZ}_n(\bm{K})}$ to implement dynamical decoupling (DD) on spectator qubits, which evolves the overall state into
\begin{equation}
    \ket{\psi(t_2)}=\ket{0}^{\otimes d} \otimes \ket{\text{MGHZ}_n(\bm{K})}\otimes \ket{00}.
\end{equation}
This aims to suppress other noises that do not result from crosstalk during the detection but from the preparation circuit and its Hermitian conjugate. The detailed deduction is discussed in \Cref{appendix: DD}.

After preparing the spectator qubits, crosstalk perturbations are launched to disrupt both data and spectator qubits, which as shown in \Cref{fig:MQC detector}. According to the condition set previously in \Cref{eq: mixing channels} and \ref{eq: total angle condition}, after a total of $m$ crosstalking gates has been executed on the action qubits, the state evolves into
\begin{equation}\label{eq: amplifying rotations}
\begin{aligned}
    &\left((1-\epsilon)e^{-imH}\rho_{D,S}e^{imH}+\epsilon \mathcal{E}_{\text{other}}(\rho_{D,S})\right)\otimes U^{\otimes m }\rho_{M}U^{\dagger\otimes m }\\
    =&(1-\epsilon)e^{-imH}\rho_{D,S}e^{imH}\otimes U^{\otimes m }\rho_{M}U^{\dagger\otimes m } + O(\epsilon),
\end{aligned}
\end{equation}
where $H=\sum\limits_{i=1}^{n+d}{\bm{k}_i\cdot \bm{\sigma}_i}=H_{T_S}+H_{T_D}$ is the single-qubit Hamiltonian generator for the rotation around $\bm{k}_i$ for each qubit. Based on our assumption of small $\epsilon$, the $O(\epsilon)$ term is neglected for ease of further computation. Therefore, \Cref{eq: amplifying rotations} can be approximated to
\begin{equation}\label{eq: amplifying rotations2}
    e^{-imH}\rho_{D,S}e^{imH}\otimes U^{\otimes m }\rho_{M}U^{\dagger\otimes m }\coloneqq\ket{\psi(t_3)}\bra{\psi(t_3)}.
\end{equation}

Consequently, the overall state after sensing crosstalk noise is given by
\begin{align}\label{eq: amplifying rotations3}
&\ket{\psi(t_3)}=\left(\bigotimes\limits_{i=n+1}^{n+d}{R_{\bm{k_i}}(\delta_i)}\right)\ket{0}^{\otimes d}\otimes\\
&\frac{\ket{\bm{k}_1^{+},\bm{k}_2^{+},\cdots,\bm{k}_n^{+}}+e^{-im\sum\limits_{i=1}^{n}{\delta_i}} \ket{\bm{k}_1^{-},\bm{k}_2^{-},\cdots,\bm{k}_n^{-}}}{\sqrt{2}}\otimes U^{m}\ket{00}.
\end{align}

To detect the amplified phase $\theta=m\sum\limits_{i=1}^{n}{\delta_i}$ as a function of the number of crosstalk perturbations $m$, we apply the inverse of the modified GHZ state preparation circuit and measure the last qubit of the spectator qubits which has its probability amplitude as a cosine function of the total rotated angle. Therefore, after applying the inverse gates, the state on the last spectator qubit with label $n$ is expressed as
\begin{equation}
\begin{aligned}
    \text{Tr}_{S_{n-1},D,M}(\rho(t_4))=\frac{1}{2}\Big(&(1+\cos(\theta))\ket{0}\bra{0}-\sin(\theta)\ket{0}\bra{1}\\
    +&\sin(\theta)\ket{1}\bra{0}+(1-\cos(\theta)\ket{1}\bra{1}\Big),
\end{aligned}
\end{equation}
with the probability of measuring 1 equal to
\begin{equation}
    P(\ket{1}_n)= \frac{1-\cos(\theta)}{2}.
\end{equation}

\section{Modified GHZ state dynamical decoupling}\label{appendix: DD}

From the stabiliser group $\mathcal{S}$ of an arbitrary GHZ state of $n$ qubits, we have $X^{\otimes n}\in \mathcal{S}$ such that
\begin{equation}\label{eq: GHZ stabilizer}
    X^{\otimes n}\ket{\text{GHZ}_n} = \ket{\text{GHZ}_n},
\end{equation}
which is also used as the pulse of dynamical decoupling (DD) in MQC \cite{MQC2}.
By inserting an identity operator $\mathcal{U}^{\dagger}\mathcal{U}=\mathds{1}$ decomposed as a product of unitary and its Hermitian conjugate, \Cref{eq: GHZ stabilizer} becomes
\begin{equation}
    X^{\otimes n}\mathcal{U}^{\dagger}\mathcal{U}\ket{\text{GHZ}_n} = \ket{\text{GHZ}_n},
\end{equation}
which simplifies down to
\begin{equation}\label{eq: GHZ stabilizer2}
     X^{\otimes n}\mathcal{U}^{\dagger}\ket{\text{MGHZ}_n(\bm{K})} = \ket{\text{GHZ}_n}.
\end{equation}
By applying the modifying unitary to both sides of \Cref{eq: GHZ stabilizer2}, we obtain
\begin{equation}\label{eq: GHZ stabilizer3}
\begin{aligned}
    \mathcal{U}X^{\otimes n}\mathcal{U}^{\dagger}&\ket{\text{MGHZ}_n(\bm{K})}\\
    =&\ket{\text{MGHZ}_n(\bm{K})}.
\end{aligned}
\end{equation}
Therefore, the transformed stabiliser as well as the DD operation for the modified GHZ state can be found as
\begin{equation}
    \Tilde{S}\coloneqq\mathcal{U}X^{\otimes n}\mathcal{U}^{\dagger}.
\end{equation}
After some simplification, the unitary operation that maps the ordinary GHZ state $\ket{\text{GHZ}_n}$ to the modified GHZ state with DD can be expressed as
\begin{equation}
    \mathcal{U}X^{\otimes n}.
\end{equation}
We also show that rotations around the axes in $\bm{K}$ are guaranteed to be decoupled by this modified stabiliser from the simple anti-commutation relation between Pauli-$X$ and $Z$ operators:
\begin{equation}
    \begin{aligned}
        \{X^{\otimes n}, Z^{\otimes n}\}&=0\\
        \mathcal{U}\{X^{\otimes n}, Z^{\otimes n}\}\mathcal{U}^{\dagger}&=0\\
        \mathcal{U}X^{\otimes n}Z^{\otimes n}\mathcal{U}^{\dagger}+\mathcal{U}Z^{\otimes n}X^{\otimes n}\mathcal{U}^{\dagger}&=0\\
        \mathcal{U}X^{\otimes n}\mathcal{U}^{\dagger}\mathcal{U}Z^{\otimes n}\mathcal{U}^{\dagger}+\mathcal{U}Z^{\otimes n}\mathcal{U}^{\dagger}\mathcal{U}X^{\otimes n}\mathcal{U}^{\dagger}&=0\\
        \{\Tilde{S},\mathcal{U}Z^{\otimes n}\mathcal{U}^{\dagger}\}&=0.
    \end{aligned}
\end{equation}
Hence, by changing the basis from $Z^{\otimes n}$ to $\bm{K}$, it is easy to show
\begin{equation}
    \{\Tilde{S},\mathcal{U}\bigotimes\limits_{i}^{n}{R_z(\delta_i)}\mathcal{U}^{\dagger}\}=\{\Tilde{S},\bigotimes\limits_{i}^{n}{R_{\bm{k}_i}(\delta_i)}\}=0.
\end{equation}
\section{Equivalence between \Cref{eq: xyz coord expected value} and \Cref{eq: xyz coord expected value reduced}}\label{appendix: single qubit Pauli operator expected value}
\begin{customthm}{C1}\label{result: C1}
\textit{The expected value of a unitary $U_i$ specifically acting on qubit with label $i$, $\text{tr}(U_i\rho)$, in a quantum system $S$ comprised of $n$ qubits is equal to its expected value with respect to the reduced density operator of the quantum system on qubit $i$, $\text{tr}(U_i\rho_i)$, where $\rho_i=\text{tr}_{S\setminus i}(\rho)$ is the reduced density operator.}
\begin{proof}[Proof.]
    We start the proof from the LHS of the equation, that is the expected value of unitary $U_i$ on qubit $i$ with respect to the quantum system $\rho$: $\text{tr}(U_i\rho)$. Expanding the density operator $\rho$ in terms of its subsequent pure states component, we have:
    \begin{equation}\label{eq: appendix C 1}
    \text{tr}\left(U_i\rho\right)=\text{tr}\left(U_i\sum\limits_{k}{p_k\ket{\psi_k}\bra{\psi_k}}\right),
    \end{equation}
    where $p_k$ is the probability of the underlying pure state $\rho_k=\ket{\psi_k}\bra{\psi_k}$ satisfying the born rule of probability $\sum\limits_{k}{p_k}=1$. In the matrix representation, the trace sums all diagonal matrix elements in the computational basis. Hence, \Cref{eq: appendix C 1} is further expanded as
    \begin{equation}
        \text{tr}\left(U_i\rho\right)=\sum\limits_{\bm{x}\in\{0,1\}^{\otimes n}}{\bra{\bm{x}}U_i\sum\limits_{k}p_k\ket{\psi_k}\bra{\psi_k}\ket{\bm{x}}},
    \end{equation}
    where $\bm{x}=(x_1,x_2,\cdots,x_n)^{T}$, $x_j\in\{0,1\}$ is the $n$-element vector representing the bitstring of a $n$-qubit basis computational state. Therefore, the summation over $\bm{x}$ on qubit $i$ and others can be dealt separately,
    \begin{equation}
        \text{tr}\left(U_i\rho\right)=\sum\limits_{k}{p_k\sum_{\hat{\bm{x}}}\sum_{x_i}{\bra{x_i}\otimes\bra{\hat{\bm{x}}}U_i\ket{\psi_k}\bra{\psi_k}\ket{\hat{\bm{x}}}}\otimes\ket{x_i}},
    \end{equation}
    where $\bm{\hat{x}}=(x_1,\cdots,x_{i-1},x_{i+1},\cdots x_n)^{T}\in\{0,1\}^{\otimes n-1}$ is reduced vector from $\bm{x}$ after removing element $x_i$. By expressing the pure state component $\ket{\psi_k}$ as the linear combination of its underlying orthonormal basis states, the inner product can be rewritten as
    \begin{equation}
    \begin{aligned}
        \bra{\hat{\bm{x}}}U_i\ket{\psi_k}&=\bra{\hat{\bm{x}}}U_i\sum_{\bm{x}'}{c_{\bm{x}'}\ket{\bm{x}'}}\\
        &=\bra{\hat{\bm{x}}}U_i\sum_{\hat{\bm{x}}'}\sum_{x'_i}{c_{\bm{x}'}\ket{\hat{\bm{x}}'}\otimes\ket{x'_i}}\\
        &=\bra{\hat{\bm{x}}}\sum_{\hat{\bm{x}}'}\sum_{x'_i}{c_{\bm{x}'}\ket{\hat{\bm{x}}'}\otimes U_i\ket{x'_i}}\\
        &=U_i\bra{\hat{\bm{x}}}\sum_{\hat{\bm{x}}'}\sum_{x'_i}{c_{\bm{x}'}\ket{\hat{\bm{x}}'}\otimes\ket{x'_i}}\\
        &=U_i\bra{\hat{\bm{x}}}\ket{\psi_k}
        ,
    \end{aligned}
    \end{equation}
    where $c_{\bm{x'}}$ is the amplitude of the component $\bm{x'}$. Since the unitary $U_i$ is uniquely acting on qubit $i$, it can be moved outside from the inner product. Therefore, the expected value of the unitary $U_i$ is simplified to
    \begin{equation}
    \begin{aligned}
    \text{tr}\left(U_i\rho\right)&=\sum_{x_i}\bra{x_i}\otimes U_i\left(\sum_{\hat{\bm{x}}}\bra{\hat{\bm{x}}}\sum\limits_{k}p_k\ket{\psi_k}\bra{\psi_k}\ket{\hat{\bm{x}}}\right)\otimes\ket{x_i}\\
    &=\sum_{x_i}\bra{x_i}\otimes U_i\left(\sum_{\hat{\bm{x}}}\bra{\hat{\bm{x}}}\rho\ket{\bm{\hat{x}}}\right)\otimes\ket{x_i}\\
    &=\sum_{x_i}\bra{x_i}U_i\rho_i\ket{x_i}\\
    &=\text{tr}\left(U_i\rho_i\right).
    \end{aligned}
    \end{equation}
    As above, we have proved the claim \Cref{result: C1}.
\end{proof}
\end{customthm}
\vspace{20pt}
\section{Total angle mismatching}\label{appendix: total angle mismatching}
\begin{figure}[h]
    \centering
    \includegraphics[width=1\linewidth]{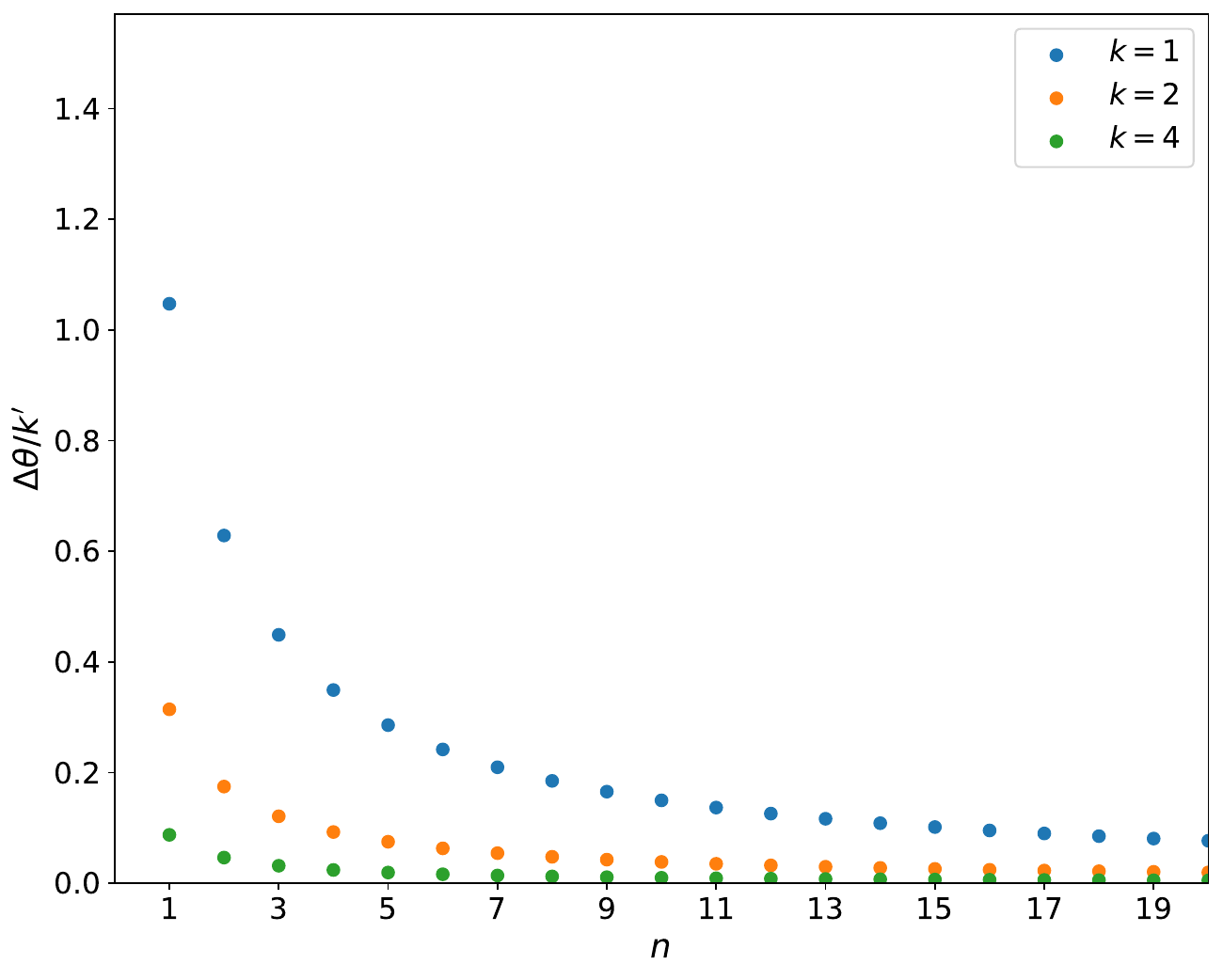}
    \caption{Graphics showing the decaying of $\Delta\theta$ as more spectator qubits are involved/single-qubit crosstalk angle reduces.}
    \label{fig: delta_theta}
\end{figure}

In practice, one should expect that the condition in \Cref{eq: total angle condition} is often not exactly satisfied due to the fact that $\pi$ does not equal the sum of the total angle rotated per crosstalk gate. However, one has the freedom of choosing which are the spectator qubits that constitute the modified GHZ state. Therefore, an upper bound of the deviation from the total rotated phase to $\pi$ is expected. In this analysis, we approach this problem by assuming the single-qubit crosstalk rotation angle is $\theta$ for each spectator qubit for simplicity. Therefore, letting $n$ denote the largest number of spectator qubits such that $\pi-n\theta>0$, the condition for a maximal difference between total rotated phase and $\pi$ (which is responsible for detecting perturbations with counts in $\{1,3,5\cdots\}$) is
\begin{equation}\label{eq: maximal angle difference condition}
    \pi - n \theta = (n+1)\theta - \pi.
\end{equation}
Hence,
\begin{equation}
    \theta=\frac{2\pi}{2n+1}
\end{equation}
The significance of the RHS of \Cref{eq: maximal angle difference condition} is that it guarantees the phase difference to $\pi$ is the same for $n$ and $n+1$ spectator qubits. Therefore, even if $\theta$ is smaller than as in \Cref{eq: maximal angle difference condition}, one could simply choose to add an extra spectator qubit so that the total rotated phase is closer to $\pi$ again. Similarly, the condition for $\theta$ when used to detecting perturbations with counts $\{2,6\cdots\},\{4,\cdots\}$ is given by
\begin{equation}\label{eq: general maximal angle difference condition}
    \begin{aligned}
        \pi - 2n \theta &= 2(n+1)\theta - \pi\\
        \pi - 4n \theta &= 4(n+1)\theta - \pi\\
        &\cdots,
    \end{aligned}
\end{equation}
where, $n$ is the largest integer such that $\pi-2n\theta>0, \pi-4n\theta>0, \cdots$.
Therefore, the general formula for $\theta$ satisfying the above conditions is 
\begin{equation}\label{eq: general angle value}
    \theta=\frac{2\pi}{2kn+1},
\end{equation}
where $k=1,2,4,\cdots$ is the smallest number of perturbations in the set that the modified GHZ state is responsible to detect for. i.e. $k=1$ for $\{1,3,5,\cdots\}$ number of perturbations, $k=2$ for $\{2,6,\cdots\}$, respectively. From above, it is easy to see that the value of $n$ depends on $k$ directly such that the product $kn$ must equal a constant integer, which satisfies $\pi-kn\theta>0$. Deriving from \Cref{eq: general maximal angle difference condition}, the formula for the deviation from the ideal phase $\Delta\theta$ is given by:
\begin{equation}\label{eq: total deviation from ideal phase}
\begin{aligned}
    \Delta\theta&=\theta_{\text{target}}-\theta_{\text{actual}}\\
    &=k'\left(\frac{\pi}{k}-n\theta\right)\\
    &=\frac{\pi k'}{k(2kn+1)},
\end{aligned}
\end{equation}
where $k'$ is the actual number of perturbations and $\theta$ is substituted from \Cref{eq: general angle value}. Therefore, given the number of perturbations $k'$ (which belongs to a particular set of counts with the smallest number of perturbations $k$), the more the spectator qubits are (or the less the angle $\theta$), the less the deviation from the ideal phase as in \Cref{fig: delta_theta}.

When the ideal phase $\theta_{\text{target}}=\pi$, the detection success $P(\ket{1})$/false negative rate $P(\ket{0})$ can be thus written as
\begin{equation}
\begin{aligned}
    P(\ket{1})&=\frac{1-\cos{(\theta_{\text{actual}})}}{2}=\frac{1-\cos{(\pi-\Delta\theta)}}{2}=\frac{1+\cos(\Delta\theta)}{2}\\
    P(\ket{0})&=1-\frac{1-\cos{(\theta_{\text{actual}})}}{2}=\frac{1+\cos{(\pi-\Delta\theta)}}{2}=\frac{1-\cos(\Delta\theta)}{2}.
\end{aligned}
\end{equation}

\section{}\label{appendix: colorplots}
In this section, we present the colour plots depicting false negative rates (P($\ket{0}$)) as functions of the ratios of the idling always-on ZZ coupling rate ($\alpha$) and the two-qubit crosstalk rate ($\phi$) to the single-qubit crosstalk rotation rate ($\theta$) in different settings. All relevant plots are referenced and discussed in the main text.

\begin{figure*}[h!]
\vspace{-10pt}
    \centering
    \subfloat[\label{fig: torino Xtalk pi4 6q colorplot}Dephasing and amplitude damping derived using $T_1$ and $T_2$ from $ibm\_torino$, $\theta=\frac{\pi}{4}$ with maximum of 4 spectator qubits]{
    \includegraphics[width=0.95\linewidth]{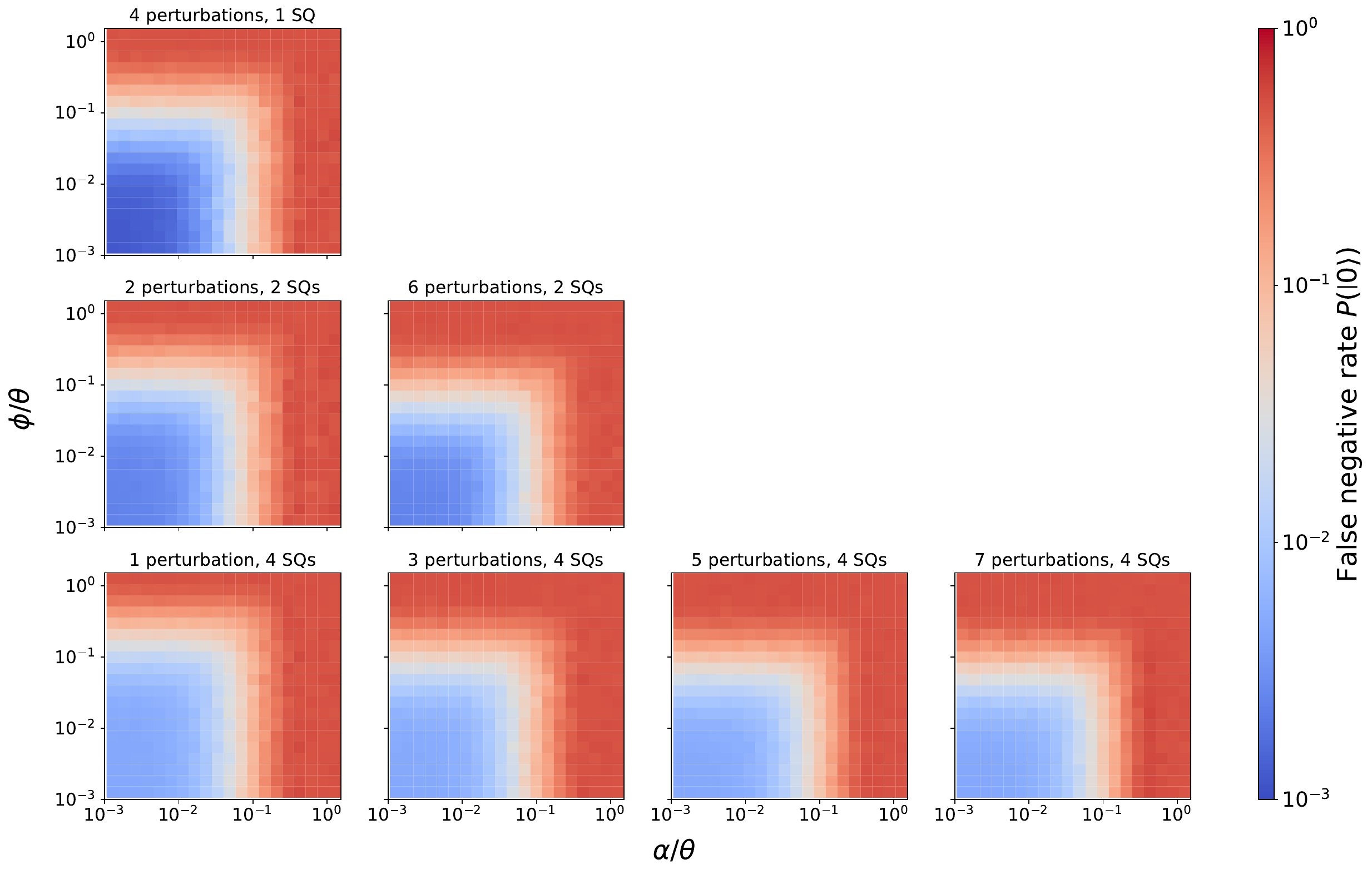}}

    \subfloat[\label{fig: sherbrooke Xtalk pi4 6q colorplot}Dephasing and amplitude damping derived using $T_1$ and $T_2$ from $ibm\_sherbrooke$, $\theta=\frac{\pi}{4}$ with maximum of 4 spectator qubits]{
    \includegraphics[width=0.95\linewidth]{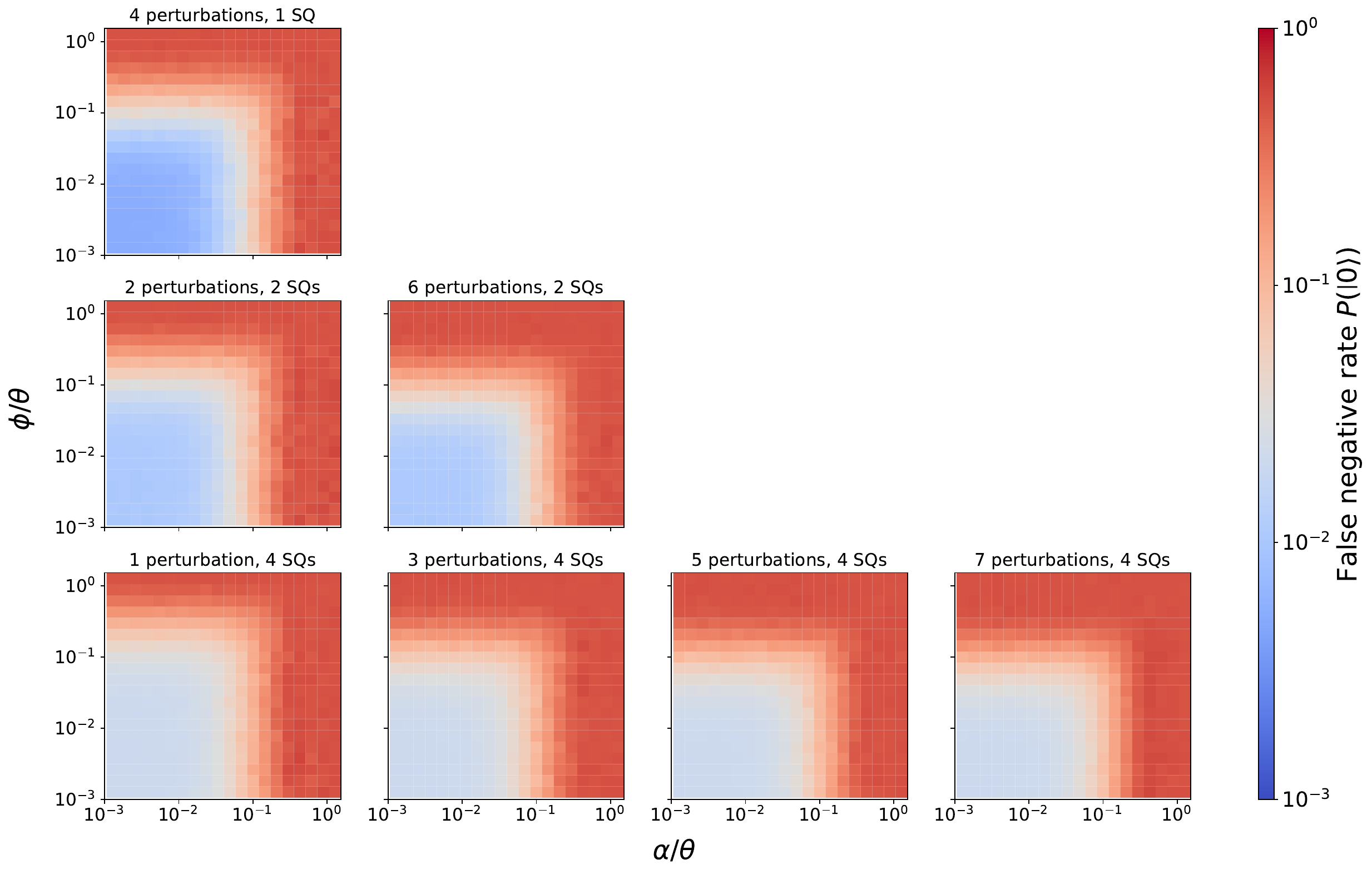}}
    \caption{}\vspace{-50pt}
\end{figure*}


\begin{figure*}[hbtp]
    \ContinuedFloat
    \centering
    \subfloat[\label{fig: pure Xtalk pi4 6q colorplot}No dephasing and amplitude damping, $\theta=\frac{\pi}{4}$ with maximum of 4 spectator qubits]{
    \includegraphics[width=0.95\linewidth]{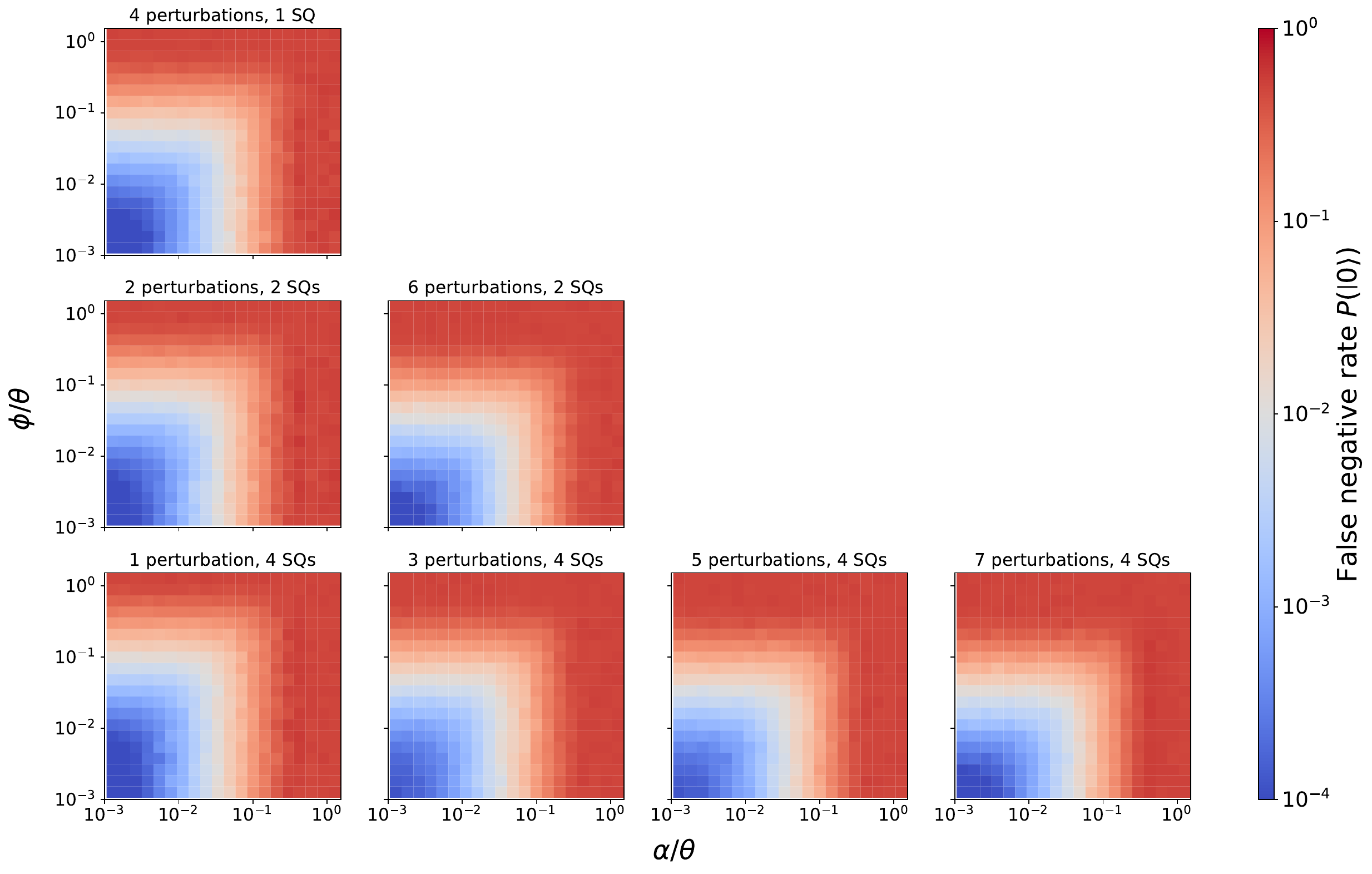}}

    \subfloat[\label{fig: detection failure rate pi4 flat vs SQs}]{
    \includegraphics[width=0.5\linewidth]{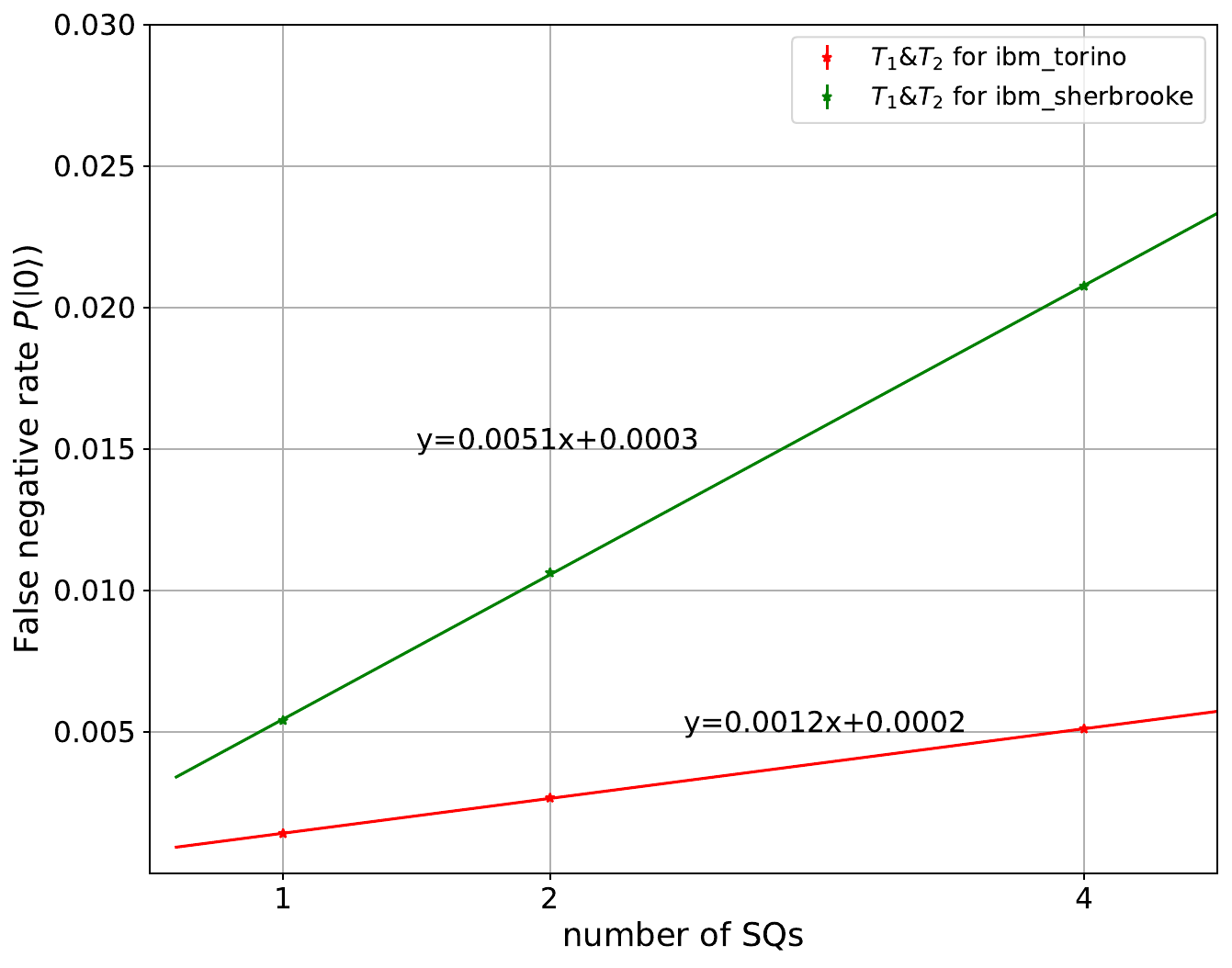}}
    
    \caption{Colour plots on different numbers of crosstalk perturbations with the ratio of ZZ coupling strength to single-qubit crosstalk angle $\theta=\frac{\pi}{4}$ and the ratio of two-qubit crosstalk rate $\phi$ to single-qubit crosstalk angle $\theta$ as the variables. The corresponding surface plots with more details in error bars are included in \cite{Sup_materials}. \textbf{(a)} Used the strengths of dephasing and amplitude damping derived from $T_1$ and $T_2$ of $ibm\_torino$. \textbf{(b)} Used the strengths of dephasing and amplitude damping derived from $T_1$ and $T_2$ of $ibm\_sherbrooke$. \textbf{(c)} No dephasing and amplitude damping channels are included. All variables are shown in a logarithmic scale. The single-qubit crosstalk angle is chosen as $\theta=\frac{\pi}{4}$ to match the largest number of spectator qubits ($m=4$). On top of this, two redundant qubits are also included in the simulation but not part of the modified GHZ throughout the process. \textbf{(d)} Mean false negative rate in the flat region (ratio for two-qubit crosstalk rate and ZZ coupling strength $<10^{-2}$) when dephasing and amplitude damping noise are included, with single-qubit crosstalk angle $\theta=\frac{\pi}{4}$.}
\end{figure*}


\begin{figure*}[hbtp]
    \centering
    \subfloat[\label{fig: torino Xtalk 2pi9 6q colorplot}$T_1$ and $T_2$ from $ibm\_torino$, $\theta=\frac{2\pi}{9}$ with maximum of 4 spectator qubits]{\includegraphics[width=0.95\linewidth]{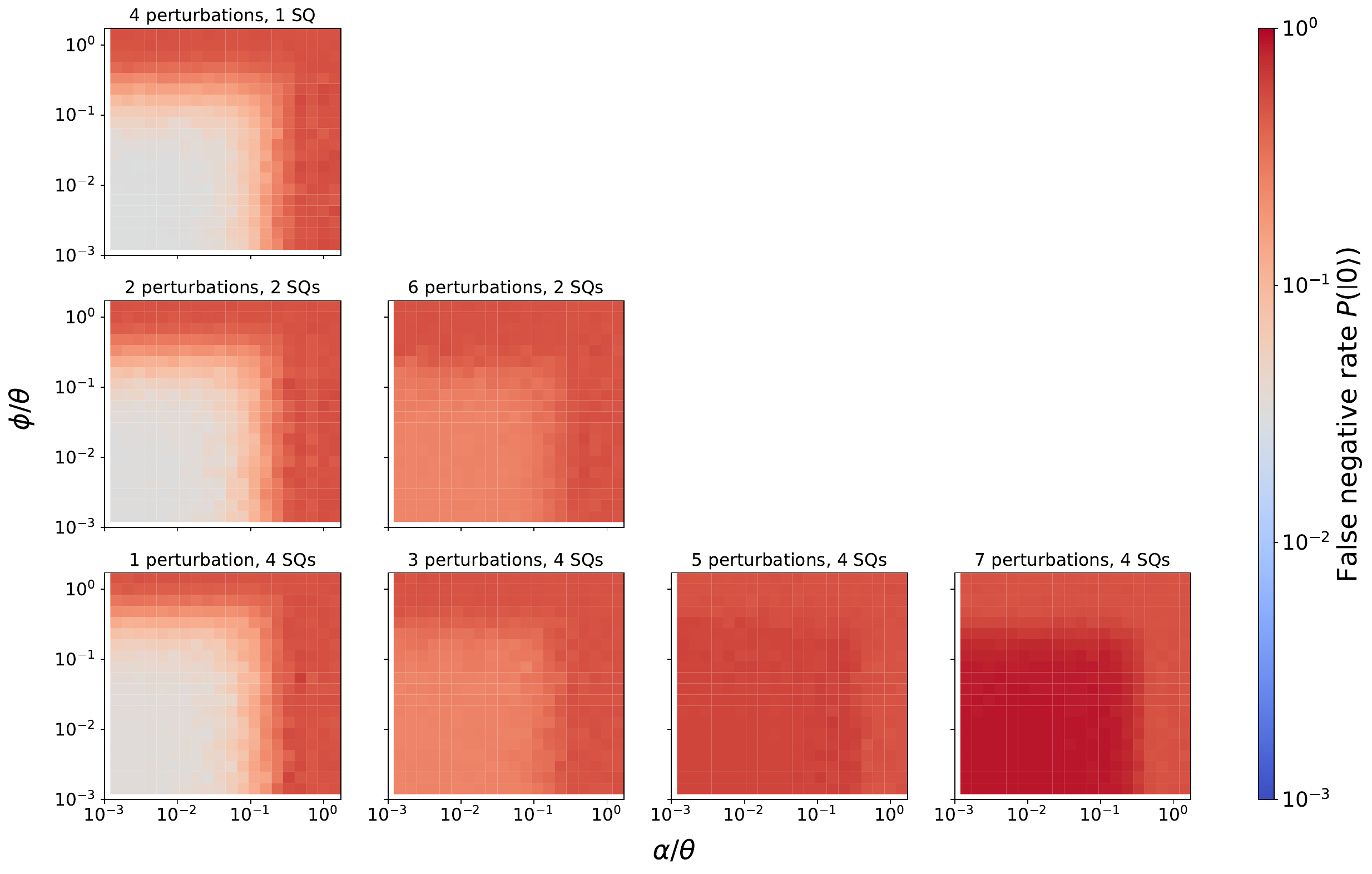}}

    \subfloat[\label{fig: sherbrooke Xtalk 2pi9 6q colorplot}$T_1$ and $T_2$ from $ibm\_sherbrooke$, $\theta=\frac{2\pi}{9}$ with maximum of 4 spectator qubits]{\includegraphics[width=0.95\linewidth]{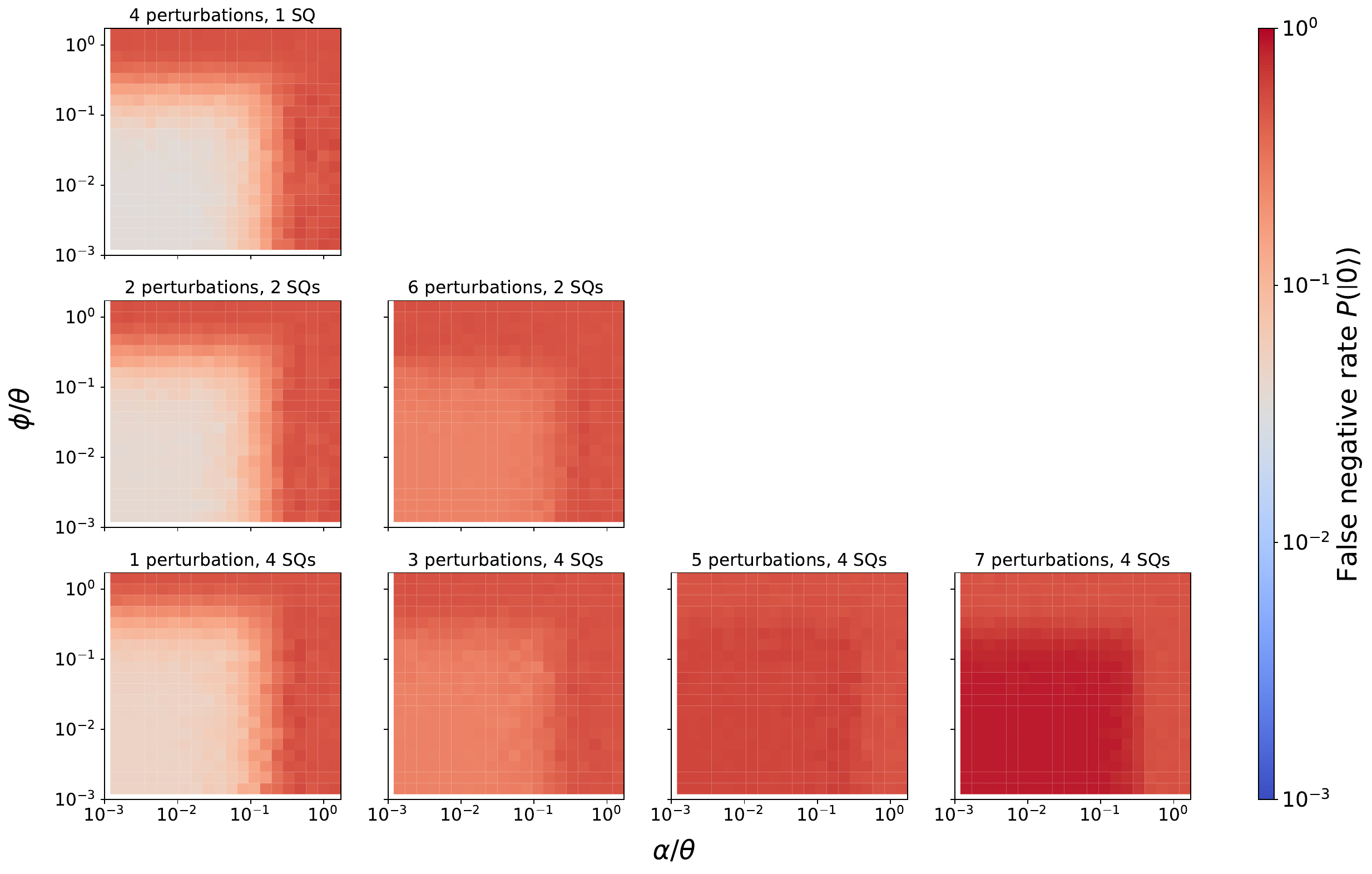}}
    \caption{}
\end{figure*}


\begin{figure*}[hbtp]
    \ContinuedFloat
    \centering
    \subfloat[\label{fig: pure Xtalk 2pi9 6q colorplot}No dephasing and amplitude damping, $\theta=\frac{2\pi}{9}$ with maximum of 4 spectator qubits]{\includegraphics[width=0.9\linewidth]{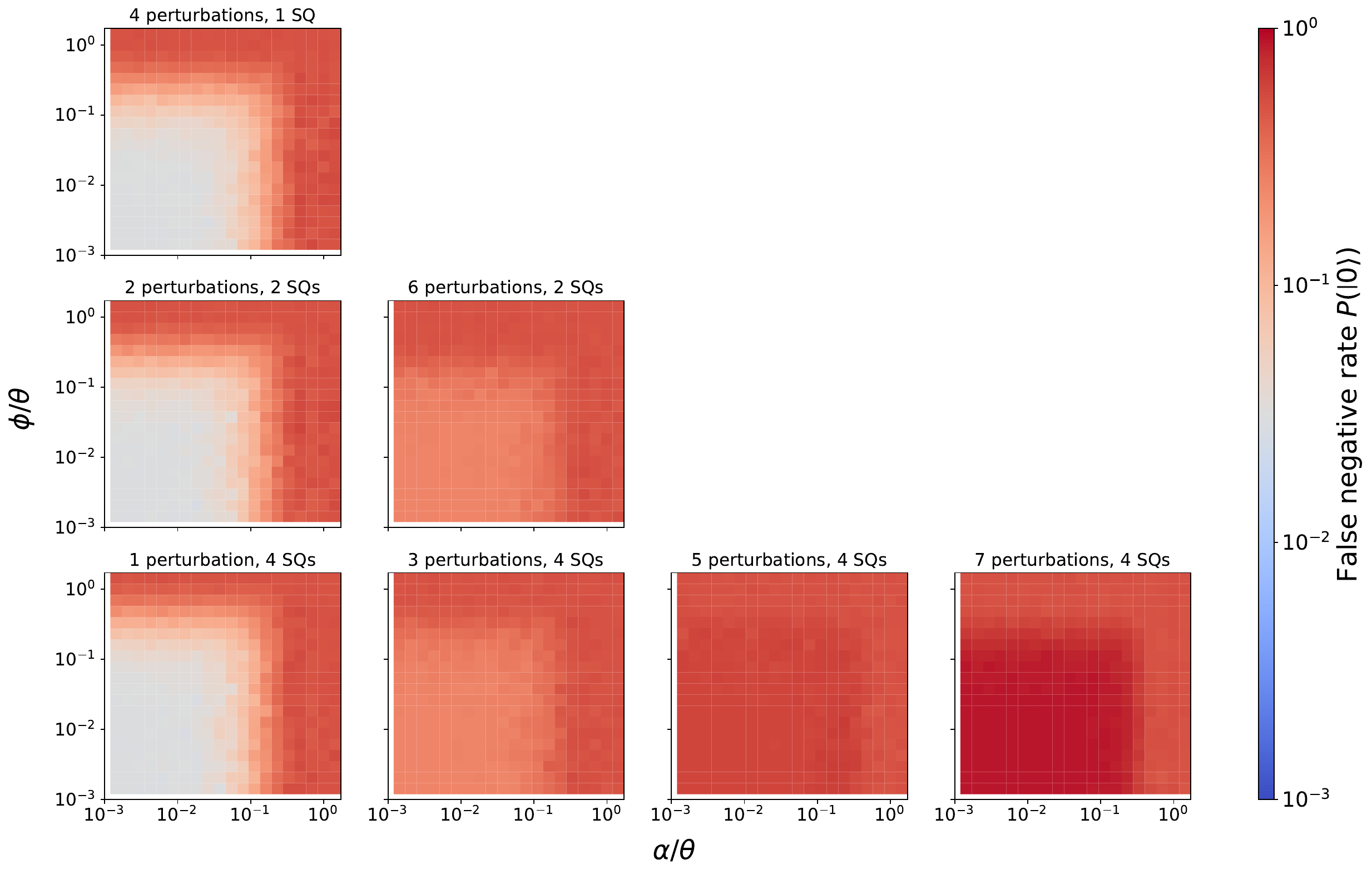}}
    \caption{Same plots as \Cref{fig: torino Xtalk pi4 6q colorplot}, \Cref{fig: sherbrooke Xtalk pi4 6q colorplot}, and \Cref{fig: pure Xtalk pi4 6q colorplot} but $\theta=\frac{2\pi}{9}$ as the angle satisfying the condition in \Cref{appendix: total angle mismatching}. The corresponding surface plots with more details in error bars are included in \cite{Sup_materials}.}
\end{figure*}

\begin{figure*}[hbtp]
    \centering
    \text{No dephasing and amplitude damping, $\theta=\frac{\pi}{8}$ with maximum of 8 spectator qubits}\par\medskip
    \includegraphics[width=0.9\linewidth]{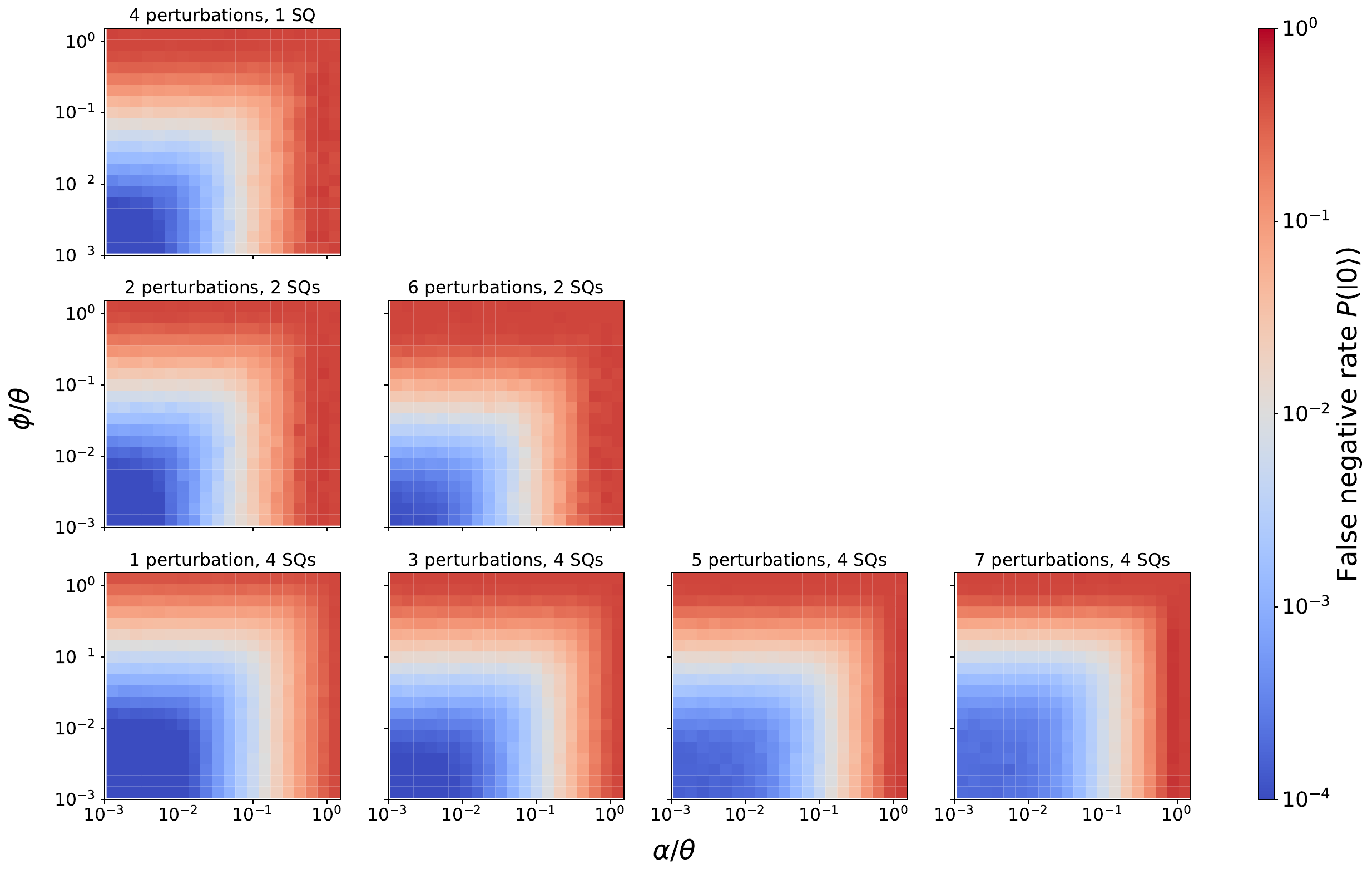}
    \caption{Same plot as \Cref{fig: pure Xtalk pi4 6q colorplot} but $\theta=\frac{\pi}{8}$ (with a maximum of 8 spectator qubits) as the angle satisfying the condition in \Cref{eq: total angle condition}. The corresponding surface plots with more details in error bars are included in \cite{Sup_materials}. }
    \label{fig: pure Xtalk pi8 8q colorplot}
    \vspace{-20pt}
\end{figure*}

\begin{figure*}[hbtp]
    \centering
    \text{No dephasing and amplitude damping, $\theta=\frac{2\pi}{17}$ with maximum of 8 spectator qubits}\par\medskip
    \includegraphics[width=0.95\linewidth]{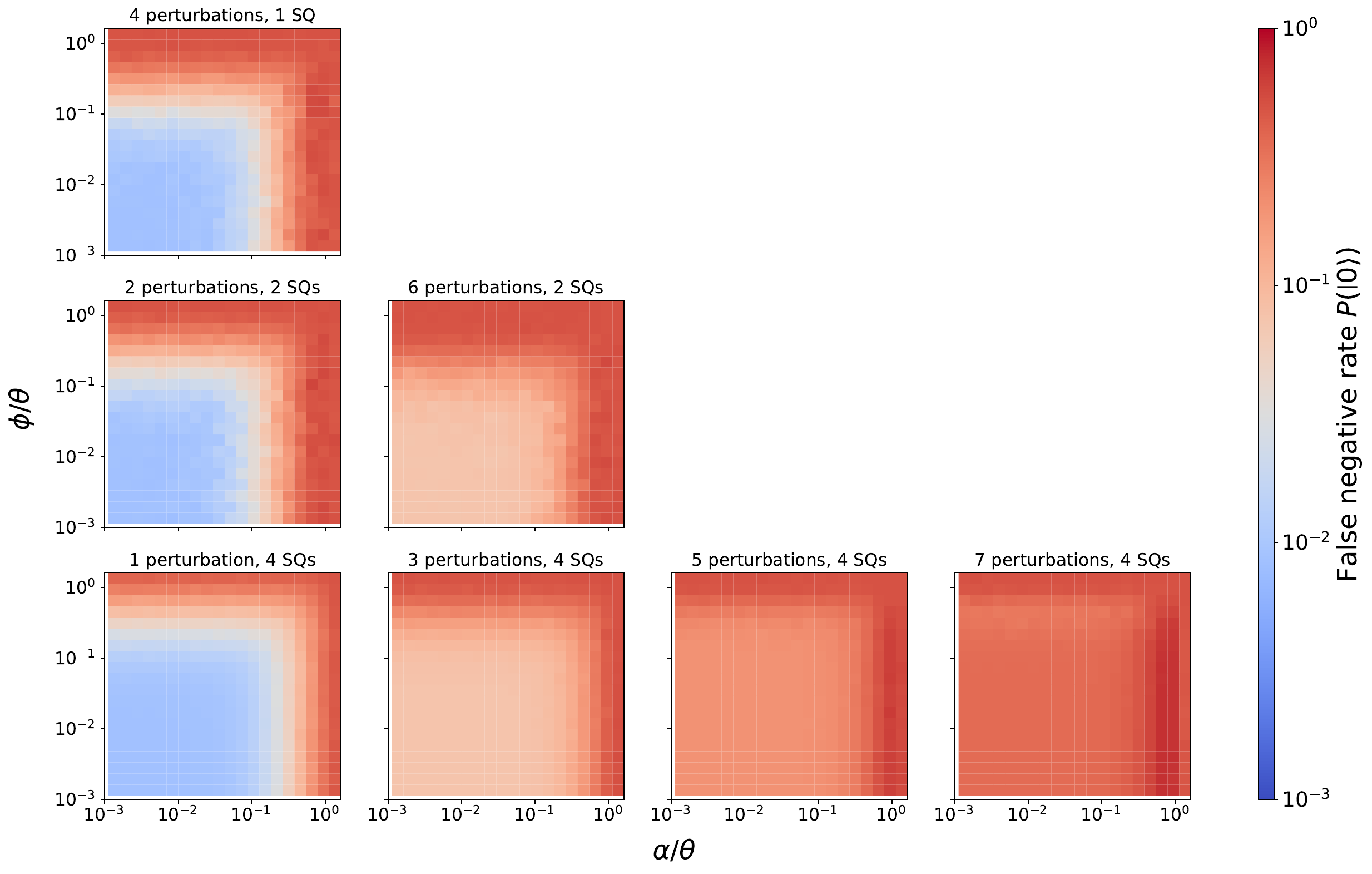}
    \caption{Same plot as \Cref{fig: pure Xtalk pi4 6q colorplot} but $\theta=\frac{2\pi}{17}$ (with a maximum of 8 spectator qubits) as the angle satisfying the worst-case scenario condition in \Cref{appendix: total angle mismatching}. The corresponding surface plots with more details in error bars are included in \cite{Sup_materials}. }
    \label{fig: pure Xtalk 2pi17 8q colorplot}
\end{figure*}

\begin{figure*}[hbtp]
    \centering
    \text{}\par\medskip
    \includegraphics[width=0.5\linewidth]{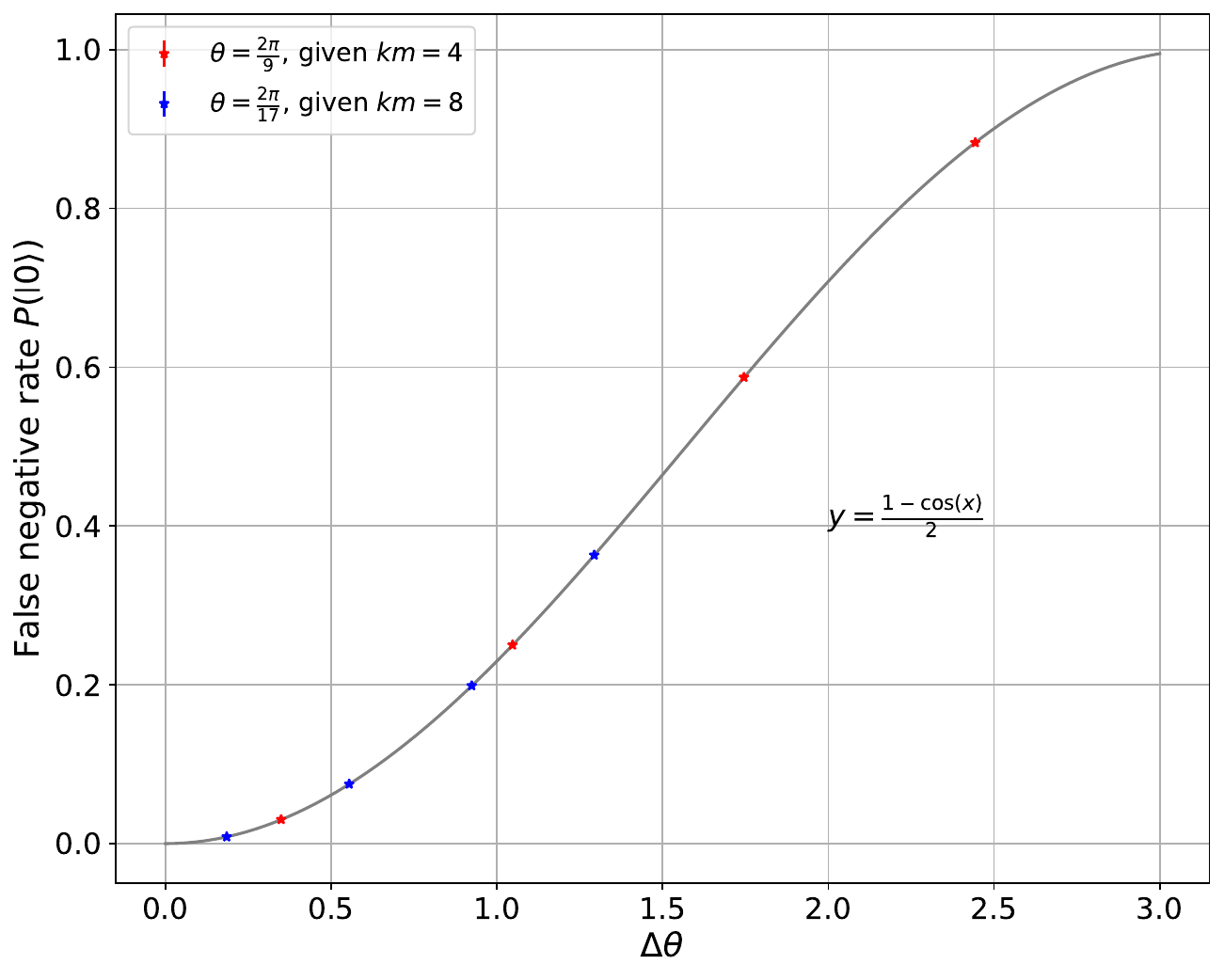}
    \caption{Mean false negative rate in the flat region (ratio for two-qubit crosstalk rate and ZZ coupling strength < $10^{-2}$) versus the deviation from the ideal phase in the worst case choice of $\theta$. Each point corresponds to the average of the false negative rate over all the data that has the same amount of deviation from the ideal phase satisfying \Cref{eq: total deviation from ideal phase}. No dephasing or amplitude-damping noise is involved.  Since the amount of deviation is inversely proportional to the number of spectator qubits, the robustness to angle mismatching increases as the number of spectator qubits increases.}
    \label{fig: detection failure rate plat vs mk}
\end{figure*}
\clearpage
\renewcommand{\href}[2]{#2}
\bibliography{main}
\end{document}